\documentclass{aa}  
\usepackage{textcomp}
\usepackage[utf8]{inputenc}
\usepackage{lscape}
\usepackage{lastpage}
\usepackage{booktabs}
\usepackage{amsmath,amssymb}
\usepackage{mathrsfs}	
\usepackage{multicol}
\usepackage{multirow}
\usepackage{afterpage}
\usepackage{graphicx}
\usepackage{txfonts}
\usepackage{hyperref}
\hypersetup{
    colorlinks   = true, 
    urlcolor     = cyan, 
    linkcolor    = blue, 
    citecolor   = blue 
}
\usepackage{placeins}
\usepackage{xcolor}
\usepackage{caption}
\usepackage{booktabs}
\usepackage[table]{xcolor}

\graphicspath{plots}

{\newif\ifnotend
\notendtrue
\def\veclist{ABCDEFGHIJKLMNOPQRSTUVWXYZabcdefghijklmnopqrstuvwxyz.}
\def\top#1#2.{#1}
\def\tail#1#2.{#2.}
\loop\expandafter\xdef\csname bb\expandafter\top\veclist\endcsname%
{{\noexpand\bf\expandafter\top\veclist}}
\edef\veclist{\expandafter\tail\veclist}
\if\veclist.\notendfalse\fi\ifnotend\repeat}

\newcommand{\kpc}   {\,{\rm kpc}}
\newcommand{\pc}    {\,{\rm pc}}

\newcommand{\Msun}  {\,M_{\odot}}
\newcommand{\Gyr}   {\,{\rm Gyr}}

\newcommand{\mas}    {\,{\rm mas}}
\newcommand{\yr}    {\,{\rm yr}}
\newcommand{\kms}   {\,{\rm km\,s^{-1}}}
\newcommand{\asec}  {\,{\rm arcsec}}
\newcommand{\amin}  {\,{\rm arcmin}}
\newcommand{\cm}    {\,{\rm cm}}

\newcommand{\GeV}   {\,{\rm GeV}}

\newcommand{\JJ}    {\boldsymbol{J}}
\newcommand{\Jr}    {J_r}
\newcommand{\Jphi}  {J_{\phi}}
\newcommand{\Jz}    {J_z}
\newcommand{\Lz}    {L_z}

\newcommand{\FF}    {\mathcal{F}}
\newcommand{\FJ}    {\mathcal{F}({\rm [Fe/H]},{\boldsymbol J})}
\newcommand{\FJMR}    {\mathcal{F}_{\rm MR}({\rm [Fe/H]},{\boldsymbol J})}
\newcommand{\FJMP}    {\mathcal{F}_{\rm MP}({\rm [Fe/H]},{\boldsymbol J})}
\newcommand{\FJi}    {\mathcal{F}_i({\rm [Fe/H]},{\boldsymbol J})}
\newcommand{\fJ}    {f({\boldsymbol J})}
\newcommand{\fJi}   {f_i({\boldsymbol J})}

\newcommand{\fsti}  {f_{\star,i}}

\newcommand{\hJi}     {h_i({\boldsymbol J})}

\newcommand{\gJi}     {g_i({\boldsymbol J})}

\newcommand{\Mst}     {M_\star}

\newcommand{\Mdm}     {M_{\rm DM}}

\newcommand{\Jsti}    {J_{\star,i}}
\newcommand{\JstMR}   {J_{\star,{\rm MR}}}
\newcommand{\JstMP}   {J_{\star,{\rm MP}}}

\newcommand{\Jtsti}   {J_{{\rm t},i}}

\newcommand{\Jcsti}   {J_{{\rm c},i}}
\newcommand{\JcstMR}  {J_{{\rm c},{\rm MR}}}
\newcommand{\JcstMP}  {J_{{\rm c},{\rm MP}}}

\newcommand{\hri}    {h_{r,i}}
\newcommand{\hrMR}   {h_{r,{\rm MR}}}
\newcommand{\hrMP}   {h_{r,{\rm MP}}}

\newcommand{\hphii}  {h_{\phi,i}}

\newcommand{\hzi}    {h_{z,i}}
\newcommand{\hzMR}   {h_{z,{\rm MR}}}
\newcommand{\hzMP}   {h_{z,{\rm MP}}}

\newcommand{\gri}    {g_{r,i}}
\newcommand{\grMR}   {g_{r,{\rm MR}}}
\newcommand{\grMP}   {g_{r,{\rm MP}}}

\newcommand{\gphii}  {g_{\phi,i}}

\newcommand{\gzi}    {g_{z,i}}
\newcommand{\gzMR}   {g_{z,{\rm MR}}}
\newcommand{\gzMP}   {g_{z,{\rm MP}}}

\newcommand{\etasti}   {\eta_{i}}
\newcommand{\etastMR}  {\eta_{\rm MR}}
\newcommand{\etastMP}  {\eta_{\rm MP}}

\newcommand{\Bi}   {B_i}
\newcommand{\BMR}  {B_{\rm MR}}
\newcommand{\BMP}  {B_{\rm MP}}

\newcommand{\Gammasti}   {\Gamma_{i}}
\newcommand{\GammastMR}  {\Gamma_{\rm MR}}
\newcommand{\GammastMP}  {\Gamma_{\rm MP}}

\newcommand{\wMR}  {w_{\rm MR}}
\newcommand{\wMP}  {w_{\rm MP}}

\newcommand{\Phitot}{\Phi_{\rm tot}}

\newcommand{\Phidm} {\Phi_{\rm DM}}
\newcommand{\Phibh} {\Phi_{\rm BH}}

\newcommand{\vsys}  {v_{\rm sys}}
\newcommand{\vlos}  {v_{\rm los}}
\newcommand{\vlosj}  {v_{{\rm los},j}}

\newcommand{\Reff}  {R_{\rm eff}}
\newcommand{\ReffMR}{R_{\rm eff, MR}}
\newcommand{\ReffMP}{R_{\rm eff, MP}}

\newcommand{\rhos}  {\rho_s}
\newcommand{\rs}    {r_s}

\newcommand{\rhodm} {\rho_{\rm DM}}
\newcommand{\rhodmcl}{\rho_{\rm DM, 150\,pc}}

\newcommand{\dd}    {{\rm d}}

\renewcommand{\log}{\mathop{\mathrm{log}}\nolimits_{10}}

\newcommand{\eeff}   {e_{\rm eff}}
\newcommand{\eeffMR} {e_{\rm eff, MR}}
\newcommand{\eeffMP} {e_{\rm eff, MP}}
\newcommand{\qeff}   {q_{\rm eff}}

\newcommand{\alim}   {a_{\rm min}}

\newcommand{\betaz}  {\beta_z}
\newcommand{\betazMP}{\beta_{z, {\rm MP}}}
\newcommand{\betazMR}{\beta_{z, {\rm MR}}}

\newcommand{\mura}  {\mu_\alpha}
\newcommand{\mudec} {\mu_\delta}
\newcommand{\rhoad} {\rho_{\alpha,\delta}}
\newcommand{\bprp}  {G_{\rm BP}-G_{\rm RP}}
\newcommand{\col}   {G-G_{\rm RP}}
\newcommand{\colo}  {(G-G_{\rm RP})_0}
\newcommand{\Gmag}  {G}
\newcommand{\Gmago} {G_0}
\newcommand{\ebmv}  {E(B-V)}

\newcommand{\Rmax}  {R_{\rm max}}
\newcommand{\DD}    {\mathcal{D}}

\newcommand{\PP}    {\mathcal{P}}
\newcommand{\PPi}   {\mathcal{P}^{\,i}}
\newcommand{\PPdsph} {\mathcal{P}^{\rm dSph}}
\newcommand{\PPC}  {\mathcal{P}^{\rm C}}

\newcommand{\BH}{M_{\rm BH}}
\newcommand{\roi}{r_{\rm roi}}

\newcommand{\logJ}{\log J_{0.5^{\circ}}}
\newcommand{\logD}{\log D_{0.5^{\circ}}}

\newcommand{\PPxyi}     {\mathcal{P}^{\,i}_{\Sigma}}
\newcommand{\PPxydsph}   {\mathcal{P}^{\rm dSph}_{\Sigma}}
\newcommand{\PPxyBG}    {\mathcal{P}^{\rm C}_{\Sigma}}

\newcommand{\PPmuprli}   {\mathcal{P}^{\,i}_{\mu\varpi}}
\newcommand{\PPmuprldsph} {\mathcal{P}^{{\rm dSph}}_{\mu\varpi}}

\newcommand{\PPmuprlBG}  {\mathcal{P}^{{\rm C}}_{\mu\varpi}}

\newcommand{\PPcmdsum}{\mathcal{P}_{\rm CMD}^{\rm dSph + C}}
\newcommand{\PPcmdi}     {\mathcal{P}^{\,i}_{\rm CMD}}

\newcommand{\PPcmdBG}    {\mathcal{P}^{\rm C}_{\rm CMD}}
\newcommand{\PPcmddsph}   {\mathcal{P}^{\rm dSph}_{\rm CMD}}

\newcommand{\GBP}  {G_{\rm BP}}
\newcommand{\GRP}  {G_{\rm RP}}

\newcommand{\GBPo} {\sigma_{G_{\rm BP},0}}
\newcommand{\GRPo} {\sigma_{G_{\rm RP},0}}
\newcommand{\Go}   {\sigma_{G_0}}

\newcommand{\FGBP} {\epsilon_{G_{\rm BP}}}
\newcommand{\FGRP} {\epsilon_{G_{\rm RP}}}
\newcommand{\FG}   {\epsilon_G}

\newcommand{\GGmuprli}    {\mathcal{N}^{\,i}_{\mu,\varpi}}
\newcommand{\GGmuprldsph}  {\mathcal{N}^{\rm dSph}_{\mu,\varpi}}
\newcommand{\GGmuprlbgi}  {\mathcal{N}^{\rm C1}_{\mu,\varpi}}
\newcommand{\GGmuprlbgii} {\mathcal{N}^{{\rm C2}}_{\mu,\varpi}}

\newcommand{\GGdSph}{\mathcal{N}^{\rm dSph}}
\newcommand{\GGBGi} {\mathcal{N}^{\rm C1}}
\newcommand{\GGBGii}{\mathcal{N}^{\rm C2}}
\newcommand{\NNi}{\mathcal{N}^i}
\newcommand{\vlosi}{v_{{\rm los}}^i}

\newcommand{\fehi}{\overline{{\rm [Fe/H]}}^i}
\newcommand{\sigmafehi}{\sigma^i_{\rm [Fe/H]}}
\newcommand{\sigmalosi}{\sigma^i_{\rm los}}

\newcommand{\mdfi}    {\varphi_i}
\newcommand{\MM}    {\mathcal{M}}
\newcommand{\MMi}   {\mathcal{M}_i}
\newcommand{\MMMR}  {\mathcal{M}_{\rm MR}}
\newcommand{\MMMP}  {\mathcal{M}_{\rm MP}}
\newcommand{\sigmai}   {\sigma_i}
\newcommand{\sigmaMR}  {\sigma_{\rm MR}}
\newcommand{\sigmaMP}  {\sigma_{\rm MP}}
\newcommand{\feh}   {{\rm [Fe/H]}}
\newcommand{\fehj}   {{\rm [Fe/H]}_j}

\newcommand{\vlosdsph}{v_{\rm los}^{\rm dSph}}
\newcommand{\vlosbgi} {v_{\rm los}^{\rm C1}}
\newcommand{\vlosbgii}{v_{\rm los}^{\rm C2}}
\newcommand{\svlosdsph}{\sigma_{\rm sys}^{\rm dSph}}
\newcommand{\slosbgi} {\sigma_{\rm los}^{\rm C1}}
\newcommand{\slosbgii}{\sigma_{\rm los}^{\rm C2}}
\newcommand{\sloso}{\sigma_{{\rm los}, 0}}
\newcommand{\slosoMR}{\sigma_{{\rm los}, 0}^{\rm MR}}
\newcommand{\slosoMP}{\sigma_{{\rm los}, 0}^{\rm MP}}
\newcommand{\slosot}{\sigma_{{\rm los}, 0}^{\rm tot}}
\newcommand{\vrotmax}{ v_{\mathrm{rot,max}}}

\newcommand{\fehdsph}{\overline{{\rm [Fe/H]}}^{\rm dSph}}
\newcommand{\sfehdsph}{\sigma_{\rm [Fe/H]}^{{\rm dSph}}}
\newcommand{\fehbgi}{\overline{{\rm [Fe/H]}}^{\rm C1}}
\newcommand{\sfehbgi}{\sigma_{\rm [Fe/H]}^{{\rm C1}}}
\newcommand{\fehbgii}{\overline{{\rm [Fe/H]}}^{\rm C2}}
\newcommand{\sfehbgii}{\sigma_{\rm [Fe/H]}^{{\rm C2}}}

\newcommand{\xj}    {x_j}
\newcommand{\yj}    {y_j}
\newcommand{\muraj} {\mu_{\alpha, j}}
\newcommand{\mudecj}{\mu_{\delta, j}}
\newcommand{\rhoadj}{\rho_{\alpha, \delta, j}}
\newcommand{\varpij}{\varpi_j}
\newcommand{\Gmagoj}{G_{0,j}}
\newcommand{\coloj} {(G - G_{\rm BR})_{0,j}}

\newcommand{\Nsamp} {N_{\rm samp}}
\newcommand{\Mvlos} {M_{\rm vlos}}
\newcommand{\Mfeh}  {M_{\rm [Fe/H]}}
\newcommand{\btheta}{\boldsymbol{\theta}}
\newcommand{\wii}   {w_1}
\newcommand{\wiii}  {w_2}
\newcommand{\xo}    {x_0}
\newcommand{\yo}    {y_0}

\newcommand{\murai}       {\mu^i_{\alpha}}
\newcommand{\muradsph}     {\mu_{\alpha}^{\rm dSph}}
\newcommand{\murabgi}     {\mu_{\alpha}^{\rm C1}}
\newcommand{\murabgii}    {\mu_{\alpha}^{\rm C2}}

\newcommand{\smurai}      {\sigma^{\,i}_{\mu_{\alpha}}}
\newcommand{\smuradsph}    {\sigma_{\mu_{\alpha}}^{\rm dSph}}
\newcommand{\smurabgi}    {\sigma_{\mu_{\alpha}}^{\rm C1}}
\newcommand{\smurabgii}   {\sigma_{\mu_{\alpha}}^{\rm C2}}

\newcommand{\mudeci}      {\mu^i_{\delta}}
\newcommand{\mudecdsph}    {\mu_{\delta}^{\rm dSph}}
\newcommand{\mudecbgi}    {\mu_{\delta}^{\rm C1}}
\newcommand{\mudecbgii}   {\mu_{\delta}^{\rm C2}}

\newcommand{\smudeci}     {\sigma^{\,i}_{\mu_{\delta}}}
\newcommand{\smudecdsph}   {\sigma_{\mu_{\delta}}^{\rm dSph}}
\newcommand{\smudecbgi}   {\sigma_{\mu_{\delta}}^{\rm BG1}}
\newcommand{\smudecbgii}  {\sigma_{\mu_{\delta}}^{\rm BG2}}

\newcommand{\prli}        {\varpi^i}
\newcommand{\prldsph}     {\varpi^{\rm dSph}}
\newcommand{\prlbgi}      {\varpi^{\rm C1}}
\newcommand{\prlbgii}     {\varpi^{\rm C2}}

\newcommand{\sprli}       {\sigma^{\,i}_{\varpi}}
\newcommand{\sprldsph}     {\sigma_{\varpi}^{\rm dSph}}
\newcommand{\sprlbgi}     {\sigma_{\varpi}^{\rm C1}}
\newcommand{\sprlbgii}    {\sigma_{\varpi}^{\rm C2}}

\newcommand{\rhoadi}      {\rho^{\,i}_{\alpha,\delta}}
\newcommand{\rhoaddsph}    {\rho_{\alpha,\delta}^{\rm dSph}}
\newcommand{\rhoadbgi}    {\rho_{\alpha,\delta}^{\rm C1}}
\newcommand{\rhoadbgii}   {\rho_{\alpha,\delta}^{\rm C2}}

\newcommand{\bSigmai}     {\boldsymbol{\Sigma}^{i}}

\newcommand{\bxi}         {{\boldsymbol\xi}}
\newcommand{\bmu}         {\boldsymbol{\mu}}
\newcommand{\mui}         {\boldsymbol{\mu}^i}

\newcommand{\Pmembj}{P_{\mathrm{memb},j}}

\newcommand{\LL}{\mathcal{L}}
\newcommand{\EE}{\mathcal{E}}

\defcitealias{Walker2023}{W23}
\defcitealias{Geha2026}{G26}

\begin{document} 

\title{Multi-component, axisymmetric dynamical models of dSphs based on distribution functions: inferences on dark matter and intermediate-mass black holes in Draco and Ursa Minor}
\titlerunning{Axisymmetric models of Draco and Ursa Minor}

\author{
R. Pascale\inst{1} \thanks{\email{raffaele.pascale@inaf.it}},
G. Battaglia \inst{2,3}, J.M. Arroyo-Polonio \inst{2,3}, E. Vasiliev\inst{4}, C. Nipoti \inst{5}, G. F. Thomas\inst{3,2}}
\institute{
INAF - Osservatorio di Astrofisica e Scienza dello Spazio di Bologna, via Gobetti 93/3, 40129 Bologna, Italy 
\and
Instituto de Astrofísica de Canarias, Calle Vía Láctea s/n E-38206, La Laguna, Santa Cruz de Tenerife, Spain.
\and
Universidad de La Laguna, Avda. Astrofísico Francisco Sánchez E-38205, La Laguna, Santa Cruz de Tenerife, Spain.
\and 
University of Surrey, Guildford, GU2 7XH, United Kingdom
\and
Dipartimento di Fisica e Astronomia “Augusto Righi”, Università di Bologna, via Gobetti 93/2, Bologna, I-40129, Italy
}
\authorrunning{Pascale et al.}
\date{Received ...; accepted ...}
 
\abstract
{Dwarf spheroidal galaxies (dSphs) are prime laboratories for studying dark matter (DM) and the black hole demographics in the low-mass regime. These systems are also often significantly flattened; nevertheless, most studies have relied on spherical models, which may affect dynamical inferences. In this work we introduce the first multi-component, axisymmetric dynamical models of dSphs based on analytic distribution functions (DFs) and apply them to the Milky Way dSphs Draco and Ursa Minor. The stellar distribution is described by chemo-dynamically distinct axisymmetric populations tracing a common spherical gravitational potential generated by a dominant DM halo and a central intermediate-mass black hole (IMBH). The models are fitted to discrete stellar data from an astrometrically selected sample of stellar coordinates based on Gaia, plus either of the two different spectroscopic datasets providing line-of-sight velocities and metallicities. This approach allows us to test the robustness of our inferences against the use of different kinematic samples. We compare the inferred DM properties under different modelling assumptions, including flattened one-component and spherical two-component models. We find that both galaxies are better described by two stellar populations with distinct chemo-dynamical properties: a metal-rich component that is kinematically colder and more centrally concentrated, and a more extended metal-poor component characterised by hotter kinematics. We detect weak signatures of rotation in both galaxies, which are dynamically unimportant and are ignored in the models.
We measure a cuspy DM density profile in Draco ($\gamma=0.98_{-0.26}^{+0.28}$), while Ursa Minor is consistent with a more cored distribution ($\gamma=0.37_{-0.24}^{+0.31}$). We find that the DM profile of Draco remains stable across all models and datasets, making its DM halo the most robustly determined among Local Group dSphs and extremely relevant in the context of indirect DM searches. Also, we show that modelling intrinsically flattened systems with spherical models can bias the DM inner slope towards cuspier values, while we find no degeneracy between inner halo density and galaxy inclination. Finally, we find no evidence for IMBHs and we place upper limits on their masses, $\log\BH[\Msun] < 5.2$ for Draco and $< 3.33$ for Ursa Minor (95\% confidence), with the latter providing a particularly stringent constraint.}



\maketitle

\section{Introduction}
\label{sec:intro}

Galaxies on the smallest scales play a key role in modern cosmology and constitute the regime in which the $\Lambda$ Cold Dark Matter ($\Lambda$CDM) paradigm is most severely tested. Indeed, although the $\Lambda$CDM model provides a highly successful description of the Universe on large scales, its predictions for the detailed properties of individual low-mass galaxies remain uncertain and are difficult to reconcile with observations. These uncertainties concern the abundance and diversity of dwarf galaxies, the physical processes governing their formation and evolution, and the internal structure of their dark matter (DM) halos \citep{Sales2022}. 

A particularly long-standing small-scale tension concerns the shape of the inner DM halo, with collisionless cosmological simulations predicting density profiles with central cusps  \citep{Dubinski1991,NavarroFrenkWhite1996,Navarro1997} and observations of low-mass galaxies often indicating cored profiles from neutral gas rotation curves \citep{deBlok2008,Read2017,Oh2015,Relatores2019}. This apparent discrepancy, commonly referred to as the core--cusp problem \citep{deBlok2009,Bullock2017,DelPopolo2022}, has motivated extensive theoretical work. Proposed solutions include baryonic processes capable of reshaping the inner halo through repeated potential fluctuations driven by star formation and/or supernova feedback \citep{Pontzen2012,Oh2011,Governato2012,Onorbe2015,Muni2025}, mergers \citep{Orkney2021}, the transfer of energy to DM particles via dynamical friction \citep{Goerdt2010,NipotiBinney2015}, as well as alternative DM models, such as self-interacting DM \citep{Elbert2015,Kaplinghat2016}, fuzzy DM \citep{Hui2017}, or DM with non-negligible thermal velocities.

Another major open question on small scales concerns the formation and demographics of black holes (BHs) in the intermediate-mass regime \citep{Greene2020}, i.e. BHs with masses in the range $\simeq10^2$–$10^5\Msun$. These intermediate-mass black holes (IMBHs) are of particular interest for two main reasons: (i) the growing observational evidence for super massive black holes (SMBHs) with masses $\gtrsim 10^{9}\Msun$ already assembled by $z \gtrsim 7$ \citep[][e.g.]{Banados2018,Kokorev2024,Bogdan2024,Akins2025} strongly suggests that a fraction of the SMBH population must have originated from comparatively massive seeds \citep{Inayoshi2022}; and (ii) BH formation models predict that IMBH should reside in present-day dwarf galaxies. Contrary to stellar-mass BHs and SMBHs, whose existence and properties are now firmly established observationally, IMBHs remain poorly constrained, despite theoretical predictions of multiple formation channels from massive seeds. These include remnants of massive Population~III stars \citep{MadauRees2001,Volonteri2003,Smith2018,Cammelli2025}, runaway stellar collisions in dense stellar systems \citep{PortegiesZwart2004,Giersz2015}, and the direct collapse of metal-poor gas \citep{Bromm2003,Lodato2007,Mayer2015,DiMatteo2017}.

One of the most powerful diagnostics for discriminating BH seed formation models is the present day BH occupation fraction -- the fraction of galaxies that host a central BH -- as a function of stellar mass  \citep{MadauRees2001,Volonteri2008,Volonteri2009,vanWassenhove2010,Greene2012,Miller2015,Greene2020}. It is predicted to remain close to unity at high masses and to decline rapidly below $\Mst \simeq 10^9\Msun$ \citep[e.g.][]{Burke2025}, with both the characteristic mass scale and the steepness of the transition depending sensitively on the underlying BH formation mechanism.

Milky Way (MW) dwarf spheroidal galaxies (dSphs) provide a uniquely clean and accessible environment in which the evolution and internal structure of low-mass galaxies can be studied. First and foremost, dSphs are strongly dominated by DM \citep{IrwinHatzidimitriou1995,Mateo1998,Tolstoy2009,Hayashi2020,BattagliaNipoti2022}, making them ideal laboratories for probing its properties. In addition, their nearness and lack of significant emission outside the optical renders them prime targets for indirect DM searches through gamma-ray emission \citep[e.g.][]{Albert2020,Abe2024,Boddy2024}. Their proximity allows individual stars to be resolved and their kinematics to be measured on a star-by-star basis (\citealt{Battaglia2006,Walker2009a,Walker2015}; \citealt[][hereafter W23]{Walker2023}; \citealt{Tolstoy2023,Tolstoy2025}; \citealt[][hereafter G26]{Geha2026}), providing finely sampled and potentially strong constraints on the underlying gravitational potential. Finally, they populate the stellar mass regime and the region of the stellar-to-halo mass relation where predictions for the BH occupation fraction, and for cuspy versus cored DM profiles within the $\Lambda$CDM, framework differ most significantly \citep[e.g.][]{DiCintio2014a,Fitts2017,Hopkins2018,Tollet2016}.

However, deriving robust dynamical constraints from dSphs remains challenging, owing to both observational uncertainties (e.g. foreground contamination and limited kinematic information) and limitations of commonly adopted dynamical modelling techniques. Existing dynamical studies of dSphs can be broadly divided into three main approaches: solving Jeans equation, orbit-superposition, and distribution-function (DF) modelling. Historically, Jeans-based models were the first to be widely implemented, largely because of their relative simplicity and low computational cost. Early applications often relied on binned kinematic data \citep{Lokas2002,Walker2009b}, reducing the effective phase-space information, the use of only the second-order velocity moment, and typically assumed spherical symmetry. Moreover, Jeans models are not guaranteed to be fully physical, as they do not rely on an underlying DF and the anisotropy is not necessarily tied to it. Nevertheless, significant efforts have been put to improve these methods, including the modelling of multiple chemo-kinematic populations \citep[e.g.][]{Battaglia2008,Zhu2016}, the inclusion of higher-order velocity moments \citep[e.g.][]{Merrifield1990,Lokas2005,Richardson2013,Read2017b,Read2019,Wardana2025}, the use of discrete kinematic data \citep{Watkins2013,Zhu2016,Banares2025,Wardana2025}, and the extension to non-spherical models \citep[specifically axisymmetric, e.g.][]{Hayashi2020}, as well as combinations of these approaches \citep{Vitral2024,Yang2025}, making modern Jeans-based methods highly competitive. Orbit-superposition techniques \citep{Schwarzschild1979} provide a physically consistent description by construction \citep{BreddelsHelmi2013} and can be extended to non-spherical geometries \citep{Jardel2012}. However, their high computational cost generally limits a comprehensive exploration of parameter space. DF-based models represent the most physically complete framework, as they ensure phase-space consistency by construction and allow for a natural treatment of observational uncertainties, selection functions, discrete data and multiple populations \citep{Pascale2018,Arroyo2025}. 
Nevertheless, all existing DF-based applications to dSphs remained restricted to spherical symmetry, which can be a severe limitation given the significant flattening of these systems \citep{Munoz2018}. Across the existing literature, different studies have focused on distinct components of dSphs, ranging from the DM distribution to multiple stellar populations or the presence of a central IMBH, but rarely within a single modelling framework that simultaneously 
accounts for all these aspects and incorporates their flattened morphology.

In this work, we introduce axisymmetric dynamical models based on analytic DFs to describe the internal dynamics of dSphs, operating directly on unbinned stellar chemo-kinematic data. The models incorporate multiple, flattened stellar populations as distinct tracers of a common spherical gravitational potential, which includes a dominant DM halo and the possible presence of a central IMBH. By exploiting the full phase-space information of bona fide member stars, the framework naturally accounts for observational uncertainties and allows the velocity anisotropy to emerge self-consistently from the DF, rather than being assumed. As far as we know, this represents the first application of a fully self-consistent DF-based dynamical framework, tailored for asymmetric systems, that simultaneously combines multiple stellar populations, a central IMBH, and a DM halo while operating on discrete stellar data. As such, our approach provides one of the most comprehensive chemo-kinematic description of a low-mass galaxy. We demonstrate its capabilities by applying it to the local dSphs galaxies Draco and Ursa Minor.

Draco and Ursa Minor are classical dSphs, satellites of the MW. They lie at comparable heliocentric distances, $\simeq 76\kpc$ \citep{Munoz2018}, and have stellar masses of $\simeq 2.9 \times 10^5\Msun$ \citep{McConnachie2012}. Recent studies show that both systems host multiple stellar populations with distinct kinematics \citep[e.g.][]{Pace2020,Yang2025} and are predominantly composed of stars spanning a metallicity range of $\feh \simeq -3$ to $-1$ dex (see also \citetalias{Walker2023}). Their star formation histories indicate that the bulk of their stellar mass formed at early times, followed by rapid quenching and little to no evidence for recent star formation \citep{Aparicio2001,Carrera2002}. For all these reasons, Draco and Ursa Minor are particularly well suited for probing the inner galaxy potential. The early quenching of their star formation, combined with the low stellar mass, suggests a limited impact on the gravitational potential of baryonic feedback, which, in standard $\Lambda$CDM scenarios, is expected to flatten central density cusps into cores only in more massive systems. Likewise, their quiet merger histories and limited mass assembly imply that they have remained close to their conditions at early cosmic times, allowing their present-day BH demographics to put constraints on models of seed formation and early growth.


This paper is organised as follows. In Section~\ref{sec:mod} we introduce the adopted framework of DF-based dynamical models, while we leave for Appendices ~\ref{app:A}, ~\ref{sec:data} and \ref{app:B} the description of the photometric and spectroscopic datasets of bona fide member stars used in our analysis. The main results of the modelling are presented in Section~\ref{sec:results}. In Section~\ref{sec:disc} we critically discuss these results and place them in the broader context of DM and IMBH studies in dSphs. Finally, Section~\ref{sec:concl} summarises our conclusions.


\section{Models}
\label{sec:mod}

We rely on analytic DFs $\fJ$ that depend on the action integrals $\JJ$ \citep{Binney2014,Posti2015,ColeBinney2017,Pascale2019,Vasiliev2019} to construct physically motivated, flattened dynamical models of dSphs. Action-based DFs have been widely employed in recent years. They have been used to build detailed models of the MW \citep{ColeBinney2017,Das2016a,Das2016b,Binney2023,Binney2024}, as well as applied to dSphs, both to constrain the inner shape of their DM density profiles \citep{Pascale2018,Pascale2025,Arroyo2025} and to search for IMBH \citep{Pascale2019,Pascale2024a}. These methods have also been applied to GCs, particularly in the context of IMBH searches \citep{DellaCroce2023}, and to the Milky Way nuclear star cluster \citep{Vasiliev2026}. Moreover, beyond their application to real systems, their performance has been extensively tested against diverse sets of mock observations -- both for spherical \citep{Read2021} and for flattened systems \citep{Gherghinescu2024,Gherghinescu2026}, in the latter cases also using mock data from cosmological simulations. Taken together, these tests have demonstrated that action-based DFs are highly powerful and consistently perform well across different regimes.

We model a dSph as a system composed of two flattened stellar populations, treated as tracers of a common spherical gravitational potential generated by a dominant DM halo and a central IMBH. In Section~\ref{subsec:stars}, we describe the stellar DFs and explain how chemical information is incorporated. Section~\ref{subsec:dm} focuses on the DM halo, while Section~\ref{subsec:imbh} describes how the central IMBH is included in the models.
\subsection{Stellar populations}
\label{subsec:stars}

While classical DFs fully specify the dynamical equilibrium of a collisionless tracer population, they are unable to capture the chemical complexity revealed by modern spectroscopic data, such as the coexistence of multiple stellar populations with distinct metallicities. To incorporate this additional information, we model the stellar component of a dSph as the superposition of two distinct populations. For each population, we adopt an extended DF (eDF) defined over the augmented space $(\JJ,\feh)$, which assigns a joint probability density to actions and metallicity, [Fe/H] being a measure of the metallicity of a star\footnote{The formalism is general and can be applied to other metallicity indicators. The use of [Fe/H] is motivated by the characteristics of the spectroscopic datasets analysed in this work.}. In our implementation, the eDF of each component, $\FJi$, is modelled as the product of an action-based DF, $\fJi$, and a metallicity distribution function (MDF), $\mdfi$ \footnote{Alternative and more complex formulations of eDFs are possible, in which the parameters of the DF vary explicitly with the additional quantity being modelled. Such extensions may involve stellar properties other than metallicity, such as age or colour \citep[e.g.][]{Das2016a}.}. Formally
\begin{equation}\label{edf}
    \FJi = \mdfi(\feh)\fJi.
\end{equation}
In the equation above and in all those that follow, the index $i$ denotes any parameter or term associated with one of the two stellar populations considered here, a Metal-Rich (MR) and Metal-Poor (MP) ones, i.e., $i$=MR, MP. The total eDF of the target dSph is the weighted superposition of two stellar components
\begin{equation}\begin{split}\label{for:tot}
    \FJ = \wMR\, \FJMR + \wMP \FJMP,
\end{split}\end{equation}
where $\wMP$ and $\wMR$ are non-negative coefficients such that $\wMP+\wMR=1$. 

The MDF assigned to each population is a Gaussian
\begin{equation}\label{for:mdf}
    \mdfi\,(\feh) = \frac{1}{\sqrt{2\pi}\sigmai} \exp\biggl[-\frac{(\feh - \MMi)^2}{2\sigmai^2}\biggr],
\end{equation}
with $\MMi$ the mean metallicity of the population and $\sigmai$ its dispersion. The action-space DF of each population is given by \footnote{More general formulations may include exponential truncation and rotation \citep{Vasiliev2019}. We neglect truncation as it is unnecessary for the present data quality and defer the inclusion of rotation to future works.}
\begin{equation}\label{for:df}
\begin{split}
 \fJi = & \fsti\frac{\Mst}{(2\pi\Jsti)^3} \biggl[1 - \beta\frac{\Jcsti}{\hJi} + \biggl(\frac{\Jcsti}{\hJi}\biggr)^2\biggr]^{-\frac{\Gammasti}{2}} \times \\
 & \biggl[1 + \biggl(\frac{\Jsti}{\hJi}\biggr)^{\etasti}\biggr]^{\frac{\Gammasti}{\etasti}} \biggl[1 + \biggl(\frac{\gJi}{\Jsti}\biggr)^{\etasti}\biggr]^{-\frac{\Bi}{\etasti}} 
\end{split}
\end{equation} 
with
\begin{equation}\begin{split}\label{for:hgo}
& \hJi = \hri\Jr+\hphii|\Jphi|+\hzi\Jz, \\
& \gJi = \gri\Jr+\gphii|\Jphi|+\gzi\Jz.
\end{split}\end{equation} 
$\Jr$, $\Jz$, and $\Jphi\equiv\Lz$ are the radial, vertical, and azimuthal actions, respectively: $\Jr$ and $\Jz$ measure the amplitude radial and vertical oscillations, and $\Jphi$ (the third component of the angular momentum, $\Lz$) quantifies the azimuthal motion. 

The DF (\ref{for:df}) can describe a broad variety of density profiles, with varying velocity anisotropy. The factor $\fsti$ ensures that the DF is normalised to the total stellar mass, denoted by $\Mst$. The parameters $\Jcsti$, $\Jsti$ are characteristic action scales that govern the transitions between different regimes within the action space: i) when $|\JJ|\ll\Jcsti$ the DF reduces to a constant; ii) when $\Jcsti<|\JJ|<\Jsti$, the DF~(\ref{for:df}) depends on $\JJ$ thorough $\hJi$, and it behaves as a power-law $\hJi^{-\Gammasti}$; iii) similarly, when $\Jsti<|\JJ|<\Jtsti$ the DF~(\ref{for:df}) depends on $\JJ$ thorough $\gJi$, and it behaves as a power-law $\gJi^{-\Bi}$.
$\Gammasti$, $\Bi$, and $\etasti$ are dimensionless parameters that control the inner and outer slopes of the DF in the action space. The dimensionless parameters $\gri$ and $\hri$ are related to the model's velocity distributions. The DF~(\ref{for:df}) generates spherically symmetric models when the total gravitational potential is spherical and the DF depends on the actions only through the magnitude of the angular momentum $L=\Jz+|\Jphi|$. This condition is satisfied when $\hphii=\hzi$ and $\gphii=\gzi$ in equations~(\ref{for:hgo}). Departures from symmetry can be introduced by assigning different weights to the vertical and azimuthal actions.

Depending on the potential, the density distributions generated by the DF~(\ref{for:df}) exhibit similar behaviour in physical space. For instance, in one-component models in which the gravitational potential is self-consistently generated by the same DF, they can feature a constant-density core of variable size in the innermost regions, followed by a first power-law regime whose slope is controlled by $\Gammasti$, and a second power-law regime with slope set by $\Bi$, with the sharpness of the transition regulated by $\etasti$ \citep{Posti2015}. In addition, the stellar components display radially varying velocity anisotropy, transitioning from one regime in the inner regions to a different regime in the outer parts.

The dimensionless parameters $\hri, \hphii, \hzi, \gri, \gphii, \gzi$ must be positive. Additionally, we adopt the gauge convention introduced by \citet{Binney2014}, which fixes 
\begin{equation}\begin{split}
 \gri + \gphii + \gzi = 3 \,\,\,\,\text{and} \,\,\,\, \hri + \hphii + \hzi = 3. \\
\end{split} \end{equation}
Using this constraint, we rewrite equations.~(\ref{for:hgo}) as
\begin{equation}\begin{split}\label{for:hg}
&\hJi = \hri \Jr + (3 - \hri - \hzi) \Jphi + \hzi \Jz \,, \\ 
&\gJi = \gri \Jr + (3 - \gri - \gzi) \Jphi + \gzi \Jz \,.
\end{split}\end{equation}
As an example, requiring $\hzi>\hphii$ and $\gzi>\gphii$ penalizes vertical orbits since the DF~(\ref{for:df}) is a decreasing function of both $\hJi$ and $\gJi$ and thereby it makes the tracer population oblate. In our application we restrict to oblate stellar populations (see also Section~\ref{subsec:logl}).

It is well established that in dSphs, stars provide a negligible contribution to the total gravitational potential at all radii, exhibiting some of the highest mass-to-light ratios among Local Group dwarf galaxies \citep{BattagliaNipoti2022}. Therefore, we treat the two stellar populations as axisymmetric tracers of a spherical total gravitational potential. Operationally, this implies that the Poisson equation is not solved for the stellar components, and that the total potential is entirely specified by a spherical DM and IMBH potentials, as described in the following Sections. Since the stellar component does not contribute to the total potential, the total stellar mass, $\Mst$, does not play a role in the model and is therefore not included among the free parameters. Note, however, that what can be constrained is the fractional mass contribution of the MR and MP populations, through the parameters $\wMR$ and $\wMP$.


\begin{figure*}[h!]
    \centering
    \includegraphics[width=1\hsize]{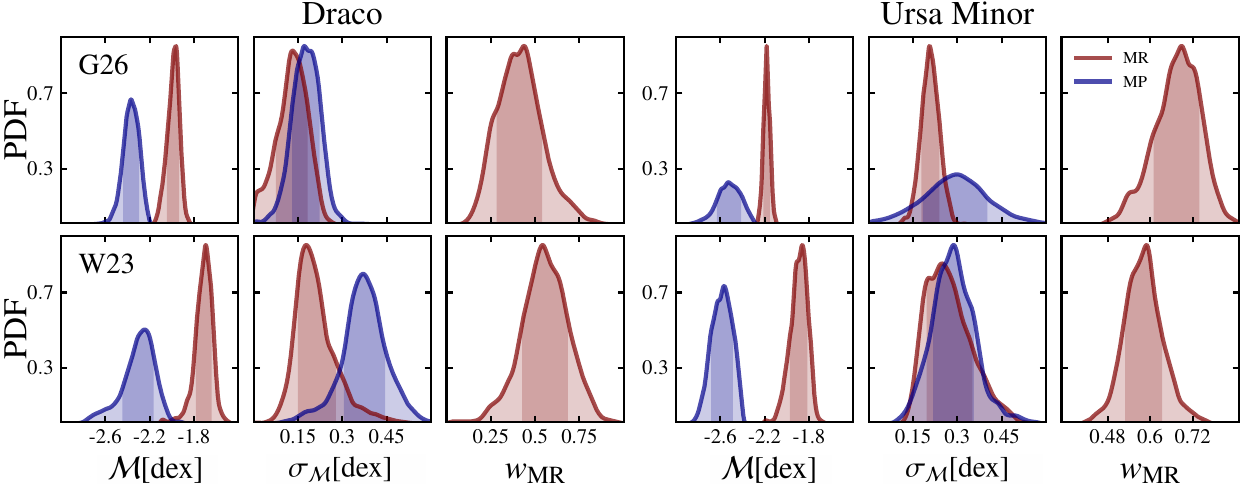} 
    \caption{Marginalised posterior distributions of the metallicity-related model parameters. The left set of panels refers to Draco, the right to Ursa Minor. Top upper panels show results from the G26 sample, while the bottom panels from the W23 sample. In each triplet, the left panel shows the one-dimensional marginalised posteriors of $\MMMP$ and $\MMMR$, the middle panel those of $\sigmaMP$ and $\sigmaMR$, and the right panel that of $\wMR$. The MP and MR population parameters are in blue and red, respectively, with the darker region indicating the $1\sigma$ credible region. In each panel, distributions are normalised to the maximum value among them.}
    \label{fig:metdis}
\end{figure*}

\subsection{DM}
\label{subsec:dm}

DM is modelled with an $\alpha\gamma\beta$ density profile, which provides a flexible parametrization capable of reproducing a wide range of density distributions. We restrict our analysis to spherical DM halos because, as shown by \cite{Gherghinescu2026}, fit to three-dimensional phase-space information alone (position on the plane of the sky and line-of-sight velocity), as in our case (see Section~\ref{subsec:logl}), does not allow the halo flattening to be constrained. The density profile is therefore given by

\begin{equation}\label{for:dm}
    \rhodm(r) = \rhos \left( \frac{r}{\rs} \right)^{-\gamma} \left[ 1 + \left( \frac{r}{\rs} \right)^{\alpha} \right]^{-\frac{\beta - \gamma}{\alpha}},
\end{equation}
where $r$ is the spherical distance from the halo centre, $\rhos$ is the characteristic density, $\rs$ is the scale radius, $\gamma$ represents the inner logarithmic slope, $\beta$ the outer slope, and $\alpha$ governs the sharpness of the transition between the two regimes. The profile asymptotically behaves as $\rhodm(r) \propto r^{-\gamma}$ for $r \ll \rs$ and $\rhodm(r) \propto r^{-\beta}$ for $r \gg \rs$. In the following, we adopt the total DM mass, $\Mdm$, as a free parameter instead of the characteristic density $\rhos$. This choice is physically motivated by the fact that we always impose an outer density slope $\beta > 3$.

\subsection{IMBH}
\label{subsec:imbh}

The IMBH is included as a Keplerian, fixed, central contribution to the total gravitational potential
\begin{equation}
    \Phitot = \Phidm + \Phibh.
\end{equation}
In the above equation, $\Phidm$ is the DM potential, and
\begin{equation}\label{for:bh}
    \Phibh = -\frac{G\BH}{r}
\end{equation}
where $G$ is the gravitational constant and $\BH$ denotes the IMBH mass, which is a free parameter of the model. In our analysis we will consider IMBH masses in the range $10^2\Msun$ -- $10^6\Msun$.

Because actions are adiabatic invariants, DFs expressed as $\fJ$ do not change if the gravitational potential evolves slowly with time. Models including a central IMBH constructed in this framework therefore can be interpreted as systems in which the BH has grown adiabatically, i.e. on a timescale longer than the dynamical time but shorter than the two-body relaxation time of the stellar system. This assumptions is likely appropriate for collisionless systems such as dSphs, which have very long relaxation times.
In spherical systems, adiabatic BH growth is expected to induce cuspy stellar profiles in the central regions \citep[e.g.][]{Young1980,Quinlan1995}, therefore we expect our models to develop shallow cusps in the innermost regions close to the IMBHs. This aspect is important in the context of our interpretations: the absence of a BH may manifest not only through the lack of high-velocity stars, but also through the absence of small-scale cusps in the central density profile. 

\subsection{The likelihood}
\label{subsec:logl}

In this Section we describe the likelihood function adopted to constrain the dynamical models. Details on the numerical implementation of the Markov Chain Monte Carlo (MCMC) sampling, and the way confidence intervals are computed are provided in Appendix~\ref{app:A}.

Here, we recall that for each galaxy we consider two independent datasets, each derived from a different spectroscopic sample. Each of these datasets, denoted by $\DD$, comprises $N$ Gaia-selected member stars, all with measured positions $(\xj, \yj)$ on the plane of the sky. For a subset of $\Mvlos$ stars, the line-of-sight velocity, $\vlosj$, is available and for a fraction of these $\Mvlos$ stars, metallicity, $\fehj$, is also measured (see Appendices~\ref{sec:data} and~\ref{app:B}). In Fig.~\ref{fig:datasets} we show the distribution of member stars of both galaxies used in this work on the colour-magnitude diagram (CMD), on the plane of the sky and in the metallicity-line-of-sight velocity plane, while Table~\ref{tab:samples} summarises the main characteristics of the resulting datasets. 

The likelihood is constructed by evaluating the total eDF~(\ref{for:tot}) at the observed metallicity and phase-space coordinates of each star, after marginalizing over missing quantities and convolving with the observational error distributions, assuming statistical independence between individual stellar measurements. Explicitly, the likelihood function is given by
\begin{equation}\label{for:logl}
    \LL(\DD \mid \btheta) =  \prod_{k=1}^{N-\Mvlos}\Sigma_\star(\DD_k \mid \btheta)\prod_{j=1}^{\Mvlos} (\FF_m\ast\EE)(\DD_j \mid \btheta).
\end{equation}
In the above equation
$\Sigma_\star$ is the projected stellar density obtained by marginalizing the eDF~(\ref{for:tot}) over the line-of-sight, velocity components and metallicity, and is used to evaluate the likelihood of the $N-\Mvlos$ stars for which only the projected position is known. The function $\FF_m$ is obtained by marginalizing the eDF over the phase-space coordinates that are not observed for a given star. In practice, these always include the line-of-sight coordinate (i.e. distance) and the components of the velocity on the plane of the sky. For stars lacking metallicity measurements, the marginalization also extends over metallicity. There are no cases in our sample for which metallicity is available in the absence of a velocity measurement; hence, $\FF_m$ always depends at least on the projected position and line-of-sight velocity, and, when available, on the stellar metallicity. The symbol $\ast$ denotes the convolution operator, and $\EE$ is the error kernel, modelled as the product of Gaussian distributions accounting for the measurement uncertainties in line-of-sight velocity and, when available, metallicity.

The inclination, $i$, of the system enters as a free parameter, since in axisymmetric models different viewing angles produce different projected configurations. The line of sight can, in general, be specified by three Euler angles relating the intrinsic $(x,y,z)$ reference frame -- where the $z$-axis is the symmetry axis -- to the observer frame $(X^{\prime},Y^{\prime},Z^{\prime})$. Here, we restrict our models to oblate configurations, so we align the projected major axis of the galaxy with the $X^{\prime}$-axis $\equiv x$, and consider as line of sight the $Z^{\prime}$ axis. In this way, the inclination of the system is the angle between the symmetry axis and the line of sight.

In this work we use the \texttt{Agama}\footnote{\url{https://github.com/GalacticDynamics-Oxford/Agama}} software package for stellar dynamics \citep{Vasiliev2019}, which provides numerical tools for the dynamical modelling of stellar systems, including efficient routines for computing gravitational potentials, evaluating action–angle coordinates, and working with $\fJ$ DFs. In our implementation, it is used to construct the DFs, compute the gravitational potential, derive observable quantities through suitable integrations of the DFs, and evaluate the likelihood. 

In particular, the computation of the likelihood for each star requires integrating the DF over the missing dimensions (distance and/or some velocity components) and convolving it with the Gaussian error distribution of the line-of-sight velocity. These integrals need to be computed with sufficient accuracy so that the total log-likelihood of all stars (\ref{for:logl}) is known with an error well below 1. In \texttt{Agama}, these integrals can be computed using the \texttt{projectedDF} routine, which relies on the multidimensional adaptive integration method implemented in the \texttt{cubature} library \citep{cubature}. However, this approach is fairly expensive. An alternative method, introduced in \cite{Gherghinescu2026}, is quasi-Monte Carlo integration with importance sampling (see Section 2.4 of that paper for details), which is particularly efficient when a large number of likelihood evaluations is required, as in MCMC sampling. We adopt this quasi-Monte Carlo scheme and use 4000 Monte Carlo samples per star in our implementation. We verified that the differences in the total likelihood, compared to the \texttt{projectedDF} routine, are at most of order unity and do not affect the posterior sampling or the convergence of the MCMC chains.


We summarise the model parameters in a dedicated table provided as supplementary material. The table also reports the priors adopted in the applications presented in this work, together with the inferred parameter estimates and their 1$\sigma$ and 2$\sigma$ credible regions. In total, the model contains 30 free parameters: 18 describing the two stellar DFs (9 for each component, equation~\ref{for:df}), 4 describing the MDFs (2 per component, equation~\ref{for:mdf}), the relative contribution of the MR population to the total ($\wMR$), 5 parameters describing the DM halo (equation~\ref{for:dm}), the mass of the central IMBH (equation~\ref{for:bh}), and the inclination of the system, which is treated as a free parameter although it remains effectively unconstrained in all applications.

\begin{figure*}[h!]
    \centering
    \includegraphics[width=1\hsize]{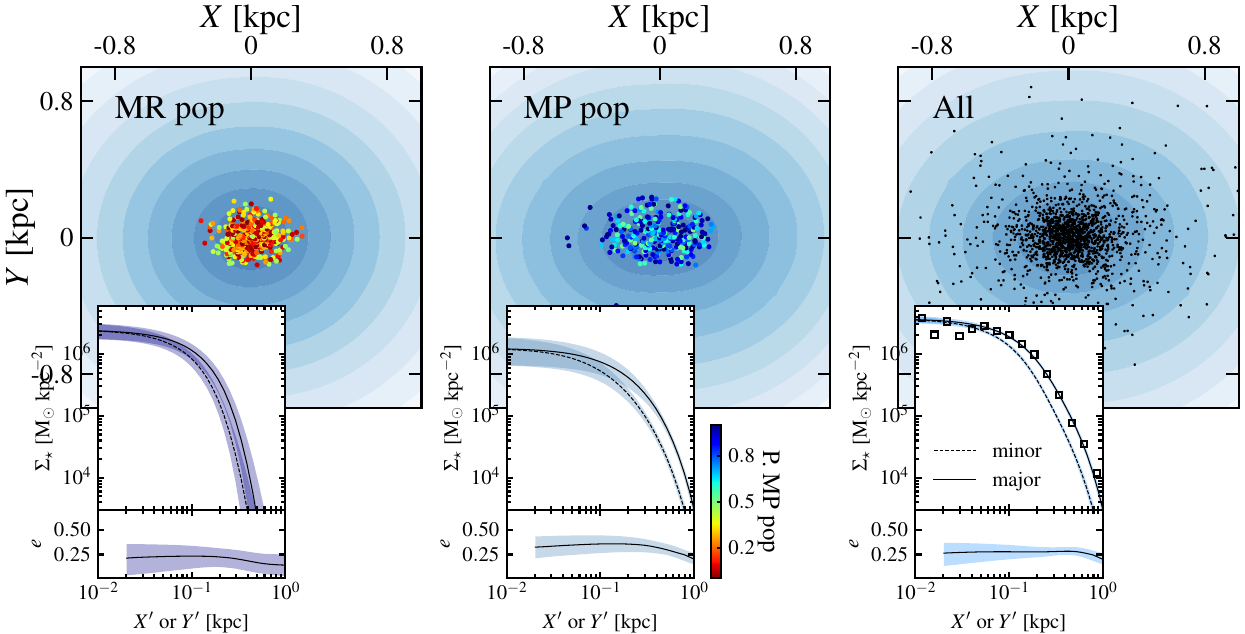} \\
    \includegraphics[width=1\hsize]{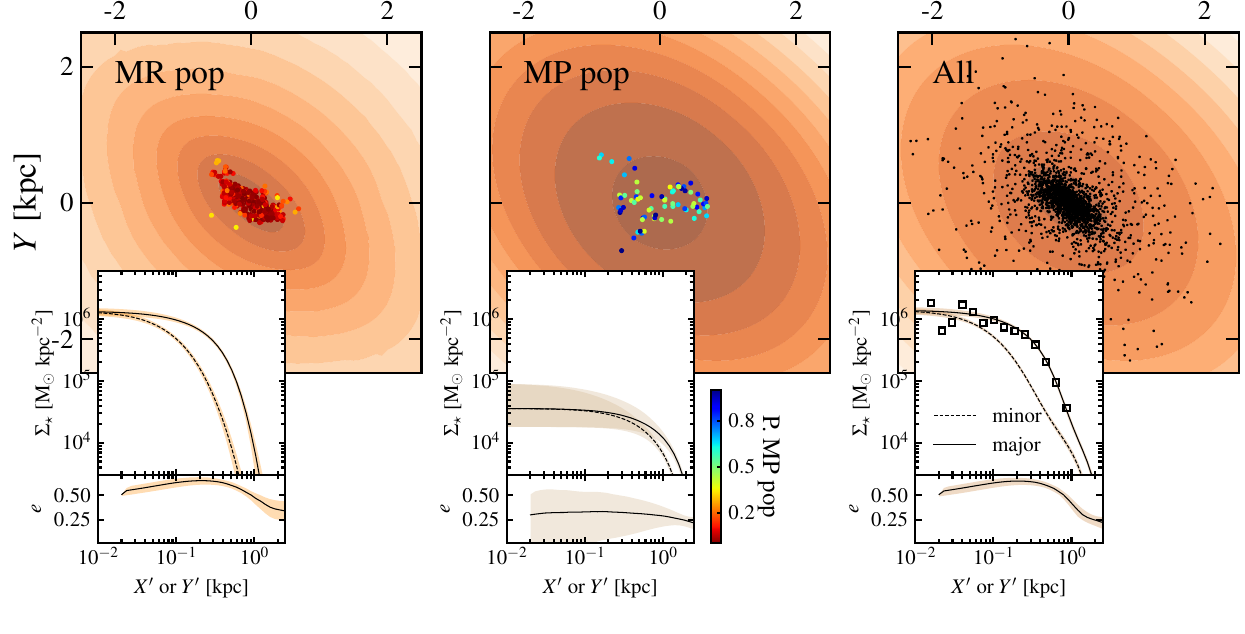} \\
    \vspace{-0.7cm}
    \caption{Structural properties of Draco (top) and Ursa Minor (bottom) inferred from the fit to the G26 sample. Left, middle and right panels show the projected stellar density maps of the MR, MP, and total components, computed from the model with the highest-posterior among those explored by the MCMC. Insets report the corresponding model-predicted density profiles as a function of semi-major (solid lines) and semi-minor (dashed lines) axes, together with the ellipticity profiles. Here, the density profiles have been scaled so that the total stellar mass of each galaxy is $\Mst=2.9\times10^5\Msun$ \citep{McConnachie2012}. The total density profile in the right inset is compared to projected number density from the data, rescaled to match the model total stellar mass. The ellipticity is defined as $1-\frac{Y^{\prime}}{X^{\prime}}$, where $X^{\prime}$ and $Y^{\prime}$ are the semi-major and semi-minor axes of the same isodensity contour. In the left and middle panels, stars with available metallicity are overplotted and assigned to each component according to their membership probabilities (see Section~\ref{subsec:stars}). In addition, they are colour-coded according to their probability of belonging to the MP population (P. MP), with blue indicating high probability and red low probability. The right panels show the full sample. Shaded regions indicate the $1\sigma$ uncertainties.}
    \label{fig:spatial}
\vspace{-0.3cm}
\end{figure*}

\section{Results}
\label{sec:results}

In this Section we present the main results of our analysis. The datasets used in this work are described in Appendices~\ref{subsec:wal} and~\ref{subsec:geha} and they are based on two different spectroscopic samples of G26 and W23. For clarity of exposition, many of the figures presented below refer to G26 data. Nevertheless, in the relevant Sections we explicitly discuss and show differences that arise when using the W23 dataset. In the tables, we report the results obtained with both datasets. 

In Section~\ref{subsec:stc} we describe the structural and kinematic properties of the stellar components of Draco (Section~\ref{subsub:drc}) and Ursa Minor (Section~\ref{subsub:umi}) inferred from our dynamical models. Section~\ref{subsec:dmprop} focuses on the properties of their DM haloes, while Section~\ref{subsec:jd} presents the resulting $J$- and $D$-factors. In Sections~\ref{subsec:onec} and~\ref{subsec:twocsph} we compare our models with flattened single-component and spherical two-component models based on $\fJ$ DFs. Finally, in Section~\ref{subsec:dscimbhs} we discuss the constraints on the possible presence of an IMBH at the centres of these systems. In each Section, our results are discussed in the context of previous studies available in the literature.

The main tables summarizing the characteristic quantities of the models are provided in Appendix~\ref{app:D}: Table~\ref{tab:modparams} reports the stellar component parameters for the different datasets, while Table~\ref{tab:dm} focuses on the DM properties.

\subsection{Structural and chemo-kinematic properties}
\label{subsec:stc}

\subsubsection{Draco}
\label{subsub:drc}

In Draco, the analysis of both datasets reveals the presence of two stellar populations with distinct chemical and kinematic properties. When adopting the G26 dataset, the two components have mean metallicities of $\MMMR=-1.98_{-0.06}^{+0.05}$ dex and $\MMMP=-2.37_{-0.08}^{+0.07}$ dex, with intrinsic dispersions of $\sigmaMR=0.13_{-0.06}^{+0.05}$ dex and $\sigmaMP=0.18_{-0.05}^{+0.05}$ dex, respectively. The corresponding posterior distributions are shown in the top left panels of Fig.~\ref{fig:metdis}. The metallicity peaks are clearly separated and constrained within the priors, while the posteriors of the metallicity dispersions partially overlap, with a mildly smaller dispersion preferred by the MR component. The inferred fraction of stars in the MR population is $0.41\pm0.13$, corresponding to a median of 660 MR stars, 403 of which have a line-of-sight velocity measurement.

Fig.~\ref{fig:spatial} shows the projected surface-density maps of the MR and MP components, and the corresponding total stellar distribution, obtained as a weighted combination of the two populations (equation~\ref{for:tot}). The maps are computed from a representative model corresponding to the one with the highest posterior among the models explored by the MCMC. Radial surface-density profiles for all components are shown in the insets. The two chemically distinct populations exhibit different spatial distributions: the MR population is more centrally concentrated, while the MP population is significantly more extended, with projected half-mass radii $\ReffMR=140_{-17}^{+19}\pc$ and $\ReffMP=299_{-31}^{+46}\pc$, respectively\footnote{The quoted projected half-mass radii are measured along the semi-major axis and are computed assuming radially constant flattening. The adopted value, denoted as $\qeff$, is defined as a mass-weighted average of the axis ratio,
\begin{equation}\label{for:qeff}
    \qeff \equiv \frac{\int_{\alim}^{\infty} \Sigma_\star(a)q(a)a\dd a}{\int_{\alim}^{\infty}\Sigma_\star(a)a\dd a},
\end{equation}
where $\Sigma_\star(a)$ is the projected surface density profile and $q(a)$ the radially varying axis ratio inferred from the model. The lower integration limit is set to $\alim=20\pc$, as the computation of the flattening becomes numerically unstable at smaller radii.}. Stars with available spectroscopic measurements are overplotted on the MR and MP surface-density maps and colour-coded by their membership probability, which further emphasises the different spatial extent of the two populations\footnote{The number of stars per population is derived following the procedure described in Appendix~\ref{app:B}: stars are ranked by membership probability, and the probability threshold is set by the membership probability of the $Nw$-th object.}. The projected half-mass radius of the overall stellar component is $\Reff=211_{-9}^{+9}\pc$. The small inset in the top right panel of Fig.~\ref{fig:spatial} further illustrates the quality of the modelling, showing that the radial surface-density profile of the global stellar component closely matches the observed stellar counts over the full radial range probed by the data.

\begin{figure*}[p]
    \centering
    \includegraphics[width=1\hsize]{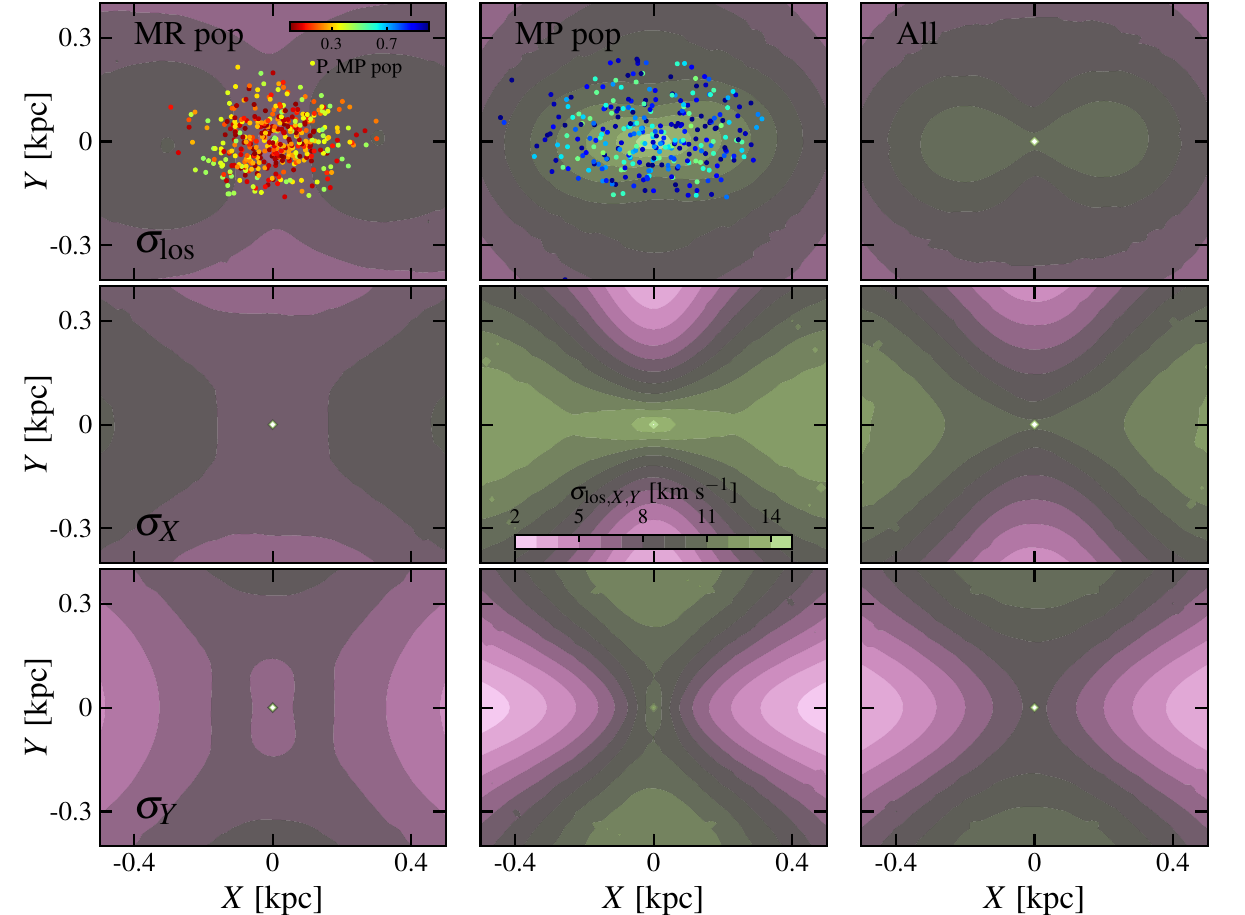} \\
    \includegraphics[width=1\hsize]{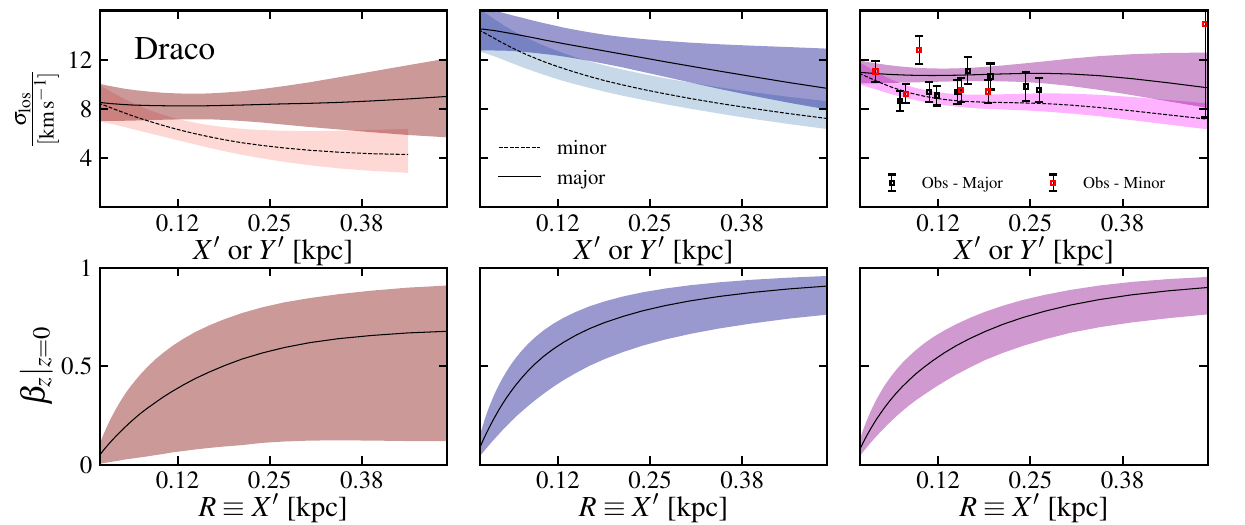} \\
    \vspace{-0.2cm}
    \caption{Kinematic properties of Draco inferred from the fit to the G26 sample. The top three rows show projected velocity-dispersion maps along the line of sight, and the prediction for projected velocity-dispersions along the $X$ and $Y$ directions. Columns correspond to the MR (left), MP (middle), and total (right) components. Maps are computed from the highest-posterior model in the MCMC. In the top-left and middle panels, kinematic subsample stars are overplotted and assigned to components based on membership probabilities (colour-coding as in Fig.~\ref{fig:spatial}). The bottom rows show the line-of-sight velocity dispersion profiles along the projected minor and major axes, and the anisotropy profile $\betaz$ along the intrinsic (and projected) major axis. Shaded regions indicate $1\sigma$ uncertainties. For the total population, data-based dispersion profiles are also shown for reference. Data are binned in the meridional plane via Voronoi tessellation ($\simeq$50 stars per bin), with dispersions computed per quadrant following \cite{Pryor93}. Red (black) points correspond to bins closer to the projected semi-minor (semi-major) axis.}\label{fig:slos}
\end{figure*}

\begin{figure*}
    \centering
    \includegraphics[width=1\hsize]{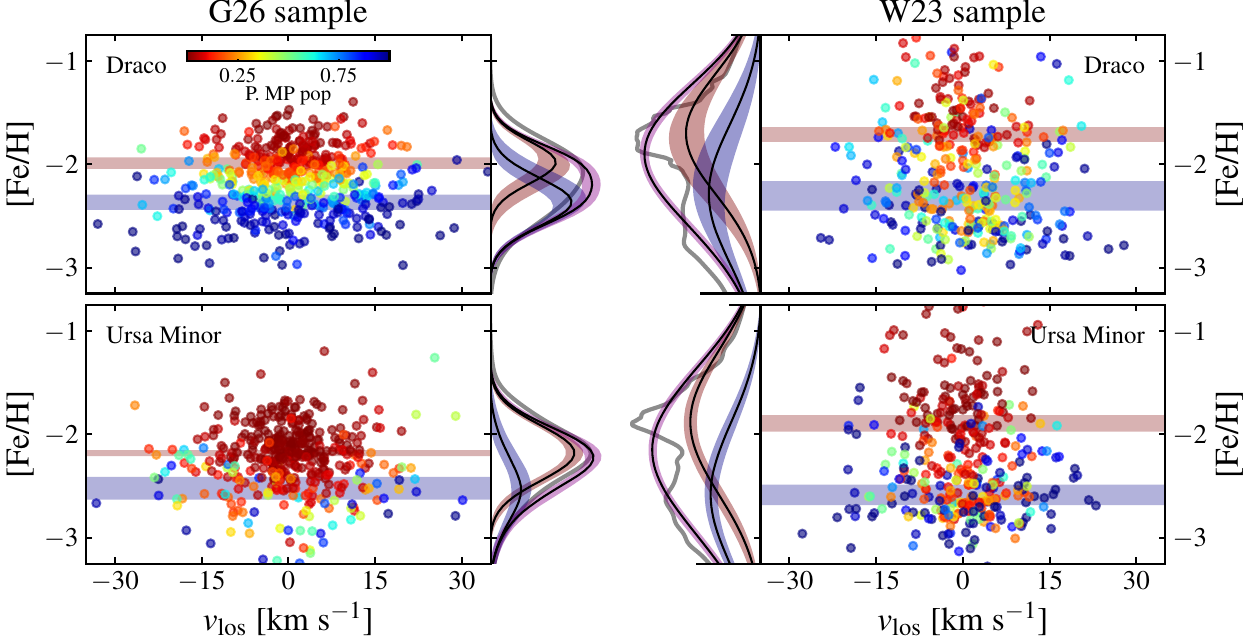}
    \caption{Distribution of stars in the metallicity–line-of-sight velocity plane for Draco (top panels) and Ursa Minor (bottom panels). The left and right panels show results based on the G26 and W23 samples, respectively. Stars are colour-coded by probability to belong to the MP population (P. MP), with red and blue indicating association with the MR and MP components, respectively. Here, the red and blue bands indicate the $1\sigma$ credible regions for the MR and MP mean metallicities (Table~\ref{tab:modparams}). The middle panels show the MDFs of the two components using the same colour coding, with the purple curve representing the total model MDF; all model distributions are convolved with the median metallicity error of the corresponding sample. The grey curve shows the distribution of the data, obtained by smoothing the measurements over their individual uncertainties with Gaussian kernels.}
    \label{fig:met}
\end{figure*}

\begin{figure*}[h!]
    \centering
    \includegraphics[width=1\hsize]{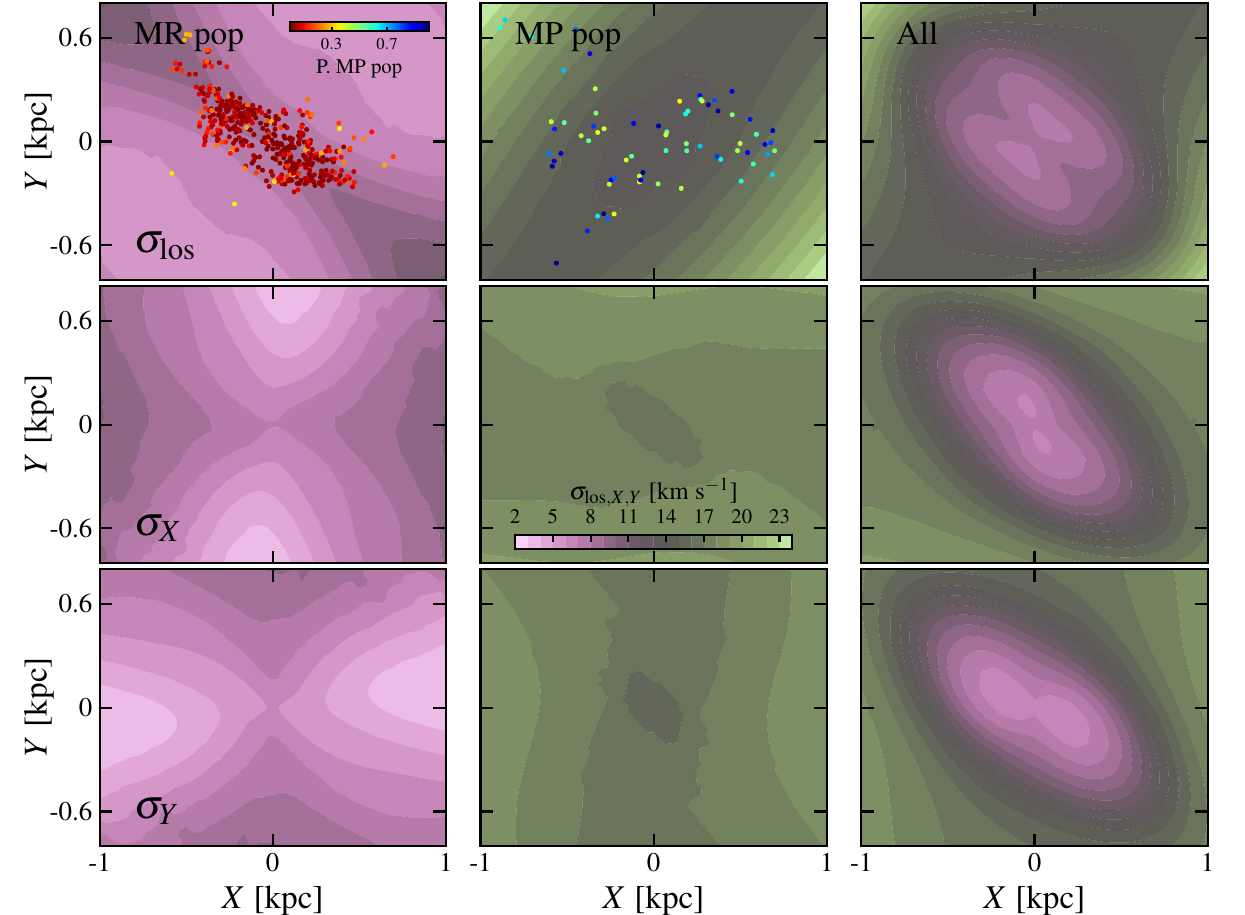} \\
    \includegraphics[width=1\hsize]{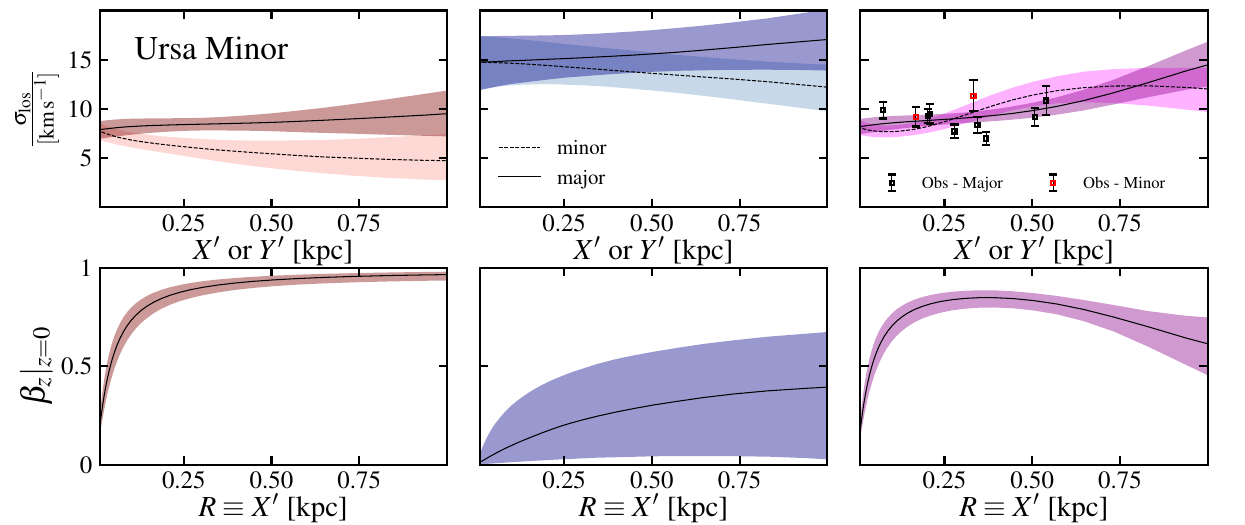} \\
    \caption{Same as Fig.~\ref{fig:slos} but for Ursa Minor.}
    \label{fig:slos2}
\end{figure*}


Across the radial range probed by the data, the two Draco populations show similar flattening profiles (bottom panels of the insets of Fig.~\ref{fig:spatial}). The effective ellipticities, defined as $\eeff\equiv 1-\qeff$ (see equation \ref{for:qeff}) are $\eeffMR=0.23_{-0.11}^{+0.09}$ for the MR component and $\eeffMP=0.33_{-0.07}^{+0.06}$ for the MP component. For comparison, the effective ellipticity of the total stellar population is $\eeff=0.28_{-0.05}^{+0.04}$.

The top panels of Fig.~\ref{fig:slos} show the two-dimensional maps of the line-of-sight velocity dispersion for the MR, MP, and total stellar populations (left, middle, and right columns, respectively), derived from the same model of Fig.~\ref{fig:spatial}. The dispersion maps indicate that the kinematics is neither spherically symmetric nor simply aligned with the stellar isodensity contours. In particular, the velocity dispersion decreases more rapidly along the minor axis than along the major axis, and the resulting iso-dispersion contours exhibit a characteristic peanut-like morphology. The MR component is dynamically colder than the MP one, with a systematically lower line-of-sight velocity dispersion across the full radial range probed by the data (see central panels of Fig.~\ref{fig:slos}). Reference central line-of-sight velocity dispersions are $\slosoMR = 8.5_{-1.5}^{+1.5}\kms$ and $\slosoMP = 14.5_{-1.8}^{+1.7}\kms$, with $\slosot = 11.0_{-0.9}^{+1.0}\kms$ for the global stellar population. All the quoted quantities are listed in Table~\ref{tab:modparams}.

The anisotropy profiles highlight the distinct dynamical structure of the two components. The anisotropy is quantified by the parameter $\betaz \equiv 1 - \sigma_z^2 / \sigma_R^2$, where $\sigma_z$ and $\sigma_R$ denote the intrinsic velocity dispersions along the symmetry and the cylindrical radial directions, respectively (see bottom panels of Fig.~\ref{fig:slos}). All components display a gentle outward increase in $\betaz$, gradually approaching unity. Evaluated at the projected half-mass radius of each stellar component, we obtain $\betazMR=0.43_{-0.35}^{+0.24}$ and $\betazMP=0.83_{-0.17}^{+0.08}$ for the MR and MP populations, respectively, with $\betaz=0.71_{-0.14}^{+0.10}$ for the global population. The latter value is consistent with that reported by \cite{Hayashi2020}, although in their work the anisotropy is assumed to be constant and treated as an input parameter of their single population, axisymmetric Jeans models, whereas in our case the anisotropy profiles are inferred self-consistently from the DFs.

The results obtained from the kinematic sample of W23 are broadly analogous. The posterior distributions of the metallicity parameters are shown in the bottom left panels of Fig.~\ref{fig:metdis}. In this case, we measure MR component with a higher mean metallicity than that inferred from the G26 sample, while displaying a comparable intrinsic dispersion. The MP component, instead, is characterised by a larger metallicity dispersion. No significant differences emerge in terms of structural properties: the projected half-mass radii, effective ellipticities, and the overall shapes of the surface-density and flattening profiles of both the MR and MP components are fully consistent, within the uncertainties, with those obtained from the G26 dataset. For this reason, we do not show the corresponding profiles. From a kinematic standpoint, the W23 sample provides a sparser coverage of the innermost regions but a better sampling at large projected radii; this feature will recur also for Ursa Minor, where the outer regions are more densely sampled. Nevertheless, the inferred kinematic trends are consistent with those from the G26 dataset: the MR population remains dynamically colder than the MP one, with a systematically lower velocity dispersion. Overall, the line-of-sight velocity dispersion inferred from the W23 sample sample is smaller, $\sloso=8.4_{-1.1}^{+1.2}\kms$, whereas the anisotropy constraints are very similar, showing no significant differences within the uncertainties. All the mentioned parameters and derived quantities are reported in Table~\ref{tab:modparams}, where their agreement can be directly and quantitatively evaluated.

Finally, Fig.~\ref{fig:met} shows the distribution of member stars with available metallicity in the metallicity–line-of-sight velocity plane for the two catalogues. The G26 and W23 samples display slightly different metallicity distributions, with the latter appearing more sparse in metallicity, reflecting the larger measurement uncertainties of that dataset. Instead, the G26 sample shows a larger spread in line-of-sight velocities, with a greater number of stars populating the high-velocity tails. This larger spread affects both stellar populations, but is more pronounced in the MP component, where, as a consequence, the (central) velocity dispersion is about $2\kms$ higher than in W23 (see Table~\ref{tab:modparams}). This broader velocity dispersion propagates into the total $\slosot$, which is correspondingly larger in G26, as anticipated. In both catalogues, the two stellar populations are not separated into clearly distinct metallicity peaks; rather, they correspond to two nearby components that overlap and jointly produce a composite MDF. 

\subsubsection{Draco -- comparison with literature}

In the literature, \citet{Yang2025} present one of the few analyses that explicitly identify multiple chemo-kinematic components in Draco, providing a useful point of comparison for our results. Using DESI spectroscopic data and two-component axisymmetric Jeans models, the authors find in Draco a MP and a MR population with mean metallicities of $-2.28_{-0.09}^{+0.11}$ dex and $-1.75\pm0.04$ dex, and dispersions of $0.23\pm0.03$ dex and $0.26\pm0.04$ dex, respectively. The properties of the MP component are in very good agreement with our estimates for both datasets. The MR component shows a somewhat larger variability, with the W23 sample being in closer agreement with \cite{Yang2025}. Despite these small differences, the overall agreement between the independent analyses underscores the robustness of the inferred chemo-kinematic decomposition, even when based on different datasets and modelling approaches. We also note that the half-mass radius and flattening of the overall stellar population are in perfect agreement with those estimated by \cite{Munoz2018}.

\subsubsection{Ursa Minor}
\label{subsub:umi}

Ursa Minor shows a behaviour qualitatively similar to Draco: two stellar populations with different metallicities and spatial distributions. The mean metallicities of the two components, based on the analysis of the G26 sample, are $\MMMR=-2.18_{-0.03}^{+0.03}$ dex and $\MMMP=-2.52_{-0.11}^{+0.11}$ dex, with intrinsic dispersions $\sigmaMR=0.21_{-0.03}^{+0.03}$ dex and $\sigmaMP=0.29_{-0.11}^{+0.11}$ dex. The posterior distributions of these quantities are shown in the top right panels of Fig. ~\ref{fig:metdis}. The two metallicity peaks are well separated, and the posterior distributions of the metallicity dispersions show some differences, with the MP component displaying a broader distribution and correspondingly larger uncertainties. In this case, the sample is divided into 1455 MR and 682 MP stars, of which 454 and 71, respectively, belong to the spectroscopic subsample.

Also in Ursa Minor, the MR population is more centrally concentrated than the MP one (see lower left panels of Fig.~\ref{fig:spatial}). The projected half-mass radii of the two components are $\ReffMR=374_{-22}^{+21}\pc$ and $\ReffMP=1.03_{-0.12}^{+0.14}\kpc$, with the half mass radius of the overall stellar population $\Reff=463_{-18}^{+20}\pc$. The MP population is not only more extended but also significantly rounder than the MR population, with effective ellipticities of $\eeffMP = 0.29_{-0.11}^{+0.09}$ and $\eeffMR = 0.61_{-0.03}^{+0.03}$, respectively. For comparison, the effective ellipticity of the total stellar population is $\eeff =0.56_{-0.03}^{+0.03}$. In addition, both the MR component and the total population exhibit a mildly radially varying ellipticity in the innermost regions. 

The MR component is dynamically colder than the MP one, with central line-of-sight velocity dispersions of $\slosoMR = 7.9_{-0.9}^{+0.9}\kms$ and $\slosoMP = 14.8_{-2.9}^{+2.6}\kms$, respectively. The iso-dispersion maps exhibit a behaviour different from that observed in Draco (see Fig.~\ref{fig:slos2}, showing the same maps and profiles as in Fig.~\ref{fig:slos}, but now for Ursa Minor). Along the projected major axis, the velocity dispersion of both stellar populations remains approximately constant with radius. Along the minor axis, instead, the dispersion of MR and MP populations gradually decline. However, within the radial extent effectively probed by the kinematic data, the global line-of-sight velocity dispersion measured along the two principal axes remain broadly comparable, with a reference central value of $\slosot = 8.2_{-0.8}^{+0.8}\kms$. 

In terms of anisotropy (bottom panels of Fig.~\ref{fig:slos2}), all components exhibit an overall outward increase in $\betaz$. The MR population rises steeply at small radii and then flattens to an approximately constant level, while the MP population shows a milder and more gradual increase. The total population follows the trend of the MP component. As seen in Fig.~\ref{fig:met}, the two components remain well differentiated in metallicity space, though not sharply separated, with the MP population extending toward lower metallicities as a prominent metal-poor tail.

\begin{figure*}
    \centering
    \includegraphics[width=0.95\hsize]{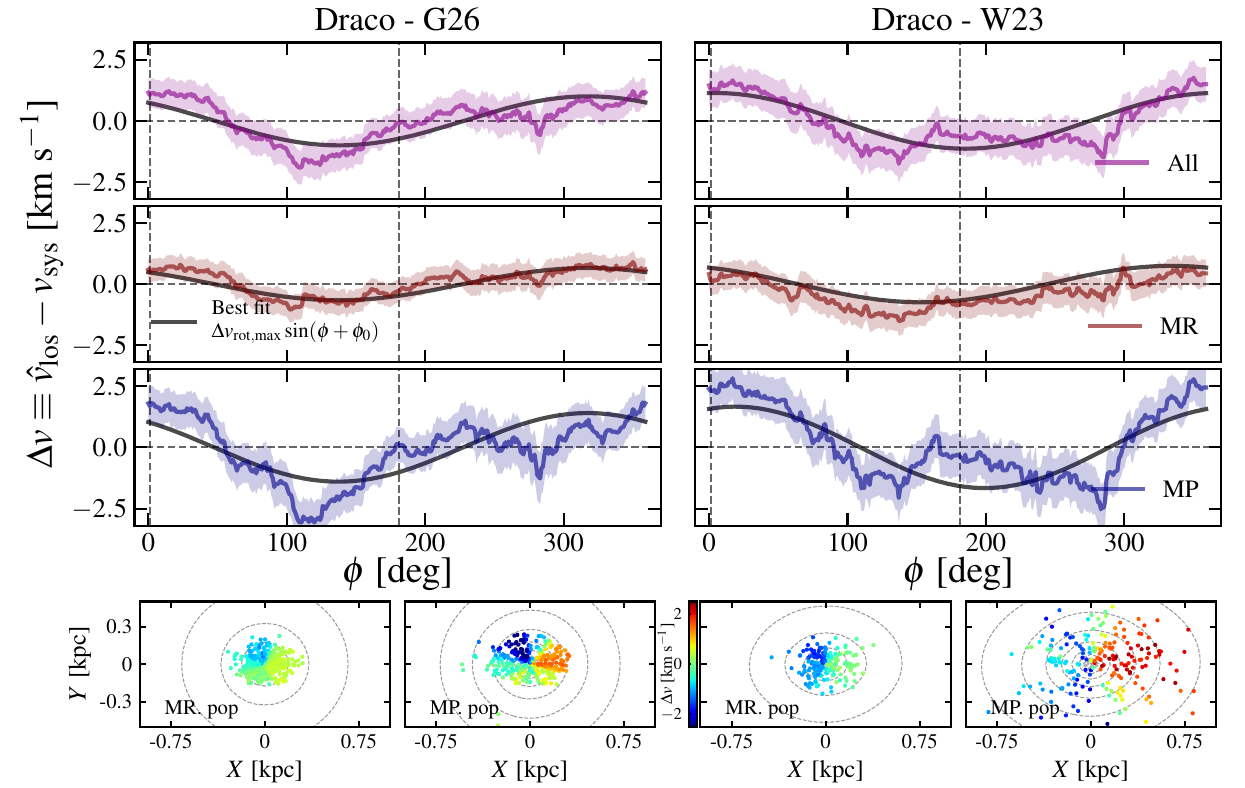}
    \caption{Difference between mean line-of-sight and systemic velocities ($\Delta v$) as a function of the azimuthal angle, $\phi$, shown for the total (purple, top), MR (red, middle), and MP (blue, bottom) populations of Draco. Left and right columns correspond to the G26 and W23 datasets, respectively. Uncertainties on the mean values, which we show with bands, are derived from the Fisher information matrix. The vertical dashed line indicates the azimuthal angle of the major axis. The black sinusoidal line is the best fit to the sinusoidal signal measured in each panel. The values of the maximum rotation variation, $|\vrotmax|$ are reported in Table~\ref{tab:modparams}. The bottom panels also include iso-density contours (grey dashed lines) of the MR and MP populations computed from the model with the highest posterior among those explored by the MCMC. The lower panels show the spatial distribution of MR and MP stars, colour-coded according to the velocity difference at each $\phi$ taken from the upper panels. }\label{fig:rotdrc}
\end{figure*}

\begin{figure*}
    \includegraphics[width=1\hsize]{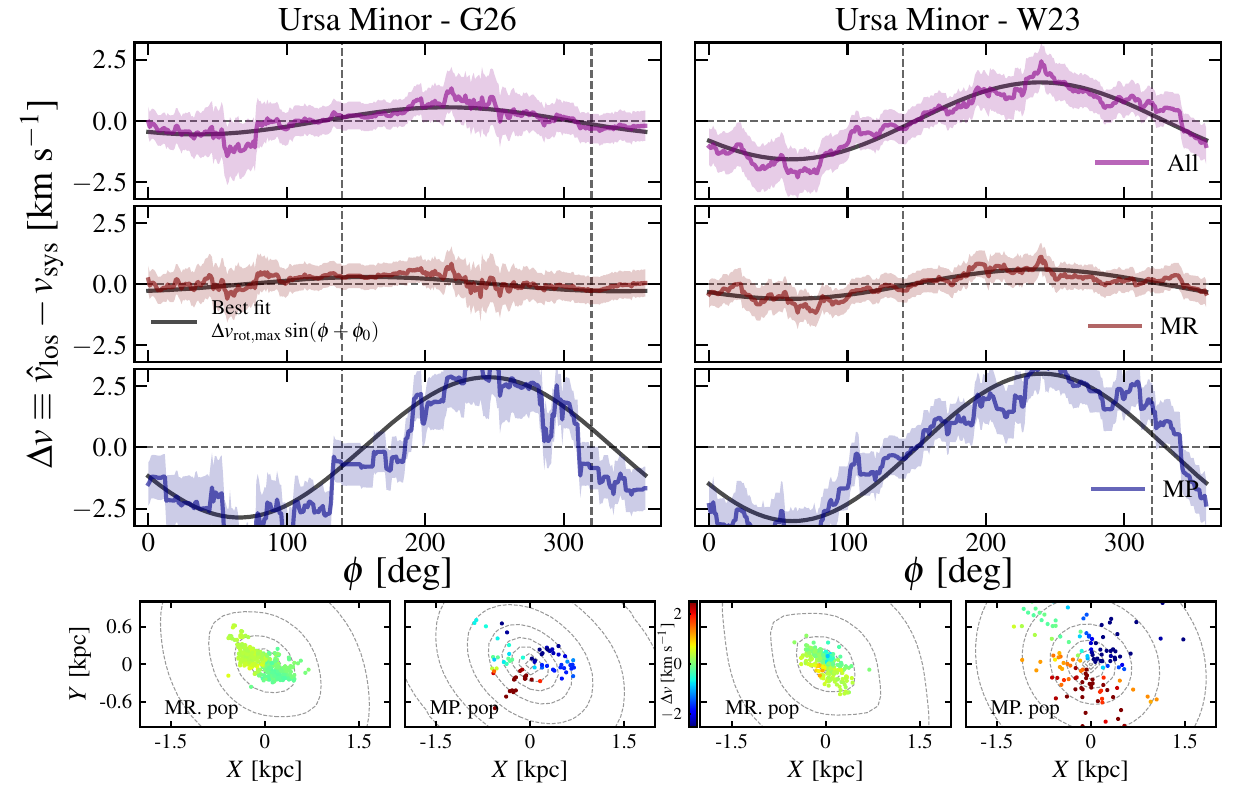}
    \caption{Same as Fig.~\ref{fig:rotdrc}, but for Ursa Minor.}\label{fig:rotumi}
\end{figure*}

\begin{figure*}[h!]
    \centering
    \includegraphics[width=1\hsize]{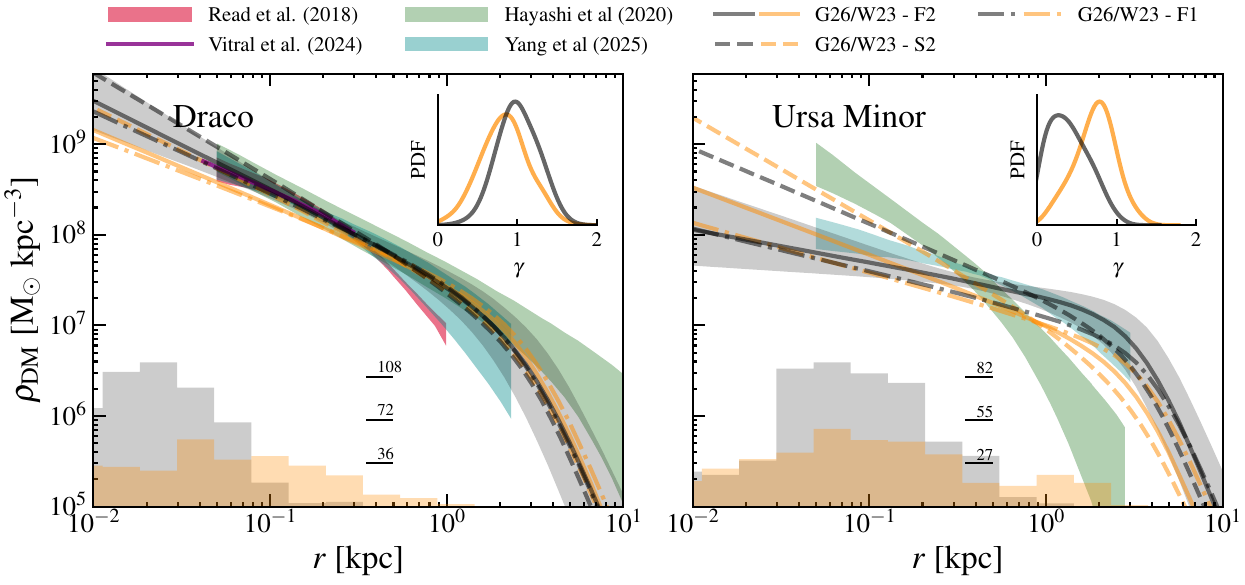}
    \caption{DM halo properties. Left panel: median DM halo density profile (solid curves) from the chemo-dynamical models of Section~\ref{sec:mod} (F2), together with estimates from the literature. Results from the analysis of the G26 and W23 samples are in grey and orange, respectively. For clarity, we show $1\sigma$ credible region only for the F2 models based on G26 (grey shaded band). We also show the median DM halo density resulting from running one-component flattened models (F1, grey and orange dashed-curves, Section~\ref{subsec:onec}) and two-component spherical models (S2, grey and orange dashed dot curves; Section~\ref{subsec:twocsph}). The histograms at the bottom show the radial distribution of the kinematic tracers (orange for G26 and grey for W23); note that it is plotted with a different, linear scale on the $y$-axis. The inset displays the posterior distribution of $\gamma$. Right panel: same as the left panel, but for Ursa Minor. Literature estimates from \citet[green]{Hayashi2020}, \citet[cyan]{Yang2025}, \citet[red]{Read2018}, and \citet[purple]{Vitral2024}.}
    \label{fig:dm}
\end{figure*}

The analysis based on the W23 sample yields qualitatively consistent results. The MR component is more centrally concentrated and exhibits a different degree of flattening than the MP one, with most of the structural parameters of both populations broadly consistent within the quoted uncertainties with those inferred from the G26 sample (see Table~\ref{tab:modparams}). We note, however, that the MR population inferred from the W23 sample is slightly more metal-rich than that derived from the G26 data,  and that the fractional contribution $\wMR$ is comparable to that of the MP component, unlike in the G26 sample (see Fig.~\ref{fig:metdis}, bottom left panels). By comparing the spatial distribution of the kinematic tracers in the W23 and G26 samples\footnote{The different radial coverage of the kinematic tracers of the two samples is further quantified in Fig.~\ref{fig:dm}, discussed in Section~\ref{subsec:dmprop}.} (bottom panels of Fig.s~\ref{fig:datasets} and \ref{fig:spatial}), it can be appreciated that two samples differ in their radial coverage. While the W23 catalogue includes fewer tracers in the innermost regions, it provides better sampling farther out, allowing for a more robust characterization of the extended MP component. The improved outer coverage in the W23 sample leads to tighter constraints on the MP spatial distribution and is probably responsible also for the smaller inferred effective radius (see Table~\ref{tab:modparams}). In terms of kinematics, the two components show broadly similar behaviour in the two catalogues. The velocity dispersion profiles of the MR component are comparable, although in the W23 sample the decline is somewhat steeper along the minor axis. The MP component, which is better sampled in the W23 data, shows a slightly different trend, with a dispersion decreasing along both axes. The overall velocity dispersion profiles of the stellar system, measured along the major and minor axes, remain nearly unchanged between the two datasets. The inferred anisotropy profiles are also consistent between the two analyses. In all cases, the MR component remains dynamically colder than the MP one. Detailed values of the kinematic parameters are reported in Table~\ref{tab:modparams}.

\subsubsection{Ursa Minor -- comparison with literature}
For Ursa Minor, our results are broadly consistent with previous studies. The mean metallicities of the MR and MP populations are in good agreement with those reported by \cite{Pace2020}. In particular, the metallicities inferred from the W23 sample are more closely aligned with their MMT/Hectochelle-based determination\footnote{The two samples share a fraction of stars, although individual measurements may differ due to the use of different spectral libraries and calibrations.}. In line with the authors, we also find that the MR and MP components exhibit different spatial distributions, with a more centrally concentrated MR population. Both our analysis and that of \cite{Pace2020} indicate a more flattened MR component, although we infer a slightly smaller degree of flattening. A more noticeable difference concerns the spatial extent of the MP population, which in our analysis appears more extended. Part of this discrepancy may stem from differences in the adopted modelling framework; for example, \cite{Pace2020} assume Plummer profiles for both components, whereas our models allow for more flexible density profiles. Additional differences may arise from the use of independent spectroscopic datasets with different spatial coverage. Indeed, as noted above, the spatial extent that we infer from the W23 data, which better samples the outer parts, is smaller than that inferred from the G26 data. Nevertheless, we note that our results are in good agreement with \cite{Pace2020} for both the half-mass radius and the flattening of the overall population, while, compared to \cite{Munoz2018}, we recover the same flattening but a slightly larger half-mass radius. A comparison with \cite{Yang2025} shows that their inferred mean metallicities for the MR and MP populations ($-1.96$ dex and $-2.32$ dex, respectively) lie between the values obtained from our two spectroscopic samples for both the MR and MP components. Also, the metallicity dispersion they report for the MP population is significantly larger than that inferred in our analysis, while the MR dispersions are more comparable. Nevertheless, the MR and MP populations exhibit distinct radial extents, in agreement with our findings. For Ursa Minor, these differences highlight a larger sensitivity of the inferred chemo-kinematic properties to both the adopted modelling assumptions and the underlying spectroscopic data. Nevertheless, the main conclusions remain robust.

\subsubsection{Rotation}
\label{subsec:rot}

The dynamical models adopted in this work neglect stellar rotation and assume that the internal kinematics of a dSph are primarily supported by velocity dispersion. This approximation is well justified in regimes where rotational support is negligible compared to stellar random motions, which happens to be the case for the majority of dSphs in which signatures of rotation have been detected \citep[e.g.][]{delPino2017,Martnez2021}. Previous studies have generally not identified significant rotational signatures in Draco, while for Ursa Minor the presence of weak prolate rotation has been suggested \citep{Pace2020}, though the significance of the signal is still not firmly established.

Here we investigate the presence of rotation in Draco and Ursa Minor and whether such signatures may depend on stellar population. We use the probability membership derived to distinguish between stellar components, allowing for a more robust assessment of any rotational trends in the stellar kinematics. This also provides an a posteriori consistency check: if rotation is not dynamically significant, the use of non-rotating dynamical models is justified and residual rotation is not expected to bias the inferred dynamical masses.

Figs~\ref{fig:rotdrc} and \ref{fig:rotumi} present the mean line-of-sight velocity offset ($\Delta v$, i.e. difference between mean and systemic line-of-sight velocities) as a function of the azimuthal angle, $\phi$, for both galaxies and for the two available datasets\footnote{We recall that line-of-sight velocities have been corrected for the perspective gradient on the plane of the sky, see Appendix~\ref{app:cfinal}.}. The azimuthal angle is measured counterclockwise from the positive $X$-axis. For each value of $\phi$, we estimate the mean line-of-sight velocity using a variant of the classical bisection line method \citep[e.g. ][]{Battaglia2008}. In our approach, we define a sector of $\pm 60^\circ$ around the chosen azimuthal angle and compare the resulting mean line-of-sight velocity to the systemic velocity. The mean velocity is obtained by maximizing a Gaussian likelihood with free mean velocity and velocity dispersion, evaluated over all stars in the sector and accounting for individual measurement uncertainties through convolution. When deriving the mean velocity of the MR and MP population, each star is weighted with its membership probability of belonging to that population. The upper panels of Figs~\ref{fig:rotdrc} and \ref{fig:rotumi} show measurements obtained using the full stellar samples with available line-of-sight velocities, while the lower panels display results obtained by separating stars into populations. 

In the case of Draco (see Fig.~\ref{fig:rotdrc}), we find evidence for weak rotation in both G26 and W23 catalogues. In both cases, the weak rotational signal observed in the total population appears to be primarily driven by the MP component and it is indicative of rotation around the minor axis. The hint of rotation is more pronounced in the W23 catalogue. The mild rotational signature of the MP population is particularly visible in the lower panels, where individual stars, divided in MR and MP, are colour-coded according to their velocity offset at fixed $\phi$. We fit a sinusoidal function to the mean velocity offset curves shown in Fig.~\ref{fig:rotdrc} in order to quantify a statistically meaningful rotational amplitude ($|\Delta\vrotmax|$). The fit is performed using a standard least-squares optimisation. The strongest rotational signal is found in the W23 sample, with $|\Delta\vrotmax| = (0.75\pm0.04)\kms$ for the MR component and $|\Delta\vrotmax| = (1.66\pm0.07)\kms$ for the MP component. All the other values (total population and G26 catalogues) are reported in Table~\ref{tab:modparams}.

Ursa Minor displays a more complex behaviour. In the G26 catalogue, the total population shows only a marginally significant rotational pattern. However, the W23 dataset reveals a clearly sinusoidal variation of the mean velocity with $\phi$, with maximum signals located at approximately $40\deg$ and $220\deg$. These directions are close to the minor axis of the system and correspond to a rotation pattern consistent with prolate rotation around the major axis. The decomposition into metallicity components provides additional insight: the MR population shows almost no evidence of rotation and remains fully consistent with the systemic velocity at all PAs. In contrast, the rotational signal is entirely driven by the MP component, which is more spatially extended. It is worth noting that the rotation signal is present in the MP population of both datasets. From the sinusoidal fit to the line-of-sight velocity offset variations of Fig.~\ref{fig:rotumi}, we infer a maximum rotational amplitude in the MP component of $|\vrotmax|=(3.01\pm0.07)\kms$, while it is smaller in the total population $(1.57\pm0.06)\kms$.

To summarise, we detect a weak signal of rotation in both galaxies, predominantly in the MP population, while the MR component remains fully dispersion-dominated in all cases. The nature of the rotation differs between the two systems: in Draco it is consistent with rotation around the minor-axis, whereas Ursa Minor exhibits a prolate rotation in both samples, confirming previous, independent, indications. Despite these signatures, rotation remains dynamically subdominant, with the ratio $\Delta \vrotmax/\sloso$ (see Table~\ref{tab:modparams}) consistent with zero in the MR component, while it reaches only modest values, up to $\simeq0.1$ in Draco and $\simeq0.2$ in Ursa Minor, and only for the MP population, which is the less dominant component. Although rotation is detectable in specific stellar populations, its overall contribution to the dynamical support of both dSphs is minor. Neglecting rotation is therefore not expected to significantly bias the inferred dynamical masses, since the associated underestimate scales with the square of the ratio of rotational velocity to velocity dispersion \citep{BinneyTremaine2008}.

We point out that the membership probabilities are computed assuming that the stellar components are oblate spheroids, whereas the measured prolate rotation signal in the MP component of Ursa Minor would be more naturally associated with a prolate or triaxial geometry. Therefore, although the inferred rotation amplitude is consistent with previous estimates in the literature, our result should be interpreted with caution, as it may reflect the kinematic complexity of the system.

\subsection{DM halo properties}
\label{subsec:dmprop}

Fig.~\ref{fig:dm} presents the DM density profiles inferred for Draco (left panel) and Ursa Minor (right panel). The solid grey and orange curves show the median DM model obtained fitting the G26 and W23 samples, respectively, with two component flattened models (F2). For the sake of clarity, we only show the $1\sigma$ credible regions for the G26 models (grey band). The F1 (flattened, one-component) and S2 (spherical, two-component) models are analysed in the following sections.

\subsubsection{Inner halo density}

In Draco, the inferred DM halo is nearly insensitive to the spectroscopic catalogue, with only minor differences between G26 and W23. The inner density profile clearly favours a cusp in both cases: the posterior distribution of the inner logarithmic slope (shown in the top-right inset of Fig.~\ref{fig:dm}) peaks at 
$\gamma=0.98_{-0.26}^{+0.28}$ for the former and $\gamma=0.82_{-0.31}^{+0.29}$ for the latter, effectively ruling out a density core with $\gamma\lesssim0.5$ at the $1\sigma$ level, and $\gamma\lesssim0.2$ at $2\sigma$ (see Table in supplementary material). Although $\gamma$ formally represents the asymptotic logarithmic slope of the DM density profile, this value remains representative over a substantial fraction of the inner radial extent of the galaxy. For instance, at $r\simeq150\pc$ (corresponding to $\simeq0.5$ times the half-mass radius), the posterior distribution of the local logarithmic slope is essentially indistinguishable from that of the asymptotic slope. In Ursa Minor, the situation is more nuanced. The use of different spectroscopic catalogues leads to quantitatively different inferences for the inner slope. The G26 sample favours a significantly shallower profile, with $\gamma=0.37_{-0.24}^{+0.31}$, closer to a constant-density core, whereas the W23 sample yields $\gamma=0.74_{-0.30}^{+0.22}$ and continues to disfavour $\gamma\simeq0$. As in Draco, the asymptotic slope remains representative of the local behaviour at $r\simeq150\pc$ (corresponding to $\simeq0.3$ times the half-mass radius). A possible origin of this discrepancy can be traced to the different radial distribution of the kinematic tracers in the two catalogues, which we show in the lower part of Fig.~\ref{fig:dm}. In both galaxies, the G26 catalogue includes a substantially larger number of tracers in the central regions. This difference in radial coverage may contribute to the different constraints on the inner slope. In particular, the improved central sampling—and possibly the higher overall data quality—of the G26 sample likely leads to a more robust determination of the inner density profile.

\subsubsection{Comparison with the literature}

Fig.~\ref{fig:dm} also overlays recent determinations of the galaxies DM profiles, allowing a direct comparison with results obtained using different datasets and modelling assumptions. For Draco, the comparison indicates broad agreement with the literature. \cite{Yang2025} employ an axisymmetric Jeans framework, considering both a single-population and a multi-population model. Despite differences in the method and dataset, for the two-population model the authors infer an inner slope of $\gamma=0.71_{-0.35}^{+0.34}$, consistent with our results in the inner regions. Similarly, \cite{Vitral2024} combine HST proper motions with line-of-sight velocities and solve the axisymmetric Jeans equations; they report a logarithmic slope averaged over the observed radial range rather than a strictly asymptotic value, but this is likewise consistent with a cusped density profile. Using axisymmetric Jeans, one-population models applied to the kinematic data available at the time, \cite{Hayashi2020} also find that Draco hosts a cusped DM halo. \cite{Read2018} solve the spherical Jeans equations including high-order moments \citep{Read2017}, inferring a DM density consistent with a cusp. 

Within the radial range probed by the data, we infer a DM density normalization that is more or less in agreement with the determinations from \cite{Hayashi2020} and \cite{Yang2025}. We provide an estimate for the DM density measured at $150\pc$, $\rhodmcl=2.10_{-0.31}^{+0.34}\times10^8\Msun\kpc^{-3}$. The outer halo is essentially unconstrained, allowing us to place only a lower limit on its scale radius, $\rs \geq 1.10\kpc$, consistent with that provided by \citet{Hayashi2020}. 

\begin{figure*}
    \centering
    \includegraphics[width=0.45\linewidth]{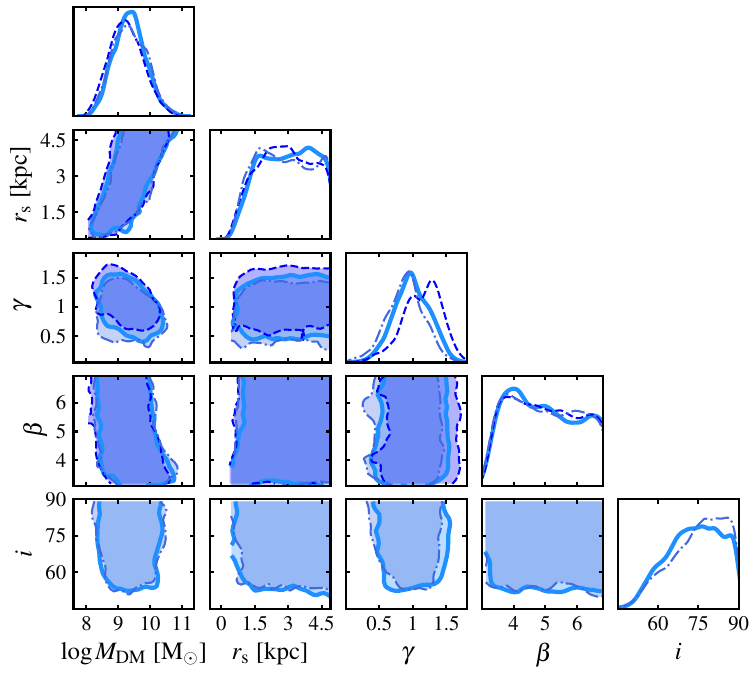}
    \includegraphics[width=0.45\linewidth]{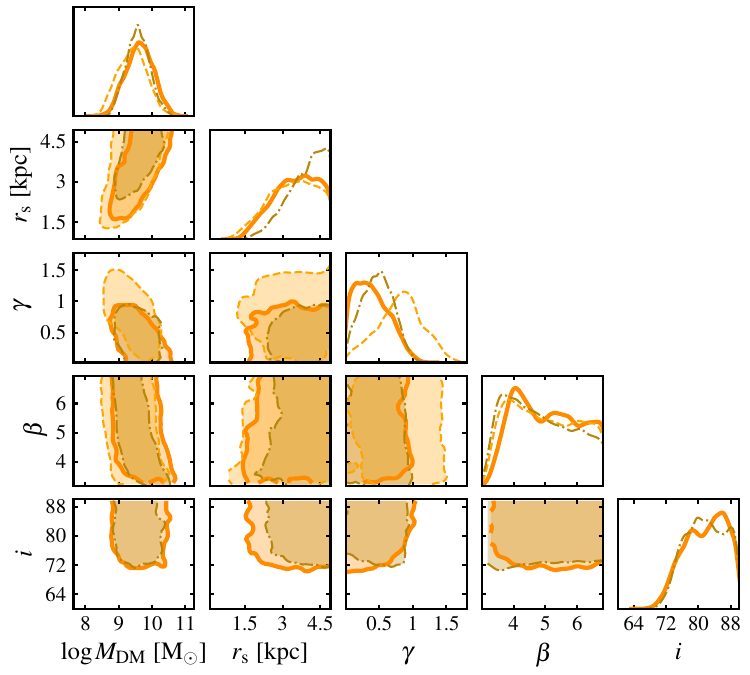}
    \caption{Marginalised one- and two-dimensional posterior distributions of the DM halo parameters of Draco (left) and Ursa Minor (right). We report the DM total mass $\log\Mdm$, the halo scale radius $\rs$, and the inner and outer slopes of the DM density profile, $\gamma$ and $\beta$ (equation~\ref{for:dm}). We also include the inclination $i$. Different colours and line styles indicate the flattened two-component models (thick solid), the flattened one-component models (dashed), and the spherical two-component model (dot-dashed). For clarity, we show only the regions enclosing the 95\% probability. In spherical two component models the inclination is not a free parameter.}
    \label{fig:corner}
    \vspace{-0.3cm}
\end{figure*}

For Ursa Minor, a relevant comparison is with \citet{Yang2025} and \citet{Hayashi2020}. The former favour a weakly cusped or partially cored DM profile, with $\gamma = 0.35^{+0.24}_{-0.17}$, while the latter predict a steeper cusp, $\gamma = 1.16^{+0.44}_{-0.66}$. The results obtained from the G26 sample are in close agreement with \citet{Yang2025}, favouring a similarly shallow inner slope. By contrast, the inference based on the W23 catalogue falls between these two estimates. In terms of normalization, our analysis applied to both the G26 and W23 datasets yield lower central DM densities compared to \citet{Yang2025}. Taking the G26 result as reference, we find $\rhodmcl=4.39_{-1.12}^{+1.33}\times10^7\Msun\kpc^{-3}$, approximately a factor of two smaller than the value reported by \citet{Yang2025}. At larger radii however, the outer density profile from G26 is fully consistent, within the uncertainties, with \citet{Yang2025}, while being significantly more extended than predicted by \citet{Hayashi2020}, whose models favour a more compact and denser DM halo. The DM profile inferred from W23 also shows an offset of about a factor of two with respect to \cite{Yang2025}; however, in this case the discrepancy persists over the full radial extent of the galaxy. This difference can be traced back to the systematic offset between W23 and G26. As discussed in Section \ref{subsub:umi}, the G26 kinematic sample exhibits a consistently higher velocity dispersion than W23 (see also Table~\ref{tab:modparams}), and this difference propagates into the inferred mass profiles, ultimately driving the discrepancy between W23 and both G26 and \cite{Yang2025}. All the relevant quantities discussed above are summarised in Table~\ref{tab:dm}, where the measurements obtained from both spectroscopic catalogues can be directly compared.

\subsubsection{Insensitivity to inclination}

Recently, \cite{Vitral2026} have highlighted the potential impact of galaxy inclination on the inferred inner DM density slope, showing that, in the case of the Sculptor dSph, a strong degeneracy can arise between the assumed inclination and the recovered DM profile. In their analysis, based on axisymmetric Jeans modelling, acceptable fits are found across a wide range of models, from cuspy to cored DM halos. We do not find evidence for such a degeneracy in our analysis. In Fig.~\ref{fig:corner} we present the marginalised one- and two-dimensional posterior distributions of the inclination $i$, and of the main DM halo parameters (the total mass, $\log\Mdm$, the scale radius, $\rs$, and the inner and outer slopes, $\gamma$ and $\beta$) for both Draco and Ursa Minor obtained with the G26 sample. Although the posterior on $i$ is skewed towards edge-on configurations -- i.e. the regime in which \cite{Vitral2026} preferentially recover steeper, cuspy profiles -- we do not observe any significant correlation between inclination and the inner density slope $\gamma$, nor with other halo parameters. In particular, although the high-probability region of the posterior of $i$ spans a broad range of inclinations, the inferred values of $\gamma$ remain remarkably stable across it. We therefore conclude that, within our modelling framework, the inference of $\gamma$ appears largely robust against uncertainties in the inclination, with no clear indication of degeneracy.

Similar results are found by \cite{Gherghinescu2026} who analyse axisymmetric mock galaxies constructed both from $N$-body simulations and from analytic DFs, using action-based DFs and Schwarzschild modelling, and show that the inclination is generally well recovered. The main effect the authors report is an increase in the uncertainties of the inferred DM halo properties towards face-on configurations, but no systematic bias with inclination. Our findings and those of \cite{Gherghinescu2026} suggest that the inclination degeneracy reported by \cite{Vitral2026} may be method-dependent and present only in specific applications of Jeans-based approaches.

\subsection{Astrophysical $J$- and $D$- factors}
\label{subsec:jd}

dSphs are among the most favourable targets for indirect DM searches via high-energy emission from particle annihilation or decay \citep{Turner1984,Albert2020,Hu2024}. This is due to their high DM content, proximity, and lack of astrophysical gamma-ray sources that could mask a DM signal. In addition, their angular size is well matched to the field of view of current gamma-ray instruments, allowing most of their halos to be probed in a single observation.

Once the DM density profile of a system is constrained through dynamical modelling, the expected gamma-ray flux from annihilation or decay can be predicted in a straightforward manner. The astrophysical contribution to this flux is fully encoded in the so-called $J$ and $D$ factors, respectively. For an observation within an angular aperture $\theta$, these quantities are defined as \citep[e.g.][]{Evans2016}
\begin{equation}
    J(\theta) = \frac{2\pi}{d^2}\int_{-\infty}^{+\infty} \dd z\int_0^{\theta d} \rhodm^2(R,z) R\dd R,
\end{equation}
and
\begin{equation}
    D(\theta) = \frac{2\pi}{d^2}\int_{-\infty}^{+\infty} \dd z\int_0^{\theta d} \rhodm(R,z) R\dd R,
\end{equation}
where $d$ is the distance to the galaxy, $R$ is the projected distance from the centre on the plane of the sky, $z$ denotes the line of sight, and $\rhodm$ is the DM density distribution. These quantities provide the essential link between the DM distribution of a galaxy and the observable gamma-ray signal and constitute a key input for all indirect detection analyses \citep{GeringerSameth2015,Evans2016}.

To date, no statistically significant gamma-ray excess has been detected from dSphs. As a consequence, current constraints on DM particle properties rely almost entirely on non-detections interpreted through robust estimates of $J$ and $D$ factors. Such analyses have been used to place upper limits on the annihilation cross section and lower limits on the decay lifetime for a wide range of DM candidates, in some cases ruling out entire classes of models \citep{GeringerSameth2015,Hoof2020,DiMauro2022}. Despite the lack of confirmed signals, dSphs remain prime targets for present and future indirect detection experiments, and accurate determinations of their $J$ and $D$- factors remain of central importance.

In this context, we compute the $J$ and $D$ factors for Draco and Ursa Minor using the DM density profiles inferred from our dynamical models. Throughout this section, we quote logarithmic values integrated within an angular aperture of $0.5\deg$, a choice commonly adopted in indirect detection studies and well suited to the angular resolution of current gamma-ray telescopes. For Draco, from the the G26 sample, we obtain\footnote{Annihilation $J$ factors are always given in units of $\GeV^2 \cm^{-5}$, while decay $D$ factors in units of $\GeV \cm^{-2}$.} $\logJ = 18.60_{-0.17}^{+0.19}$ and $\logD = 18.88_{-0.14}^{+0.21}$. For Ursa Minor, our estimates are $\logJ = 18.51_{-0.18}^{+0.18}$ and $\logD = 18.10_{-0.13}^{+0.15}$. Using the W23 sample, we obtain essentially identical results for Draco, as expected given the close similarity of the inferred DM density profiles. For Ursa Minor, instead, the two catalogues yield more noticeable differences, with the estimates based on W23 generally leading to slightly lower $J$ and $D$ factors (see Table~\ref{tab:dm}).

Comparing the results from the G26 sample with \cite{Hayashi2020}, for Draco, the authors report $\logJ = 19.03_{-0.28}^{+0.37}$ and $\logD = 18.86_{-0.19}^{+0.21}$, while for Ursa Minor $\logJ = 18.59_{-0.26}^{+0.44}$ and $\logD = 18.09_{-0.12}^{+0.17}$. Our estimates are consistent with these values within the quoted uncertainties, except for the $J$ factor of Draco, which is only marginally compatible. A similar comparison with \citet{Yang2025} also yields broad agreement. In this case, the $J$ factors are consistent within the uncertainties, whereas the main discrepancy concerns the $D$ factor of Draco, for which our models predict a comparatively larger value. It is noteworthy that, despite the overall similarity of the inferred DM density profiles, the derived $J$ and $D$ factors for Draco still exhibit non-negligible uncertainties and residual discrepancies. As discussed by \citet{Yang2025}, such differences may primarily arise from alternative parameterisations and priors of the -- unconstrained -- outer DM halo, which can significantly affect the line-of-sight integrals entering the determination of these quantities.

Several additional determinations of $J$ and $D$ factors for Draco and Ursa Minor exist in the literature. An exhaustive comparison is beyond the scope of this work; here we restrict the discussion to studies adopting symmetry assumptions most similar to ours. Further estimates can be found, for example, in \cite{GeringerSameth2015} and \cite{Evans2016}.

\begin{figure*}
\begin{minipage}{0.7\textwidth}
    \includegraphics[width=0.5\linewidth]{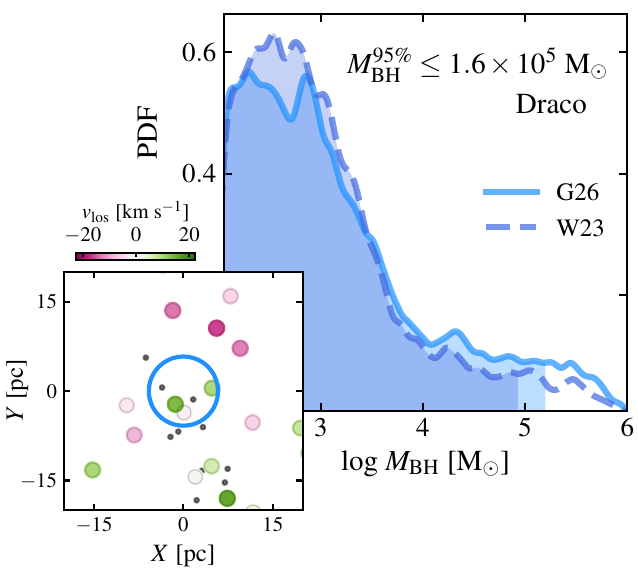}
    \includegraphics[width=0.5\linewidth]{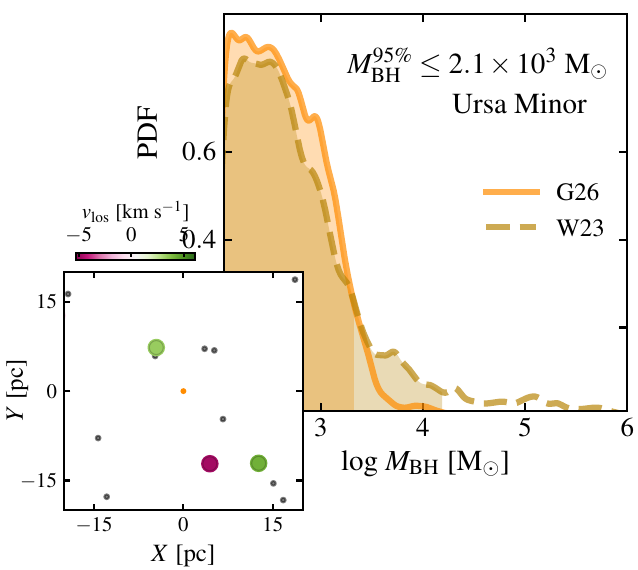}
\end{minipage}
\hfill\hfill
\begin{minipage}{0.28\textwidth}
\caption{IMBH properties, with the left and right pairs of panels referring to Draco and Ursa Minor, respectively. In each pair, the large panel shows the posterior distribution of $\log\BH$. The shaded regions indicate the $2\sigma$ upper limit, computed as HPD (see Appendix~\ref{app:A}). The smaller panel shows the spatial distribution of the tracers, colour-coded according to their line-of-sight velocity, when available. The coloured circle is centred on the galaxy centre and has a radius equal to the upper limit on the radius of influence, $\roi$.}
\label{fig:imbhs}
\end{minipage}
\vspace{-0.4cm}
\end{figure*}

\subsection{Comparison with one-component models}
\label{subsec:onec}

 
To quantify possible systematic differences and assess the importance of explicitly modelling multiple stellar populations, we complement our analysis with a set of flattened one-component models. In these models, the stellar body of each galaxy is described as a single tracer population, without exploiting metallicity information. As a consequence, the eDF~(\ref{for:tot}) reduces to the DF ~(\ref{for:df}). All other aspects of the modelling framework do not change. By construction, this simplification leads to a substantial reduction in the number of free parameters, as it removes those associated with the decomposition into MR and MP populations.

Remarkably, the inner slopes inferred for Draco with flattened one-component models are consistent across the G26 ($\gamma=0.90_{-0.29}^{+0.27}$) and W23 ($\gamma=0.75_{-0.36}^{+0.32}$) samples, and these values are fully consistent with those obtained from the corresponding two-component models. Ursa Minor shows larger variability. The slope inferred from the G26 sample ($\gamma = 0.48_{-0.25}^{+0.23}$) remains consistent, within the uncertainties, with the two-component models, while the W23 sample ($\gamma = 0.55_{-0.28}^{+0.26}$) shows a slightly larger discrepancy. In this case, flattened one-component models yield a slope closer to that inferred from the G26 sample. All these values are reported in Table~\ref{tab:dm}.

In Fig.~\ref{fig:corner} we show the marginalised one- and two-dimensional posterior distributions of the main DM halo parameters for the flattened one-component models, and for the corresponding flattened two-component models for comparison. For simplicity, the figure is limited to results obtained from the G26 spectroscopic dataset. For Draco, the remaining halo parameters are consistent across the different modelling assumptions. The posterior distributions (see Fig.~\ref{fig:corner}) are nearly identical, and this stability is reflected in Fig.~\ref{fig:dm}, where the median DM density profiles of the flattened one-component models applied to the two datasets closely overlap. In contrast, Ursa Minor shows a different behaviour. The flattened one-component models yield DM density profiles that differ from their two-component counterparts, although, as anticipated above, the one-component profiles are more similar to each other than the corresponding two-component models. The DM density at $150\pc$ is reported in Table~\ref{tab:dm}. In all cases, however, the posterior distributions of $\rs$ and of the outer halo slope remain largely unconstrained.

\subsection{Comparison with spherical models}
\label{subsec:twocsph}

To assess the impact of the assumption of spherical symmetry (often adopted in previous works), we also consider spherical two-component models. In practice, spherical symmetry is enforced by requiring the DFs to depend on the actions only through the modulus of the angular momentum. This is achieved by imposing that the coefficients of $\Jphi$ and $\Jz$ in equation~(\ref{for:hg}) are equal. All other elements of the modelling framework remain unchanged.

The comparison between spherical and flattened two-component models reveals different behaviours in the two galaxies. In Draco, the inner DM slope inferred from the G26 and W23 catalogues is very similar and favours  cuspy halos. We measure slopes of $\gamma=1.17_{-0.31}^{+0.25}$  and $\gamma=0.95_{-0.32}^{+0.37}$ for the G26 and W23 samples, respectively. These values are remarkably consistent with those obtained with the flattened two-component models, although the latter tend to prefer slightly shallower inner slopes. Ursa Minor exhibits a more complex behaviour. In this case, spherical models systematically favour steeper inner slopes in both datasets ($\gamma=0.83_{-0.37}^{+0.32}$ of G26 and $\gamma=1.13_{-0.32}^{+0.27}$ for W23). This difference is also visible in Fig.~\ref{fig:dm}, which shows the corresponding spherical DM density profiles. While the profiles inferred for Draco remain largely stable across the different modelling assumptions, those of Ursa Minor display substantially larger variations. Inspection of the posterior distributions (Fig.~\ref{fig:corner}) indicates that most halo parameters remain weakly constrained, with the inner slope $\gamma$ primarily driving the differences between the models. Estimates of the remaining halo parameters are reported in Table~\ref{tab:dm}.

\subsection{Upper limits on IMBHs}
\label{subsec:dscimbhs}

To date, little attention has been devoted to the search for IMBHs in dSphs, largely because detecting their dynamical signatures is intrinsically difficult. Unlike GCs, which are nearby and dominated by their stellar component, dSphs are typically more distant and embedded in gravitational potentials overwhelmingly dominated by DM. As a result, even a central IMBH is expected to leave only a weak dynamical imprint: the characteristic rise in velocity dispersion is strongly suppressed \citep{Pascale2019}, with detectable signatures confined to subtle features in the high-velocity tails of the stellar velocity distributions.

Despite these challenges, interest in IMBHs in dSphs has recently increased. This renewed attention has been driven in part by a claimed detection of SMBH in Leo I \citep{Bustamante2021}, which were later shown to be unsupported by the data \citep{Pascale2024a}, highlighting both the difficulty and the importance of robust dynamical modelling. Previously, only a limited number of studies have addressed this problem, placing upper limits on putative IMBHs masses in classical dSphs, most notably in Ursa Minor, using either X-ray observations \citep{Nucita2013} or stellar kinematics \citep{Lora2009,Manni2015}. Additional constraints have also been reported for other systems, such as Fornax \citep{Jardel2012}. More recently, \citet{Aditya2026} derived systematic upper limits on IMBH masses in classical dSphs using DF-based dynamical models.

The two sets of panels in Fig.~\ref{fig:imbhs} show the constraints resulting from our models regarding the presence of IMBHs in Draco (left) and Ursa Minor (right). In each group, the main panel shows the posterior distributions of $\log\BH$ obtained with the two samples. The smaller inset displays the projected spatial distribution of member stars from the G26 datasets, which are the ones that provide a larger number of kinematic tracers in the centre. Here, the circle at the centre marks the 95\% HPD upper limit on the IMBH radius of influence, $\roi$\footnote{The $\roi$ is the distance along the semi-major axis where the stellar line-of-sight velocity dispersion equals the circular velocity of the BH potential \citep{BinneyTremaine2008}, approximately indicating the radius within which the IMBH dominates the stellar potential.} 

The posterior distribution of $\log\BH$ indicates that the IMBH mass is not constrained toward low values. Consequently, within the adopted priors, only an upper limit can be placed on the IMBH mass in either galaxy, which we quote as the $2\sigma$ (95\% HPD, see Appendix~\ref{app:A}) bound. Adopting the G26 sample we get a $2\sigma$ limit of $\log\BH[\Msun]\leq 5.20$ for Draco and $\log\BH[\Msun]\leq 3.33$ for Ursa Minor, with similar limits obtained from the W23 samples. As shown by the posterior distribution in Fig.~\ref{fig:imbhs}, the constraints for Ursa Minor are remarkably stringent, effectively excluding IMBHs with masses $\ge10^4\Msun$, particularly with the G26 catalogue. In both galaxies, the inferred $2\sigma$ upper limit on the $\roi$ is very small (of less than one parsec). 
Inspection of the spatial distributions in Fig.~\ref{fig:imbhs} clarifies the origin of these constraints. As discussed in Section~\ref{subsec:imbh}, in $\fJ$ models a central BH growing adiabatically is expected to induce a mild cusp in the stellar distribution on scales comparable to its radius of influence. In our models, therefore, both the spatial distribution of the stars and the kinematics contribute to constraining the presence of a central BH. As shown in Fig.~\ref{fig:imbhs}, the innermost regions are sparsely sampled, so that our constraints mainly reflect the absence of high velocity stars and mild cusps, effectively ruling out large IMBH masses.


Recently, \cite{Aditya2026} provided a systematic compilation of IMBH mass estimates (upper limits) in classical dSphs. Their models are likewise based on DFs, but of the \citep{Osipkov1979,Merritt1985} type, and are applied to kinematic datasets largely drawn from the same W23 samples used here, enabling a direct comparison. At the 95\% confidence level, \cite{Aditya2026} report upper limits on the IMBH masses $\log\BH [\Msun]<5.46$ for Draco and $\log\BH [\Msun]<5.34$ for Ursa Minor. At the same statistical significance, our upper limits are a factor 2 smaller in Draco and almost two orders of magnitude smaller in Ursa Minor, much more stringent. We attribute this discrepancy mainly to differences in the adopted methodology. While the underlying datasets are broadly similar, \cite{Aditya2026} fit binned velocity-dispersion profiles and provide the spatial distribution of the tracers as an input of the models, whereas our analysis operates directly on individual stellar phase-space positions. In binned approaches, the kinematic information carried by individual stars is averaged through radial bins. The finite width of these bins -- and the uncertainty associated with it -- is typically not propagated into the modelling. As a consequence, the effective spatial scale at which the kinematic constraints apply is set by the mean distance of the innermost bin, rather than by the innermost stars that would otherwise provide the strongest constraint on the potential in the central regions. These results therefore underscore the critical importance of incorporating individual stellar phase-space position when searching for IMBHs. Finally, we note that our upper limits are significantly more restrictive than those proposed by \cite{Nucita2013}, while remaining fully consistent with the constraints reported by \cite{Lora2009}.


\section{Discussion}
\label{sec:disc}

\subsection{Effect of chemo-kinematic complexity and symmetry assumptions on DM inference}

Early observations of dSphs lacked the precision and sample sizes needed to fully resolve their internal structure. As a result, they were often described and modelled as relatively simple stellar systems, primarily characterised by large dynamical mass-to-light ratios. With increasingly large and precise photometric and kinematic datasets, together with improved theoretical understanding of galaxy formation and evolution, a more complex picture has gradually emerged. Many dSphs host chemo-kinematically distinct stellar populations with different spatial distributions \citep{Tolstoy2004,Battaglia2006}, and several exhibit significant flattening \citep{Munoz2018} despite frequently being modelled under the assumption of spherical symmetry. 
These complexities highlight a practical limitation: in many cases it is not yet possible to incorporate all these aspects simultaneously in dynamical models, and it remains unclear how strongly the inferred DM properties may depend on them being neglected. Our analysis, although applied to a limited number of test cases, provides a first step toward addressing this issue, as the modelling framework presented allows these complexities to be accounted for and treated in a physically consistent way. Interpreting the results nevertheless requires careful consideration, as each galaxy presents its own peculiarities.

In the case of Draco, the inferred DM profile remains remarkably stable across different datasets and modelling assumptions. Using multiple spectroscopic datasets, dividing or not between stellar populations, and exploring different symmetries (spherical and axisymmetric), we consistently recover the same properties of the underlying DM halo (within uncertainties). This agreement strongly indicates that the inferred halo structure is highly robust and, in this sense, we can now argue that Draco stands out as the classical dSphs for which the DM density profile is currently best constrained. This robustness reinforces its importance even more as a target for indirect DM searches. Draco is frequently identified as the most promising system for gamma-ray observations aimed at detecting signals of annihilation or decay of DM particles \citep[e.g.][who provided a compilation of $J$ and $D$ factors for all dSphs]{Hayashi2020}, and our results indicate that the inferred astrophysical factors are not only among the largest, but also relatively insensitive to modelling assumptions.



Ursa Minor highlights a different situation. In this system, the inferred DM structure shows a stronger dependence on the adopted modelling assumptions. In particular, spherical models systematically favour steeper inner density profiles than axisymmetric models. This behaviour is present, albeit to an extent that can be accommodated within the uncertainties, also in Draco and points to a potential source of systematic bias. Correctly modelling the intrinsic flattening of dSphs is therefore particularly important, as these systems are often significantly non-spherical, with Ursa Minor being an extreme example in this respect, since it is one of the most flattened classical dSphs.

\subsection{Implications for the core--cusp problem}
\label{subsec:corecusp}

Hydrodynamical simulations have shown that stellar feedback can potentially transform cuspy DM profiles into cores \citep[e.g.][]{Pontzen2012}. The efficiency of this mechanism depends strongly on both the stellar-to-halo mass ratio and the total stellar mass \citep{DiCintio2014a}, with galaxies below a certain threshold mass starting to provide less to insufficient feedback energy to significantly alter the DM distribution. However, the specific value of this threshold is not well established, with different simulations suggesting values around $\Mst\simeq10^6\Msun$ with some scatter \citep{Onorbe2015,Fitts2017}. Also, it has been shown that the detailed star formation history plays a key role, with core formation occurring in low-mass systems, provided that star formation is sufficiently extended and bursty \citep{Onorbe2015,Read2018}.

Draco and Ursa Minor share remarkably similar global properties. In particular, both have very low stellar masses among dSphs, $\Mst \simeq 3 \times 10^5\Msun$, with variations up to a factor of two \citep[e.g.][]{McConnachie2012,Kirby2013}, thus probing a similar stellar-mass regime. Both systems are strongly DM dominated (in Table~\ref{tab:dm}, we report the enclosed DM mass within some characteristic radii for comparison), with Ursa Minor showing a higher enclosed DM mass. Beyond their mass similarity, they exhibit broadly comparable star formation histories, with most of their stellar populations formed at early times  \citep[$\gtrsim 10,\Gyr$, e.g.][]{Aparicio2001,Sato2025}, although differences in their detailed evolution may still be present.

In this context, our inference on the inner halo density of Draco are broadly consistent with theoretical expectations: given its low stellar mass and large DM content, a cuspy halo naturally falls within the regime where feedback-driven core formation is expected to be inefficient, as also noted by several authors \citep{Read2018,Vitral2024}. Although similar results might be expected for Ursa Minor, interpreting our DM estimates is more challenging for two main reasons. First, the inference in the inner structure is not stable across the different datasets used. Second, our results exclude profiles that are as steep as the one inferred for Draco. A possible key ingredient in understanding the origin of a shallower profile, and differences with respect to Draco, is the fraction of star formation that occurred before and after reionization ($z \simeq 6$). In the dwarf-galaxy regime, systems with a larger fraction of late-time star formation (i.e. after reionization) tend to be more depleted in DM in their central regions, while systems dominated by early star formation are less efficient in this process \citep{Muni2025}. In this sense, differences in the timing of star formation between Draco and Ursa Minor could naturally contribute to explaining their different inner DM structures. However, although in both galaxies the bulk of star formation occurred at lookback times $\gtrsim 10$ Gyr (corresponding to $z \gtrsim 1.7$), current determinations of their star formation histories do not provide sufficient temporal resolution to accurately quantify the fraction of star formation occurring before and after reionization. Nonetheless, the available evidence suggests that Ursa Minor may have quenched star formation later than Draco \citep{Weisz2014}.

Ursa Minor represents a particularly complex system, and several of its structural and dynamical peculiarities may contribute to the scatter observed in the inferred DM halo properties. These include the presence of localised kinematic substructures and/or off-centred stellar overdensities \citep[e.g.][]{Kleyna2003,Pace2014}, and evidence for tidal disturbances, such as extra-tidal stars and tidal extensions, as well as possible velocity gradients induced by the interaction with the Milky Way \citep{Martinez2001,Martinez2023}. Taken together, these features suggest that Ursa Minor may depart from strict dynamical equilibrium, and therefore equilibrium-based dynamical models may require some caution in interpretation.

\subsection{Implications of stringent upper limits on $\BH$}
\label{subsec:imbhimpl}

Since massive BHs are thought to grow from lower-mass seeds formed at high redshift, a relic population of their progenitors is expected to survive in the present-day Universe. Nearby dwarf galaxies are particularly valuable in this context \citep{Natarajan2021}, as their relatively quiet merger histories and limited late-time growth may preserve a clearer imprint of the original seed population. As a result, the demographics of BHs in dwarfs can provide direct constraints on seed formation and early growth. However, observational constraints at dwarf-galaxy scales remain challenging. Direct BH mass measurements are scarce \citep{denBrok2015,Nguyen2018,Nguyen2019}, since dynamical detections become increasingly difficult at low masses: the sphere of influence rapidly falls below current instrumental resolution, and detections rely on a small number of stars at the centre that may remain unobserved. These difficulties are compounded by the expected decline of the BH occupation fraction toward low galaxy masses, implying that a significant fraction of dwarfs may be intrinsically BH-free \citep{vanWassenhove2010,Volonteri2016,Haidar2022}.

In more massive galaxies, the empirical correlation between central BH mass and stellar velocity dispersion, the $\BH-\sigma$ relation, provides a key benchmark for BH–galaxy coevolution. This relation is well described by a power law with relatively small intrinsic scatter at the high-mass end \citep{Ferrarese2000,Gebhardt2000,Merritt2001}, and has been progressively refined through larger and more homogeneous samples, and improved treatments of uncertainties and systematics \citep{Tremaine2002,Gultekin2009,McConnell2013,Kormendy2013,Saglia2016}. Its applicability in the dwarf-galaxy regime, however, remains uncertain. Different seeding channels can produce BHs spanning several orders of magnitude in mass, from light Population III remnants ($\simeq100\Msun$) to massive direct-collapse seeds ($\simeq10^4$--$10^6\Msun$), suggesting that the intrinsic scatter of low-mass scaling relations may be large \citep[e.g.][]{Bromm2003,Lodato2006,Begelman2006,Greene2020}. Nevertheless, such extrapolations can still provide a useful order-of-magnitude reference. Using the $\BH$--$\sigma$ relation of \cite{vandenBosch2016} and extrapolating it down to velocity dispersions typical of Draco and Ursa Minor ($\sigma \simeq 7$--$10\kms$), we obtain expected BH masses of $\lesssim 3 \times 10^3\Msun$. These values are consistent with those inferred from more recent calibrations \citep[e.g.][]{Aditya2026}, and are in full agreement with the upper limits derived in this work.

These arguments highlight the value of stringent dynamical constraints in nearby dSphs. In the absence of secure IMBH detections at $\sigma\simeq7$--$10\kms$, robust upper limits still provide key information where BHs retain the strongest memory of their initial seed masses. In their work, \cite{vanWassenhove2010} follow the formation and evolution of BH population in a MW-like galaxy. In their model they include two different formation mechanisms, seeds from population III stars and via gas dynamical collapse, and predict the present-day demographics of massive BHs in dwarf satellite galaxies. Their predictions differ markedly depending on the seeding channel: in massive-seed scenarios (gas-dynamical collapse) the present-day BH population is less than a few per cent in typical dwarf galaxies, although the few that exist may be massive enough to be detectable in the brightest systems. Conversely, Population III remnants yield higher occupation fractions but much smaller BH masses, making dynamical detection challenging even in the Local Group. Our upper limits in Draco and Ursa Minor ($\BH\leq1.6\times10^5\Msun$ and $\BH\leq2.1\times10^3\Msun$) are consistent with limited BH growth in dwarf galaxies and effectively rule out the presence of a central BH with $\BH\gtrsim10^5\Msun$ in either system. In the context of \cite{vanWassenhove2010}, models involving massive seeds predict BH masses that, if present, would exceed our upper limits, thereby strongly disfavouring this formation channel in classical dSphs at $\Mst\simeq3\times10^5\Msun$. However, with only two galaxies, these non-detections can only suggest that the BH occupation fraction is low rather than provide a robust constraint. Extending similarly deep dynamical constraints to a larger sample of dSphs will therefore be essential to place quantitative constraints on the BH occupation fraction at the low-mass end and, ultimately, to discriminate between seeding channels.

\section{Summary and conclusions}
\label{sec:concl}

In this work we apply dynamical models based on analytic DFs that depend on the actions to study the internal dynamics of the dSphs Draco and Ursa Minor. Our models account for multiple stellar populations, their flattening, a flexible spherical DM halo, and a possible central IMBH. We construct homogeneous samples of likely member stars using Gaia photometric and astrometric data, combined with two independent spectroscopic datasets, allowing us to test the robustness of our results against the choice of kinematic data. We further assess the stability of our inferences by exploring different modelling assumptions, including single-component and spherical two-component configurations. By fitting the chemo-dynamical properties of individual stars (no binning required) our approach provides a self-consistent description of the stellar phase-space distribution and simultaneously constrains the structure of the stellar components, the DM halo, and any central BH. To our knowledge, this is the first dynamical analysis of these systems based on fully self-consistent, axisymmetric models based on DFs.

Our main results can be summarised as follows: In both Draco and Ursa Minor the data require two chemo-dynamically distinct stellar populations. The MR component is more centrally concentrated and dynamically colder, while the MP component is more spatially extended and characterised by a larger velocity dispersion. In agreement with previous studies, we also find that in Ursa Minor the MP population is significantly more flattened than the MR one, indicating a more complex internal structure.

Our results also highlight a complex projected kinematic structure. In particular, the maps of projected velocity dispersion show patterns that do not simply follow the assumed axial symmetry of the models, suggesting that additional structural complexity may be present in the data.

We investigate possible rotational signatures. In Draco, the detection of rotation is weak, with a mild signal of rotation around the minor axis found in both samples. The rotation is apparent in the overall stellar distribution and it is mostly driven by the MP population. In Ursa Minor, we consistently detect a prolate rotation signal in the MP component in both datasets. The MR population is always fully pressure supported. When detected, the inferred rotational support is always small ($\Delta \vrotmax/\sloso \lesssim 0.1$ in Draco and $\lesssim 0.2$ in Ursa Minor), thus dynamically subdominant, indicating that rotation does not significantly contribute to the overall dynamical support of either galaxy, and that neglecting rotation in the models does not bias the inferred dynamical masses, since the underestimate is proportional to the square of the ratio of rotational velocity to velocity dispersion.

For Draco we infer a cuspy inner DM density profile, with slope $\gamma=0.98_{-0.26}^{+0.28}$. For Ursa Minor we instead obtain a shallower inner profile, with $\gamma=0.37_{-0.24}^{+0.31}$, more compatible with a core-like distribution, although the constraints are weaker.

The inference of a cuspy halo in Draco proves to be remarkably robust. We explicitly tested the stability of our results against several modelling assumptions, including two-component and one-component axisymmetric models as well as spherical two-component models. To our knowledge, this systematic exploration of modelling choices has not been performed before for this system. The consistency of the inferred inner slope across these different models strongly suggests that Draco is currently the dSph with the best determined DM halo. In contrast, for Ursa Minor the inferred DM properties still show a stronger dependence on the adopted modelling assumptions.

This result has very important implications for indirect DM detection experiments. Draco already stands out as one of the most promising targets due to its large annihilation $J$- and decay $D$-factors, and our analysis indicates that these quantities are also among the most robustly determined, remaining stable across a wide range of dynamical modelling assumptions.

We do not find evidence for a significant degeneracy between inclination and the inferred DM halo properties. Although the inclination is largely unconstrained and slightly skewed towards large values, the recovered inner slope $\gamma$ remains stable across different inclinations, indicating that our results are insensitive to uncertainties in the viewing angle.

We find no dynamical evidence for a central IMBH in either galaxy. The posterior distributions provide only upper limits on the IMBH mass, $\log\BH[\Msun]<5.2$ for Draco and $\log\BH[\Msun]<3.33$ for Ursa Minor (95\% confidence). These limits are among the most stringent dynamical constraints currently available for classical dSphs. In particular, the limit obtained for Ursa Minor is extremely restrictive, effectively ruling out BHs more massive than $\simeq2\times10^4\Msun$.

\begin{acknowledgements}
The research activities described in this paper have been co-funded by the European Union -- NextGeneration EU within PRIN 2022 project n.20229YBSAN -- Globular clusters in cosmological simulations and in lensed fields: from their birth to the present epoch. RP acknowledges the support to this study by the INAF Mini Grant 2025 (Ob.Fu.1.05.24.07.05, CUP C33C24001390005). GB, JMAP and GFT acknowledge support from the Agencia Estatal de Investigación del Ministerio de Ciencia, Innovación y Universidades (MCIU/AEI) under grant EN LA FRONTERA DE LA ARQUEOLOGÍA GALÁCTICA: EVOLUCIÓN DE LA MATERIA LUMINOSA Y OSCURA DE LA VÍA LÁCTEA Y LAS GALAXIAS ENANAS DEL GRUPO LOCAL EN LA ERA DE GAIA. (FOGALERA) and the European Regional Development Fund (ERDF) with reference PID2023-150319NB-C21 / 10.13039/501100011033. GFT acknowledges support from the Agencia Estatal de Investigaci\'on del Ministerio de Ciencia en Innovaci\'on (AEI-MICIN) under grant number PID2020-118778GB-I00/10.13039/501100011033 and the grant RYC2024-051016-I funded by MCIN/AEI/10.13039/501100011033 and by the European Social Fund Plus. EV acknowledges support from an STFC Ernest Rutherford fellowship (ST/X004066/1)

\end{acknowledgements}


\bibliographystyle{aa}
\bibliography{main}

@ARTICLE{Turner1984,
       author = {{Turner}, M.~S. and {Steigman}, G. and {Krauss}, L.~M.},
        title = "{Flatness of the Universe: Reconciling Theoretical Prejudices with Observational Data}",
      journal = {\prl},
         year = 1984,
        month = jun,
       volume = {52},
       number = {23},
        pages = {2090-2093},
          doi = {10.1103/PhysRevLett.52.2090},
       adsurl = {https://ui.adsabs.harvard.edu/abs/1984PhRvL..52.2090T}
}

@ARTICLE{Arroyo2024,
       author = {{Arroyo-Polonio}, Jos{\'e} Mar{\'\i}a and {Battaglia}, Giuseppina and {Thomas}, Guillaume F. and {Pascale}, Raffaele and {Tolstoy}, Eline and {Nipoti}, Carlo},
        title = "{Chemo-dynamics of the stellar component of the Sculptor dwarf galaxy: I. Observed properties}",
      journal = {\aap},
         year = 2024,
        month = dec,
       volume = {692},
          eid = {A195},
        pages = {A195},
          doi = {10.1051/0004-6361/202451102},
archivePrefix = {arXiv},
       eprint = {2411.07283},
 primaryClass = {astro-ph.GA},
       adsurl = {https://ui.adsabs.harvard.edu/abs/2024A&A...692A.195A}
}

@ARTICLE{Arroyo2025,
        author = {{Arroyo-Polonio}, Jos{\'e} Mar{\'\i}a and {Pascale},
Raffaele and {Battaglia}, Giuseppina and {Thomas}, Guillaume F. and
{Nipoti}, Carlo and {Vasiliev}, Eugene and {Tolstoy}, Eline},
         title = "{Chemo-dynamics of the stellar component of the
Sculptor dwarf galaxy: II. Dynamical properties and dark matter halo
density}",
       journal = {\aap},
          year = 2025,
         month = jul,
        volume = {699},
           eid = {A347},
         pages = {A347},
           doi = {10.1051/0004-6361/202554826},
archivePrefix = {arXiv},
        eprint = {2506.11845},
  primaryClass = {astro-ph.GA},
        adsurl = {https://ui.adsabs.harvard.edu/abs/2025A&A...699A.347A}
}

@ARTICLE{Arroyo2026,
       author = {{Arroyo-Polonio}, Jos{\'e} Mar{\'\i}a and {Battaglia}, Giuseppina and {Thomas}, Guillaume F.},
        title = "{Estimating the dynamical masses of dwarf galaxies in the presence of binary-star contamination}",
      journal = {arXiv e-prints},
         year = 2026,
        month = mar,
          eid = {arXiv:2603.03129},
        pages = {arXiv:2603.03129},
          doi = {10.48550/arXiv.2603.03129},
archivePrefix = {arXiv},
       eprint = {2603.03129},
 primaryClass = {astro-ph.GA},
       adsurl = {https://ui.adsabs.harvard.edu/abs/2026arXiv260303129A}
}

@ARTICLE{Abe2024,
       author = {{Abe}, S. and {Abhir}, J. and {Abhishek}, A. and {Acero}, F. and {Acharyya}, A. and {Adam}, R. and {Aguasca-Cabot}, A. and {Agudo}, I. and {Aguirre-Santaella}, A. and {Alfaro}, J. and {Alfaro}, R. and {Alvarez-Crespo}, N. and {Alves Batista}, R. and {Amans}, J.-P. and {Amato}, E. and {Ambrosi}, G. and {Angel}, L. and {Aramo}, C. and {Arcaro}, C. and {Arnesen}, T.~T.~H. and {Arrabito}, L. and {Asano}, K. and {Ascasibar}, Y. and {Aschersleben}, J. and {Ashkar}, H. and {Backes}, M. and {Baktash}, A. and {Balazs}, C. and {Balbo}, M. and {Baquero Larriva}, A. and {Barbosa Martins}, V. and {Barres de Almeida}, U. and {Barrio}, J.~A. and {Batkovi{\'c}}, I. and {Batzofin}, R. and {Baxter}, J. and {Becerra Gonz{\'a}lez}, J. and {Beck}, G. and {Benbow}, W. and {Berge}, D. and {Bernardini}, E. and {Bernete}, J. and {Bernl{\"o}hr}, K. and {Berti}, A. and {Bertucci}, B. and {Bhattacharjee}, P. and {Bhattacharyya}, S. and {Bigongiari}, C. and {Biland}, A. and {Bissaldi}, E. and {Biteau}, J. and {Blanch}, O. and {Blazek}, J. and {Bocchino}, F. and {Boisson}, C. and {Bolmont}, J. and {Bonnoli}, G. and {Bonollo}, A. and {Bordas}, P. and {Bosnjak}, Z. and {Bottacini}, E. and {B{\"o}ttcher}, M. and {Bringmann}, T. and {Bronzini}, E. and {Brose}, R. and {Brown}, A.~M. and {Brunelli}, G. and {Bulgarelli}, A. and {Bulik}, T. and {Burelli}, I. and {Burmistrov}, L. and {Burton}, M. and {Buscemi}, M. and {Bylund}, T. and {Cailleux}, J. and {Campoy-Ordaz}, A. and {Cantlay}, B.~K. and {Capasso}, G. and {Caproni}, A. and {Capuzzo-Dolcetta}, R. and {Caraveo}, P. and {Caroff}, S. and {Carosi}, A. and {Carosi}, R. and {Carquin}, E. and {Carrasco}, M.-S. and {Cassol}, F. and {Castaldini}, L. and {Castrejon}, N. and {Castro-Tirado}, A.~J. and {Cerasole}, D. and {Cerruti}, M. and {Chadwick}, P.~M. and {Chaty}, S. and {Chen}, A.~W. and {Chernyakova}, M. and {Chiavassa}, A. and {Chudoba}, J. and {Chytka}, L. and {Cicciari}, G.~M. and {Cifuentes}, A. and {Coimbra Araujo}, C.~H. and {Colapietro}, M. and {Conforti}, V. and {Conte}, F. and {Contreras}, J.~L. and {Costa}, A. and {Costantini}, H. and {Cotter}, G. and {Cristofari}, P. and {Cuevas}, O. and {Curtis-Ginsberg}, Z. and {D'Amico}, G. and {D'Ammando}, F. and {Dai}, S. and {Dalchenko}, M. and {Dazzi}, F. and {De Angelis}, A. and {de Bony de Lavergne}, M. and {De Caprio}, V. and {de Gouveia Dal Pino}, E.~M. and {De Lotto}, B. and {De Lucia}, M. and {de Menezes}, R. and {de Naurois}, M. and {de Souza}, V. and {del Peral}, L. and {del Valle}, M.~V. and {Delgado Giler}, A.~G. and {Delgado Mengual}, J. and {Delgado}, C. and {Dell'aiera}, M. and {della Volpe}, D. and {Depaoli}, D. and {Di Girolamo}, T. and {Di Piano}, A. and {Di Pierro}, F. and {Di Tria}, R. and {Di Venere}, L. and {D{\'\i}az}, C. and {Diebold}, S. and {Dinesh}, A. and {Djuvsland}, J. and {Dominik}, R.~M. and {Dominis Prester}, D. and {Donini}, A. and {Dorner}, D. and {D{\"o}rner}, J. and {Doro}, M. and {Dournaux}, J.-L. and {Duangchan}, C. and {Dubos}, C. and {Ducci}, L. and {Dwarkadas}, V.~V. and {Ebr}, J. and {Eckner}, C. and {Egberts}, K. and {Einecke}, S. and {Els{\"a}sser}, D. and {Emery}, G. and {Errando}, M. and {Escanuela}, C. and {Escarate}, P. and {Escobar Godoy}, M. and {Escudero}, J. and {Esposito}, P. and {Ettori}, S. and {Falceta-Goncalves}, D. and {Fedorova}, E. and {Fegan}, S. and {Feng}, Q. and {Ferrand}, G. and {Ferrarotto}, F. and {Fiandrini}, E. and {Fiasson}, A. and {Filipovic}, M. and {Fioretti}, V. and {Fiori}, M. and {Foffano}, L. and {Font Guiteras}, L. and {Fontaine}, G. and {Fr{\"o}se}, S. and {Fukazawa}, Y. and {Fukui}, Y. and {Furniss}, A. and {Galanti}, G. and {Galaz}, G. and {Galelli}, C. and {Gallozzi}, S. and {Gammaldi}, V. and {Garczarczyk}, M. and {Gasbarra}, C. and {Gasparrini}, D. and {Ghalumyan}, A. and {Gianotti}, F. and {Giarrusso}, M. and {Giesbrecht Formiga Paiva}, J.~G. and {Giglietto}, N. and {Giordano}, F. and {Giuffrida}, R.},
        title = "{Dark matter line searches with the Cherenkov Telescope Array}",
      journal = {\jcap},
         year = 2024,
        month = jul,
       volume = {2024},
       number = {7},
          eid = {047},
        pages = {047},
          doi = {10.1088/1475-7516/2024/07/047},
archivePrefix = {arXiv},
       eprint = {2403.04857},
 primaryClass = {hep-ph},
       adsurl = {https://ui.adsabs.harvard.edu/abs/2024JCAP...07..047A}
}

@ARTICLE{Aditya2026,
       author = {{Aditya}, K. and {Mangalam}, A.},
        title = "{Can dwarf spheroidal galaxies host a central black hole ?}",
      journal = {arXiv e-prints},
         year = 2025,
        month = dec,
          eid = {arXiv:2512.14146},
        pages = {arXiv:2512.14146},
          doi = {10.48550/arXiv.2512.14146},
archivePrefix = {arXiv},
       eprint = {2512.14146},
 primaryClass = {astro-ph.GA},
       adsurl = {https://ui.adsabs.harvard.edu/abs/2025arXiv251214146A}
}

@ARTICLE{Albert2020,
       author = {{Albert}, A. and {Alfaro}, R. and {Alvarez}, C. and {Arteaga-Vel{\'a}zquez}, J.~C. and {Arunbabu}, K.~P. and {Avila Rojas}, D. and {Ayala Solares}, H.~A. and {Belmont-Moreno}, E. and {BenZvi}, S.~Y. and {Brisbois}, C. and {Caballero-Mora}, K.~S. and {Capistr{\'a}n}, T. and {Carrami{\~n}ana}, A. and {Casanova}, S. and {Cotti}, U. and {Cotzomi}, J. and {Couti{\~n}o de Le{\'o}n}, S. and {De la Fuente}, E. and {Dichiara}, S. and {Dingus}, B.~L. and {DuVernois}, M.~A. and {D{\'\i}az-V{\'e}lez}, J.~C. and {Engel}, K. and {Espinoza}, C. and {Fraija}, N. and {Galv{\'a}n-G{\'a}mez}, A. and {Garc{\'\i}a-Gonz{\'a}lez}, J.~A. and {Garfias}, F. and {Gonz{\'a}lez}, M.~M. and {Goodman}, J.~A. and {Harding}, J.~P. and {Hernandez}, S. and {Hona}, B. and {Huang}, D. and {Huentemeyer}, P. and {Hueyotl-Zahuantitla}, F. and {Iriarte}, A. and {Joshi}, V. and {Lara}, A. and {Le{\'o}n Vargas}, H. and {Linnemann}, J.~T. and {Longinotti}, A.~L. and {Luis-Raya}, G. and {Lundeen}, J. and {Malone}, K. and {Marinelli}, S.~S. and {Martinez}, O. and {Martinez-Castellanos}, I. and {Mart{\'\i}nez-Castro}, J. and {Matthews}, J.~A. and {Miranda-Romagnoli}, P. and {Morales-Soto}, J.~A. and {Moreno}, E. and {Mostaf{\'a}}, M. and {Nayerhoda}, A. and {Nellen}, L. and {Newbold}, M. and {Nisa}, M.~U. and {Noriega-Papaqui}, R. and {Peisker}, A. and {P{\'e}rez-P{\'e}rez}, E.~G. and {Rho}, C.~D. and {Rosa-Gonz{\'a}lez}, D. and {Rosenberg}, M. and {Salazar}, H. and {Salesa Greus}, F. and {Sandoval}, A. and {Schneider}, M. and {Springer}, R.~W. and {Tabachnick}, E. and {Tibolla}, O. and {Tollefson}, K. and {Torres}, I. and {Torres-Escobedo}, R. and {Weisgarber}, T. and {Wood}, J. and {Zepeda}, A. and {Zhou}, H. and {de Le{\'o}n}, C.},
        title = "{Search for gamma-ray spectral lines from dark matter annihilation in dwarf galaxies with the High-Altitude Water Cherenkov observatory}",
      journal = {\prd},
         year = 2020,
        month = may,
       volume = {101},
       number = {10},
          eid = {103001},
        pages = {103001},
          doi = {10.1103/PhysRevD.101.103001},
archivePrefix = {arXiv},
       eprint = {1912.05632},
 primaryClass = {astro-ph.HE},
       adsurl = {https://ui.adsabs.harvard.edu/abs/2020PhRvD.101j3001A}
}

@ARTICLE{Aparicio2001,
       author = {{Aparicio}, Antonio and {Carrera}, Ricardo and {Mart{\'\i}nez-Delgado}, David},
        title = "{The Star Formation History and Morphological Evolution of the Draco Dwarf Spheroidal Galaxy}",
      journal = {\aj},
         year = 2001,
        month = nov,
       volume = {122},
       number = {5},
        pages = {2524-2537},
          doi = {10.1086/323535},
archivePrefix = {arXiv},
       eprint = {astro-ph/0108159},
 primaryClass = {astro-ph},
       adsurl = {https://ui.adsabs.harvard.edu/abs/2001AJ....122.2524A}
}

@ARTICLE{Banados2018,
       author = {{Ba{\~n}ados}, Eduardo and {Venemans}, Bram P. and {Mazzucchelli}, Chiara and {Farina}, Emanuele P. and {Walter}, Fabian and {Wang}, Feige and {Decarli}, Roberto and {Stern}, Daniel and {Fan}, Xiaohui and {Davies}, Frederick B. and {Hennawi}, Joseph F. and {Simcoe}, Robert A. and {Turner}, Monica L. and {Rix}, Hans-Walter and {Yang}, Jinyi and {Kelson}, Daniel D. and {Rudie}, Gwen C. and {Winters}, Jan Martin},
        title = "{An 800-million-solar-mass black hole in a significantly neutral Universe at a redshift of 7.5}",
      journal = {\nat},
         year = 2018,
        month = jan,
       volume = {553},
       number = {7689},
        pages = {473-476},
          doi = {10.1038/nature25180},
archivePrefix = {arXiv},
       eprint = {1712.01860},
 primaryClass = {astro-ph.GA},
       adsurl = {https://ui.adsabs.harvard.edu/abs/2018Natur.553..473B}
}

@ARTICLE{Banares2025,
       author = {{Ba{\~n}ares-Hern{\'a}ndez}, Andr{\'e}s and {Read}, Justin I. and {J{\'u}lio}, Mariana P.},
        title = "{GravSphere2: A higher-order Jeans method for mass-modeling spherical stellar systems}",
      journal = {arXiv e-prints},
         year = 2025,
        month = sep,
          eid = {arXiv:2509.24103},
        pages = {arXiv:2509.24103},
          doi = {10.48550/arXiv.2509.24103},
archivePrefix = {arXiv},
       eprint = {2509.24103},
 primaryClass = {astro-ph.GA},
       adsurl = {https://ui.adsabs.harvard.edu/abs/2025arXiv250924103B}
}

@ARTICLE{Battaglia2006,
   author = {{Battaglia}, G. and {Tolstoy}, E. and {Helmi}, A. and {Irwin}, M.~J. and 
	{Letarte}, B. and {Jablonka}, P. and {Hill}, V. and {Venn}, K.~A. and 
	{Shetrone}, M.~D. and {Arimoto}, N. and {Primas}, F. and {Kaufer}, A. and 
	{Francois}, P. and {Szeifert}, T. and {Abel}, T. and {Sadakane}, K.
	},
    title = "{The DART imaging and CaT survey of the Fornax dwarf spheroidal galaxy}",
  journal = {\aap},
   eprint = {astro-ph/0608370},
     year = 2006,
    month = nov,
   volume = 459,
    pages = {423-440},
      doi = {10.1051/0004-6361:20065720},
   adsurl = {http://adsabs.harvard.edu/abs/2006A%26A...459..423B}
}

@ARTICLE{Battaglia2008,
   author = {{Battaglia}, G. and {Helmi}, A. and {Tolstoy}, E. and {Irwin}, M. and 
	{Hill}, V. and {Jablonka}, P.},
    title = "{The Kinematic Status and Mass Content of the Sculptor Dwarf Spheroidal Galaxy}",
  journal = {\apjl},
archivePrefix = "arXiv",
   eprint = {0802.4220},
     year = 2008,
    month = jul,
   volume = 681,
    pages = {L13},
      doi = {10.1086/590179},
   adsurl = {http://adsabs.harvard.edu/abs/2008ApJ...681L..13B}
}

@ARTICLE{Battaglia2022,
       author = {{Battaglia}, G. and {Taibi}, S. and {Thomas}, G.~F. and {Fritz}, T.~K.},
        title = "{Gaia early DR3 systemic motions of Local Group dwarf galaxies and orbital properties with a massive Large Magellanic Cloud}",
      journal = {\aap},
         year = 2022,
        month = jan,
       volume = {657},
          eid = {A54},
        pages = {A54},
          doi = {10.1051/0004-6361/202141528},
archivePrefix = {arXiv},
       eprint = {2106.08819},
 primaryClass = {astro-ph.GA},
       adsurl = {https://ui.adsabs.harvard.edu/abs/2022A&A...657A..54B}
}

@ARTICLE{BattagliaNipoti2022,
       author = {{Battaglia}, Giuseppina and {Nipoti}, Carlo},
        title = "{Stellar dynamics and dark matter in Local Group dwarf galaxies}",
      journal = {Nature Astronomy},
         year = 2022,
        month = may,
       volume = {6},
        pages = {659-672},
          doi = {10.1038/s41550-022-01638-7},
archivePrefix = {arXiv},
       eprint = {2205.07821},
 primaryClass = {astro-ph.GA},
       adsurl = {https://ui.adsabs.harvard.edu/abs/2022NatAs...6..659B}
}

@ARTICLE{Begelman2006,
       author = {{Begelman}, Mitchell C. and {Volonteri}, Marta and {Rees}, Martin J.},
        title = "{Formation of supermassive black holes by direct collapse in pre-galactic haloes}",
      journal = {\mnras},
         year = 2006,
        month = jul,
       volume = {370},
       number = {1},
        pages = {289-298},
          doi = {10.1111/j.1365-2966.2006.10467.x},
archivePrefix = {arXiv},
       eprint = {astro-ph/0602363},
 primaryClass = {astro-ph},
       adsurl = {https://ui.adsabs.harvard.edu/abs/2006MNRAS.370..289B}
}

@ARTICLE{Binney2014,
   author = {{Binney}, J.},
    title = "{Self-consistent flattened isochrones}",
  journal = {\mnras},
archivePrefix = "arXiv",
   eprint = {1402.2512},
     year = 2014,
    month = may,
   volume = 440,
    pages = {787-798},
      doi = {10.1093/mnras/stu297},
   adsurl = {http://adsabs.harvard.edu/abs/2014MNRAS.440..787B}
}

@ARTICLE{Binney2023,
       author = {{Binney}, James and {Vasiliev}, Eugene},
        title = "{Self-consistent models of our Galaxy}",
      journal = {\mnras},
         year = 2023,
        month = apr,
       volume = {520},
       number = {2},
        pages = {1832-1847},
          doi = {10.1093/mnras/stad094},
archivePrefix = {arXiv},
       eprint = {2206.03523},
 primaryClass = {astro-ph.GA},
       adsurl = {https://ui.adsabs.harvard.edu/abs/2023MNRAS.520.1832B}
}

@ARTICLE{Binney2024,
       author = {{Binney}, James and {Vasiliev}, Eugene},
        title = "{Chemodynamical models of our Galaxy}",
      journal = {\mnras},
         year = 2024,
        month = jan,
       volume = {527},
       number = {2},
        pages = {1915-1934},
          doi = {10.1093/mnras/stad3312},
archivePrefix = {arXiv},
       eprint = {2306.11602},
 primaryClass = {astro-ph.GA},
       adsurl = {https://ui.adsabs.harvard.edu/abs/2024MNRAS.527.1915B}
}

@BOOK{BinneyTremaine2008,
   author = {{Binney}, J. and {Tremaine}, S.},
    title = "{Galactic Dynamics: Second Edition}",
booktitle = {Galactic Dynamics: Second Edition, by James Binney and Scott Tremaine.~ISBN 978-0-691-13026-2 (HB).~Published by Princeton University Press, Princeton, NJ USA, 2008.},
     year = 2008,
publisher = {Princeton University Press},
   adsurl = {http://adsabs.harvard.edu/abs/2008gady.book.....B}
}

@ARTICLE{Boddy2024,
       author = {{Boddy}, Kimberly K. and {Carter}, Zachary J. and {Kumar}, Jason and {Rufino}, Luis and {Sandick}, Pearl and {Tapia-Arellano}, Natalia},
        title = "{New dark matter analysis of milky way dwarf satellite galaxies with MADHATV2}",
      journal = {\prd},
         year = 2024,
        month = may,
       volume = {109},
       number = {10},
          eid = {103007},
        pages = {103007},
          doi = {10.1103/PhysRevD.109.103007},
archivePrefix = {arXiv},
       eprint = {2401.05327},
 primaryClass = {hep-ph},
       adsurl = {https://ui.adsabs.harvard.edu/abs/2024PhRvD.109j3007B}
}

@ARTICLE{Bogdan2024,
       author = {{Bogd{\'a}n}, {\'A}kos and {Goulding}, Andy D. and {Natarajan}, Priyamvada and {Kov{\'a}cs}, Orsolya E. and {Tremblay}, Grant R. and {Chadayammuri}, Urmila and {Volonteri}, Marta and {Kraft}, Ralph P. and {Forman}, William R. and {Jones}, Christine and {Churazov}, Eugene and {Zhuravleva}, Irina},
        title = "{Evidence for heavy-seed origin of early supermassive black holes from a z {\ensuremath{\approx}} 10 X-ray quasar}",
      journal = {Nature Astronomy},
         year = 2024,
        month = jan,
       volume = {8},
       number = {1},
        pages = {126-133},
          doi = {10.1038/s41550-023-02111-9},
archivePrefix = {arXiv},
       eprint = {2305.15458},
 primaryClass = {astro-ph.GA},
       adsurl = {https://ui.adsabs.harvard.edu/abs/2024NatAs...8..126B}
}

@ARTICLE{BreddelsHelmi2013,
   author = {{Breddels}, M.~A. and {Helmi}, A.},
    title = "{Model comparison of the dark matter profiles of Fornax, Sculptor, Carina and Sextans}",
  journal = {\aap},
archivePrefix = "arXiv",
   eprint = {1304.2976},
     year = 2013,
    month = oct,
   volume = 558,
      eid = {A35},
    pages = {A35},
      doi = {10.1051/0004-6361/201321606},
   adsurl = {http://adsabs.harvard.edu/abs/2013A%26A...558A..35B}
}

@ARTICLE{Akins2025,
       author = {{Akins}, Hollis B. and {Casey}, Caitlin M. and {Lambrides}, Erini and {Allen}, Natalie and {Andika}, Irham T. and {Brinch}, Malte and {Champagne}, Jaclyn B. and {Cooper}, Olivia and {Ding}, Xuheng and {Drakos}, Nicole E. and {Faisst}, Andreas and {Finkelstein}, Steven L. and {Franco}, Maximilien and {Fujimoto}, Seiji and {Gentile}, Fabrizio and {Gillman}, Steven and {Gozaliasl}, Ghassem and {Harish}, Santosh and {Hayward}, Christopher C. and {Hirschmann}, Michaela and {Ilbert}, Olivier and {Kartaltepe}, Jeyhan S. and {Kocevski}, Dale D. and {Koekemoer}, Anton M. and {Kokorev}, Vasily and {Liu}, Daizhong and {Long}, Arianna S. and {McCracken}, Henry Joy and {McKinney}, Jed and {Onoue}, Masafusa and {Paquereau}, Louise and {Renzini}, Alvio and {Rhodes}, Jason and {Robertson}, Brant E. and {Shuntov}, Marko and {Silverman}, John D. and {Tanaka}, Takumi S. and {Toft}, Sune and {Trakhtenbrot}, Benny and {Valentino}, Francesco and {Zavala}, Jorge},
        title = "{COSMOS-Web: The Overabundance and Physical Nature of ``Little Red Dots''{\textemdash}Implications for Early Galaxy and SMBH Assembly}",
      journal = {\apj},
         year = 2025,
        month = sep,
       volume = {991},
       number = {1},
          eid = {37},
        pages = {37},
          doi = {10.3847/1538-4357/ade984},
archivePrefix = {arXiv},
       eprint = {2406.10341},
 primaryClass = {astro-ph.GA},
       adsurl = {https://ui.adsabs.harvard.edu/abs/2025ApJ...991...37A}
}

@ARTICLE{Bromm2003,
       author = {{Bromm}, Volker and {Loeb}, Abraham},
        title = "{Formation of the First Supermassive Black Holes}",
      journal = {\apj},
         year = 2003,
        month = oct,
       volume = {596},
       number = {1},
        pages = {34-46},
          doi = {10.1086/377529},
archivePrefix = {arXiv},
       eprint = {astro-ph/0212400},
 primaryClass = {astro-ph},
       adsurl = {https://ui.adsabs.harvard.edu/abs/2003ApJ...596...34B}
}

@ARTICLE{Bullock2017,
   author = {{Bullock}, J.~S. and {Boylan-Kolchin}, M.},
    title = "{Small-Scale Challenges to the {$\Lambda$}CDM Paradigm}",
  journal = {\araa},
archivePrefix = "arXiv",
   eprint = {1707.04256},
     year = 2017,
    month = aug,
   volume = 55,
    pages = {343-387},
      doi = {10.1146/annurev-astro-091916-055313},
   adsurl = {http://adsabs.harvard.edu/abs/2017ARA%26A..55..343B}
}

@ARTICLE{Burke2025,
       author = {{Burke}, Colin J. and {Natarajan}, Priyamvada and {Baldassare}, Vivienne F. and {Geha}, Marla},
        title = "{Multiwavelength Constraints on the Local Black Hole Occupation Fraction}",
      journal = {\apj},
         year = 2025,
        month = jan,
       volume = {978},
       number = {1},
          eid = {77},
        pages = {77},
          doi = {10.3847/1538-4357/ad94d9},
archivePrefix = {arXiv},
       eprint = {2410.11177},
 primaryClass = {astro-ph.GA},
       adsurl = {https://ui.adsabs.harvard.edu/abs/2025ApJ...978...77B}
}

@ARTICLE{Bustamante2021,
       author = {{Bustamante-Rosell}, M.~J. and {Noyola}, Eva and {Gebhardt}, Karl and {Fabricius}, Maximilian H. and {Mazzalay}, Ximena and {Thomas}, Jens and {Zeimann}, Greg},
        title = "{Dynamical Analysis of the Dark Matter and Central Black Hole Mass in the Dwarf Spheroidal Leo I}",
      journal = {\apj},
         year = 2021,
        month = nov,
       volume = {921},
       number = {2},
          eid = {107},
        pages = {107},
          doi = {10.3847/1538-4357/ac0c79},
archivePrefix = {arXiv},
       eprint = {2111.04770},
 primaryClass = {astro-ph.GA},
       adsurl = {https://ui.adsabs.harvard.edu/abs/2021ApJ...921..107B}
}

@ARTICLE{Cammelli2025,
       author = {{Cammelli}, Vieri and {Monaco}, Pierluigi and {Tan}, Jonathan C. and {Singh}, Jasbir and {Fontanot}, Fabio and {De Lucia}, Gabriella and {Hirschmann}, Michaela and {Xie}, Lizhi},
        title = "{The formation of supermassive black holes from Population III.1 seeds. III. Galaxy evolution and black hole growth from semi-analytic modelling}",
      journal = {\mnras},
         year = 2025,
        month = jan,
       volume = {536},
       number = {1},
        pages = {851-870},
          doi = {10.1093/mnras/stae2663},
archivePrefix = {arXiv},
       eprint = {2407.09949},
 primaryClass = {astro-ph.GA},
       adsurl = {https://ui.adsabs.harvard.edu/abs/2025MNRAS.536..851C}
}

@ARTICLE{Carrera2002,
       author = {{Carrera}, Ricardo and {Aparicio}, Antonio and {Mart{\'\i}nez-Delgado}, David and {Alonso-Garc{\'\i}a}, Javier},
        title = "{The Star Formation History and Spatial Distribution of Stellar Populations in the Ursa Minor Dwarf Spheroidal Galaxy}",
      journal = {\aj},
         year = 2002,
        month = jun,
       volume = {123},
       number = {6},
        pages = {3199-3209},
          doi = {10.1086/340702},
archivePrefix = {arXiv},
       eprint = {astro-ph/0203300},
 primaryClass = {astro-ph},
       adsurl = {https://ui.adsabs.harvard.edu/abs/2002AJ....123.3199C}
}

@ARTICLE{ColeBinney2017,
   author = {{Cole}, D.~R. and {Binney}, J.},
    title = "{A centrally heated dark halo for our Galaxy}",
  journal = {\mnras},
archivePrefix = "arXiv",
   eprint = {1610.07818},
     year = 2017,
    month = feb,
   volume = 465,
    pages = {798-810},
      doi = {10.1093/mnras/stw2775},
   adsurl = {http://adsabs.harvard.edu/abs/2017MNRAS.465..798C}
}

@misc{cubature,
  title = {Multi-dimensional adaptive integration in {C}: The {Cubature} package},
  author = {Steven G. Johnson},
  year = {2005},
  howpublished = {\url{https://github.com/stevengj/cubature}}
}

@ARTICLE{Das2016a,
   author = {{Das}, P. and {Binney}, J.},
    title = "{Characterizing stellar halo populations - I. An extended distribution function for halo K giants}",
  journal = {\mnras},
archivePrefix = "arXiv",
   eprint = {1603.09332},
     year = 2016,
    month = aug,
   volume = 460,
    pages = {1725-1738},
      doi = {10.1093/mnras/stw744},
   adsurl = {http://adsabs.harvard.edu/abs/2016MNRAS.460.1725D}
}

@ARTICLE{Das2016b,
       author = {{Das}, Payel and {Williams}, Angus and {Binney}, James},
        title = "{Characterizing stellar halo populations II: the age gradient in blue horizontal-branch stars}",
      journal = {\mnras},
         year = "2016",
        month = "Dec",
       volume = {463},
       number = {3},
        pages = {3169-3185},
          doi = {10.1093/mnras/stw2167},
archivePrefix = {arXiv},
       eprint = {1608.07297},
 primaryClass = {astro-ph.GA},
       adsurl = {https://ui.adsabs.harvard.edu/abs/2016MNRAS.463.3169D}
}

@ARTICLE{deBlok2008,
       author = {{de Blok}, W.~J.~G. and {Walter}, F. and {Brinks}, E. and {Trachternach}, C. and {Oh}, S.-H. and {Kennicutt}, Jr., R.~C.},
        title = "{High-Resolution Rotation Curves and Galaxy Mass Models from THINGS}",
      journal = {\aj},
         year = 2008,
        month = dec,
       volume = {136},
       number = {6},
        pages = {2648-2719},
          doi = {10.1088/0004-6256/136/6/2648},
archivePrefix = {arXiv},
       eprint = {0810.2100},
 primaryClass = {astro-ph},
       adsurl = {https://ui.adsabs.harvard.edu/abs/2008AJ....136.2648D}
}

@ARTICLE{deBlok2009,
   author = {{de Blok}, W.~J.~G.},
    title = "{The Core-Cusp Problem}",
  journal = {Advances in Astronomy},
archivePrefix = "arXiv",
   eprint = {0910.3538},
     year = 2010,
   volume = 2010,
      eid = {789293},
    pages = {789293},
      doi = {10.1155/2010/789293},
   adsurl = {http://adsabs.harvard.edu/abs/2010AdAst2010E...5D}
}

@ARTICLE{denBrok2015,
       author = {{den Brok}, Mark and {Seth}, Anil C. and {Barth}, Aaron J. and {Carson}, Daniel J. and {Neumayer}, Nadine and {Cappellari}, Michele and {Debattista}, Victor P. and {Ho}, Luis C. and {Hood}, Carol E. and {McDermid}, Richard M.},
        title = "{Measuring the Mass of the Central Black Hole in the Bulgeless Galaxy NGC 4395 from Gas Dynamical Modeling}",
      journal = {\apj},
         year = 2015,
        month = aug,
       volume = {809},
       number = {1},
          eid = {101},
        pages = {101},
          doi = {10.1088/0004-637X/809/1/101},
archivePrefix = {arXiv},
       eprint = {1507.04358},
 primaryClass = {astro-ph.GA},
       adsurl = {https://ui.adsabs.harvard.edu/abs/2015ApJ...809..101D}
}

@ARTICLE{DellaCroce2023,
       author = {{Della Croce}, A. and {Pascale}, R. and {Giunchi}, E. and {Nipoti}, C. and {Cignoni}, M. and {Dalessandro}, E.},
        title = "{The most stringent upper limit set on the mass of a central black hole in 47 Tucanae using dynamical models}",
      journal = {\aap},
         year = 2024,
        month = feb,
       volume = {682},
          eid = {A22},
        pages = {A22},
          doi = {10.1051/0004-6361/202347569},
archivePrefix = {arXiv},
       eprint = {2310.15221},
 primaryClass = {astro-ph.GA},
       adsurl = {https://ui.adsabs.harvard.edu/abs/2024A&A...682A..22D}
}

@ARTICLE{delPino2017,
       author = {{del Pino}, Andr{\'e}s and {Aparicio}, Antonio and {Hidalgo}, Sebastian L. and {{\L}okas}, Ewa L.},
        title = "{Rotating stellar populations in the Fornax dSph galaxy}",
      journal = {\mnras},
         year = 2017,
        month = mar,
       volume = {465},
       number = {3},
        pages = {3708-3723},
          doi = {10.1093/mnras/stw3016},
archivePrefix = {arXiv},
       eprint = {1605.09414},
 primaryClass = {astro-ph.GA},
       adsurl = {https://ui.adsabs.harvard.edu/abs/2017MNRAS.465.3708D}
}

@ARTICLE{DelPopolo2022,
       author = {{Del Popolo}, Antonino and {Le Delliou}, Morgan},
        title = "{Review of Solutions to the Cusp-Core Problem of the {\ensuremath{\Lambda}}CDM Model}",
      journal = {Galaxies},
         year = 2021,
        month = dec,
       volume = {9},
       number = {4},
          eid = {123},
        pages = {123},
          doi = {10.3390/galaxies9040123},
archivePrefix = {arXiv},
       eprint = {2209.14151},
 primaryClass = {astro-ph.CO},
       adsurl = {https://ui.adsabs.harvard.edu/abs/2021Galax...9..123D}
}

@ARTICLE{DiCintio2014a,
       author = {{Di Cintio}, Arianna and {Brook}, Chris B. and {Macci{\`o}}, Andrea V. and {Stinson}, Greg S. and {Knebe}, Alexander and {Dutton}, Aaron A. and {Wadsley}, James},
        title = "{The dependence of dark matter profiles on the stellar-to-halo mass ratio: a prediction for cusps versus cores}",
      journal = {\mnras},
         year = 2014,
        month = jan,
       volume = {437},
       number = {1},
        pages = {415-423},
          doi = {10.1093/mnras/stt1891},
archivePrefix = {arXiv},
       eprint = {1306.0898},
 primaryClass = {astro-ph.CO},
       adsurl = {https://ui.adsabs.harvard.edu/abs/2014MNRAS.437..415D}
}

@ARTICLE{DiMatteo2017,
       author = {{Di Matteo}, Tiziana and {Croft}, Rupert A.~C. and {Feng}, Yu and {Waters}, Dacen and {Wilkins}, Stephen},
        title = "{The origin of the most massive black holes at high-z: BlueTides and the next quasar frontier}",
      journal = {\mnras},
         year = 2017,
        month = jun,
       volume = {467},
       number = {4},
        pages = {4243-4251},
          doi = {10.1093/mnras/stx319},
archivePrefix = {arXiv},
       eprint = {1606.08871},
 primaryClass = {astro-ph.GA},
       adsurl = {https://ui.adsabs.harvard.edu/abs/2017MNRAS.467.4243D}
}

@ARTICLE{DiMauro2022,
       author = {{Di Mauro}, Mattia and {Stref}, Martin and {Calore}, Francesca},
        title = "{Investigating the effect of Milky Way dwarf spheroidal galaxies extension on dark matter searches with Fermi-LAT data}",
      journal = {\prd},
         year = 2022,
        month = dec,
       volume = {106},
       number = {12},
          eid = {123032},
        pages = {123032},
          doi = {10.1103/PhysRevD.106.123032},
archivePrefix = {arXiv},
       eprint = {2212.06850},
 primaryClass = {astro-ph.HE},
       adsurl = {https://ui.adsabs.harvard.edu/abs/2022PhRvD.106l3032D}
}

@ARTICLE{Dubinski1991,
       author = {{Dubinski}, John and {Carlberg}, R.~G.},
        title = "{The Structure of Cold Dark Matter Halos}",
      journal = {\apj},
         year = 1991,
        month = sep,
       volume = {378},
        pages = {496},
          doi = {10.1086/170451},
       adsurl = {https://ui.adsabs.harvard.edu/abs/1991ApJ...378..496D}
}

@ARTICLE{Elbert2015,
       author = {{Elbert}, Oliver D. and {Bullock}, James S. and {Garrison-Kimmel}, Shea and {Rocha}, Miguel and {O{\~n}orbe}, Jose and {Peter}, Annika H.~G.},
        title = "{Core formation in dwarf haloes with self-interacting dark matter: no fine-tuning necessary}",
      journal = {\mnras},
         year = 2015,
        month = oct,
       volume = {453},
       number = {1},
        pages = {29-37},
          doi = {10.1093/mnras/stv1470},
archivePrefix = {arXiv},
       eprint = {1412.1477},
 primaryClass = {astro-ph.GA},
       adsurl = {https://ui.adsabs.harvard.edu/abs/2015MNRAS.453...29E}
}

@ARTICLE{Evans2016,
   author = {{Evans}, N.~W. and {Sanders}, J.~L. and {Geringer-Sameth}, A.
	},
    title = "{Simple J-factors and D-factors for indirect dark matter detection}",
  journal = {\prd},
archivePrefix = "arXiv",
   eprint = {1604.05599},
     year = 2016,
    month = may,
   volume = 93,
   number = 10,
      eid = {103512},
    pages = {103512},
      doi = {10.1103/PhysRevD.93.103512},
   adsurl = {http://adsabs.harvard.edu/abs/2016PhRvD..93j3512E}
}

@ARTICLE{Evans2018,
       author = {{Evans}, D.~W. and {Riello}, M. and {De Angeli}, F. and {Carrasco}, J.~M. and {Montegriffo}, P. and {Fabricius}, C. and {Jordi}, C. and {Palaversa}, L. and {Diener}, C. and {Busso}, G. and {Cacciari}, C. and {van Leeuwen}, F. and {Burgess}, P.~W. and {Davidson}, M. and {Harrison}, D.~L. and {Hodgkin}, S.~T. and {Pancino}, E. and {Richards}, P.~J. and {Altavilla}, G. and {Balaguer-N{\'u}{\~n}ez}, L. and {Barstow}, M.~A. and {Bellazzini}, M. and {Brown}, A.~G.~A. and {Castellani}, M. and {Cocozza}, G. and {De Luise}, F. and {Delgado}, A. and {Ducourant}, C. and {Galleti}, S. and {Gilmore}, G. and {Giuffrida}, G. and {Holl}, B. and {Kewley}, A. and {Koposov}, S.~E. and {Marinoni}, S. and {Marrese}, P.~M. and {Osborne}, P.~J. and {Piersimoni}, A. and {Portell}, J. and {Pulone}, L. and {Ragaini}, S. and {Sanna}, N. and {Terrett}, D. and {Walton}, N.~A. and {Wevers}, T. and {Wyrzykowski}, {\L}.},
        title = "{Gaia Data Release 2. Photometric content and validation}",
      journal = {\aap},
         year = 2018,
        month = aug,
       volume = {616},
          eid = {A4},
        pages = {A4},
          doi = {10.1051/0004-6361/201832756},
archivePrefix = {arXiv},
       eprint = {1804.09368},
 primaryClass = {astro-ph.IM},
       adsurl = {https://ui.adsabs.harvard.edu/abs/2018A&A...616A...4E}
}

@ARTICLE{Ferrarese2000,
       author = {{Ferrarese}, Laura and {Merritt}, David},
        title = "{A Fundamental Relation between Supermassive Black Holes and Their Host Galaxies}",
      journal = {\apjl},
         year = 2000,
        month = aug,
       volume = {539},
       number = {1},
        pages = {L9-L12},
          doi = {10.1086/312838},
archivePrefix = {arXiv},
       eprint = {astro-ph/0006053},
 primaryClass = {astro-ph},
       adsurl = {https://ui.adsabs.harvard.edu/abs/2000ApJ...539L...9F}
}

@ARTICLE{Fitts2017,
       author = {{Fitts}, Alex and {Boylan-Kolchin}, Michael and {Elbert}, Oliver D. and {Bullock}, James S. and {Hopkins}, Philip F. and {O{\~n}orbe}, Jose and {Wetzel}, Andrew and {Wheeler}, Coral and {Faucher-Gigu{\`e}re}, Claude-Andr{\'e} and {Kere{\v{s}}}, Du{\v{s}}an and {Skillman}, Evan D. and {Weisz}, Daniel R.},
        title = "{fire in the field: simulating the threshold of galaxy formation}",
      journal = {\mnras},
         year = 2017,
        month = nov,
       volume = {471},
       number = {3},
        pages = {3547-3562},
          doi = {10.1093/mnras/stx1757},
archivePrefix = {arXiv},
       eprint = {1611.02281},
 primaryClass = {astro-ph.GA},
       adsurl = {https://ui.adsabs.harvard.edu/abs/2017MNRAS.471.3547F}
}

@ARTICLE{ForemanMackey2013,
       author = {{Foreman-Mackey}, Daniel and {Hogg}, David W. and {Lang}, Dustin and {Goodman}, Jonathan},
        title = "{emcee: The MCMC Hammer}",
      journal = {\pasp},
         year = 2013,
        month = mar,
       volume = {125},
       number = {925},
        pages = {306},
          doi = {10.1086/670067},
archivePrefix = {arXiv},
       eprint = {1202.3665},
 primaryClass = {astro-ph.IM},
       adsurl = {https://ui.adsabs.harvard.edu/abs/2013PASP..125..306F}
}

@ARTICLE{Gaia2021,
       author = {{Gaia Collaboration} and {Brown}, A.~G.~A. and {Vallenari}, A. and {Prusti}, T. and {de Bruijne}, J.~H.~J. and {Babusiaux}, C. and {Biermann}, M. and {Creevey}, O.~L. and {Evans}, D.~W. and {Eyer}, L. and {Hutton}, A. and {Jansen}, F. and {Jordi}, C. and {Klioner}, S.~A. and {Lammers}, U. and {Lindegren}, L. and {Luri}, X. and {Mignard}, F. and {Panem}, C. and {Pourbaix}, D. and {Randich}, S. and {Sartoretti}, P. and {Soubiran}, C. and {Walton}, N.~A. and {Arenou}, F. and {Bailer-Jones}, C.~A.~L. and {Bastian}, U. and {Cropper}, M. and {Drimmel}, R. and {Katz}, D. and {Lattanzi}, M.~G. and {van Leeuwen}, F. and {Bakker}, J. and {Cacciari}, C. and {Casta{\~n}eda}, J. and {De Angeli}, F. and {Ducourant}, C. and {Fabricius}, C. and {Fouesneau}, M. and {Fr{\'e}mat}, Y. and {Guerra}, R. and {Guerrier}, A. and {Guiraud}, J. and {Jean-Antoine Piccolo}, A. and {Masana}, E. and {Messineo}, R. and {Mowlavi}, N. and {Nicolas}, C. and {Nienartowicz}, K. and {Pailler}, F. and {Panuzzo}, P. and {Riclet}, F. and {Roux}, W. and {Seabroke}, G.~M. and {Sordo}, R. and {Tanga}, P. and {Th{\'e}venin}, F. and {Gracia-Abril}, G. and {Portell}, J. and {Teyssier}, D. and {Altmann}, M. and {Andrae}, R. and {Bellas-Velidis}, I. and {Benson}, K. and {Berthier}, J. and {Blomme}, R. and {Brugaletta}, E. and {Burgess}, P.~W. and {Busso}, G. and {Carry}, B. and {Cellino}, A. and {Cheek}, N. and {Clementini}, G. and {Damerdji}, Y. and {Davidson}, M. and {Delchambre}, L. and {Dell'Oro}, A. and {Fern{\'a}ndez-Hern{\'a}ndez}, J. and {Galluccio}, L. and {Garc{\'\i}a-Lario}, P. and {Garcia-Reinaldos}, M. and {Gonz{\'a}lez-N{\'u}{\~n}ez}, J. and {Gosset}, E. and {Haigron}, R. and {Halbwachs}, J. -L. and {Hambly}, N.~C. and {Harrison}, D.~L. and {Hatzidimitriou}, D. and {Heiter}, U. and {Hern{\'a}ndez}, J. and {Hestroffer}, D. and {Hodgkin}, S.~T. and {Holl}, B. and {Jan{\ss}en}, K. and {Jevardat de Fombelle}, G. and {Jordan}, S. and {Krone-Martins}, A. and {Lanzafame}, A.~C. and {L{\"o}ffler}, W. and {Lorca}, A. and {Manteiga}, M. and {Marchal}, O. and {Marrese}, P.~M. and {Moitinho}, A. and {Mora}, A. and {Muinonen}, K. and {Osborne}, P. and {Pancino}, E. and {Pauwels}, T. and {Petit}, J. -M. and {Recio-Blanco}, A. and {Richards}, P.~J. and {Riello}, M. and {Rimoldini}, L. and {Robin}, A.~C. and {Roegiers}, T. and {Rybizki}, J. and {Sarro}, L.~M. and {Siopis}, C. and {Smith}, M. and {Sozzetti}, A. and {Ulla}, A. and {Utrilla}, E. and {van Leeuwen}, M. and {van Reeven}, W. and {Abbas}, U. and {Abreu Aramburu}, A. and {Accart}, S. and {Aerts}, C. and {Aguado}, J.~J. and {Ajaj}, M. and {Altavilla}, G. and {{\'A}lvarez}, M.~A. and {{\'A}lvarez Cid-Fuentes}, J. and {Alves}, J. and {Anderson}, R.~I. and {Anglada Varela}, E. and {Antoja}, T. and {Audard}, M. and {Baines}, D. and {Baker}, S.~G. and {Balaguer-N{\'u}{\~n}ez}, L. and {Balbinot}, E. and {Balog}, Z. and {Barache}, C. and {Barbato}, D. and {Barros}, M. and {Barstow}, M.~A. and {Bartolom{\'e}}, S. and {Bassilana}, J. -L. and {Bauchet}, N. and {Baudesson-Stella}, A. and {Becciani}, U. and {Bellazzini}, M. and {Bernet}, M. and {Bertone}, S. and {Bianchi}, L. and {Blanco-Cuaresma}, S. and {Boch}, T. and {Bombrun}, A. and {Bossini}, D. and {Bouquillon}, S. and {Bragaglia}, A. and {Bramante}, L. and {Breedt}, E. and {Bressan}, A. and {Brouillet}, N. and {Bucciarelli}, B. and {Burlacu}, A. and {Busonero}, D. and {Butkevich}, A.~G. and {Buzzi}, R. and {Caffau}, E. and {Cancelliere}, R. and {C{\'a}novas}, H. and {Cantat-Gaudin}, T. and {Carballo}, R. and {Carlucci}, T. and {Carnerero}, M.~I. and {Carrasco}, J.~M. and {Casamiquela}, L. and {Castellani}, M. and {Castro-Ginard}, A. and {Castro Sampol}, P. and {Chaoul}, L. and {Charlot}, P. and {Chemin}, L. and {Chiavassa}, A. and {Cioni}, M. -R.~L. and {Comoretto}, G. and {Cooper}, W.~J. and {Cornez}, T. and {Cowell}, S. and {Crifo}, F. and {Crosta}, M. and {Crowley}, C. and {Dafonte}, C. and {Dapergolas}, A. and {David}, M. and {David}, P.},
        title = "{Gaia Early Data Release 3. Summary of the contents and survey properties}",
      journal = {\aap},
         year = 2021,
        month = may,
       volume = {649},
          eid = {A1},
        pages = {A1},
          doi = {10.1051/0004-6361/202039657},
archivePrefix = {arXiv},
       eprint = {2012.01533},
 primaryClass = {astro-ph.GA},
       adsurl = {https://ui.adsabs.harvard.edu/abs/2021A&A...649A...1G}
}

@ARTICLE{Geha2026,
       author = {{Geha}, Marla and {Pelliccia}, Debora and {Prochaska}, J. Xavier and {Cerny}, William and {Davies}, Frederick B. and {Hennawi}, Joseph and {Holden}, Brad and {Reichwein}, Dusty and {Westfall}, Kyle B.},
        title = "{The Keck/DEIMOS Stellar Archive. I. Uniform Velocities and Metallicities for 78 Milky Way Dwarf Galaxies and Globular Clusters}",
      journal = {\apj},
         year = 2026,
        month = mar,
       volume = {999},
       number = {1},
          eid = {140},
        pages = {140},
          doi = {10.3847/1538-4357/ae290d},
archivePrefix = {arXiv},
       eprint = {2602.10200},
 primaryClass = {astro-ph.GA},
       adsurl = {https://ui.adsabs.harvard.edu/abs/2026ApJ...999..140G}
}

@ARTICLE{GeringerSameth2015,
       author = {{Geringer-Sameth}, Alex and {Koushiappas}, Savvas M. and
         {Walker}, Matthew},
        title = "{Dwarf Galaxy Annihilation and Decay Emission Profiles for Dark Matter Experiments}",
      journal = {\apj},
         year = "2015",
        month = "Mar",
       volume = {801},
       number = {2},
          eid = {74},
        pages = {74},
          doi = {10.1088/0004-637X/801/2/74},
archivePrefix = {arXiv},
       eprint = {1408.0002},
 primaryClass = {astro-ph.CO},
       adsurl = {https://ui.adsabs.harvard.edu/abs/2015ApJ...801...74G}
}

@ARTICLE{Gebhardt2000,
       author = {{Gebhardt}, Karl and {Bender}, Ralf and {Bower}, Gary and {Dressler}, Alan and {Faber}, S.~M. and {Filippenko}, Alexei V. and {Green}, Richard and {Grillmair}, Carl and {Ho}, Luis C. and {Kormendy}, John and {Lauer}, Tod R. and {Magorrian}, John and {Pinkney}, Jason and {Richstone}, Douglas and {Tremaine}, Scott},
        title = "{A Relationship between Nuclear Black Hole Mass and Galaxy Velocity Dispersion}",
      journal = {\apjl},
         year = 2000,
        month = aug,
       volume = {539},
       number = {1},
        pages = {L13-L16},
          doi = {10.1086/312840},
archivePrefix = {arXiv},
       eprint = {astro-ph/0006289},
 primaryClass = {astro-ph},
       adsurl = {https://ui.adsabs.harvard.edu/abs/2000ApJ...539L..13G}
}

@ARTICLE{Giersz2015,
       author = {{Giersz}, Mirek and {Leigh}, Nathan and {Hypki}, Arkadiusz and {L{\"u}tzgendorf}, Nora and {Askar}, Abbas},
        title = "{MOCCA code for star cluster simulations - IV. A new scenario for intermediate mass black hole formation in globular clusters}",
      journal = {\mnras},
         year = 2015,
        month = dec,
       volume = {454},
       number = {3},
        pages = {3150-3165},
          doi = {10.1093/mnras/stv2162},
archivePrefix = {arXiv},
       eprint = {1506.05234},
 primaryClass = {astro-ph.GA},
       adsurl = {https://ui.adsabs.harvard.edu/abs/2015MNRAS.454.3150G}
}

@ARTICLE{Gultekin2009,
       author = {{G{\"u}ltekin}, Kayhan and {Richstone}, Douglas O. and {Gebhardt}, Karl and {Lauer}, Tod R. and {Tremaine}, Scott and {Aller}, M.~C. and {Bender}, Ralf and {Dressler}, Alan and {Faber}, S.~M. and {Filippenko}, Alexei V. and {Green}, Richard and {Ho}, Luis C. and {Kormendy}, John and {Magorrian}, John and {Pinkney}, Jason and {Siopis}, Christos},
        title = "{The M-{\ensuremath{\sigma}} and M-L Relations in Galactic Bulges, and Determinations of Their Intrinsic Scatter}",
      journal = {\apj},
         year = 2009,
        month = jun,
       volume = {698},
       number = {1},
        pages = {198-221},
          doi = {10.1088/0004-637X/698/1/198},
archivePrefix = {arXiv},
       eprint = {0903.4897},
 primaryClass = {astro-ph.GA},
       adsurl = {https://ui.adsabs.harvard.edu/abs/2009ApJ...698..198G}
}

@ARTICLE{Goerdt2010,
   author = {{Goerdt}, T. and {Moore}, B. and {Read}, J.~I. and {Stadel}, J.
	},
    title = "{Core Creation in Galaxies and Halos Via Sinking Massive Objects}",
  journal = {\apj},
archivePrefix = "arXiv",
   eprint = {0806.1951},
     year = 2010,
    month = dec,
   volume = 725,
    pages = {1707-1716},
      doi = {10.1088/0004-637X/725/2/1707},
   adsurl = {http://adsabs.harvard.edu/abs/2010ApJ...725.1707G}
}

@ARTICLE{Goodman2010,
       author = {{Goodman}, Jonathan and {Weare}, Jonathan},
        title = "{Ensemble samplers with affine invariance}",
      journal = {Communications in Applied Mathematics and Computational Science},
         year = 2010,
        month = jan,
       volume = {5},
       number = {1},
        pages = {65-80},
          doi = {10.2140/camcos.2010.5.65},
       adsurl = {https://ui.adsabs.harvard.edu/abs/2010CAMCS...5...65G}
}

@ARTICLE{Governato2012,
   author = {{Governato}, F. and {Zolotov}, A. and {Pontzen}, A. and {Christensen}, C. and 
	{Oh}, S.~H. and {Brooks}, A.~M. and {Quinn}, T. and {Shen}, S. and 
	{Wadsley}, J.},
    title = "{Cuspy no more: how outflows affect the central dark matter and baryon distribution in {$\Lambda$} cold dark matter galaxies}",
  journal = {\mnras},
archivePrefix = "arXiv",
   eprint = {1202.0554},
     year = 2012,
    month = may,
   volume = 422,
    pages = {1231-1240},
      doi = {10.1111/j.1365-2966.2012.20696.x},
   adsurl = {http://adsabs.harvard.edu/abs/2012MNRAS.422.1231G}
}

@ARTICLE{Greene2012,
       author = {{Greene}, Jenny E.},
        title = "{Low-mass black holes as the remnants of primordial black hole formation}",
      journal = {Nature Communications},
         year = 2012,
        month = dec,
       volume = {3},
          eid = {1304},
        pages = {1304},
          doi = {10.1038/ncomms2314},
archivePrefix = {arXiv},
       eprint = {1211.7082},
 primaryClass = {astro-ph.CO},
       adsurl = {https://ui.adsabs.harvard.edu/abs/2012NatCo...3.1304G}
}

@ARTICLE{Green2018,
       author = {{Green}, {Gregory M.}},
        title = "{dustmaps: A Python interface for maps of interstellar dust}",
      journal = {The Journal of Open Source Software},
         year = "2018",
        month = "Jun",
       volume = {3},
       number = {26},
        pages = {695},
          doi = {10.21105/joss.00695},
       adsurl = {https://ui.adsabs.harvard.edu/abs/2018JOSS....3..695M}
}

@ARTICLE{Greene2020,
       author = {{Greene}, Jenny E. and {Strader}, Jay and {Ho}, Luis C.},
        title = "{Intermediate-Mass Black Holes}",
      journal = {\araa},
         year = 2020,
        month = aug,
       volume = {58},
        pages = {257-312},
          doi = {10.1146/annurev-astro-032620-021835},
archivePrefix = {arXiv},
       eprint = {1911.09678},
 primaryClass = {astro-ph.GA},
       adsurl = {https://ui.adsabs.harvard.edu/abs/2020ARA&A..58..257G}
}

@ARTICLE{Hoof2020,
       author = {{Hoof}, Sebastian and {Geringer-Sameth}, Alex and {Trotta}, Roberto},
        title = "{A global analysis of dark matter signals from 27 dwarf spheroidal galaxies using 11 years of Fermi-LAT observations}",
      journal = {\jcap},
         year = 2020,
        month = feb,
       volume = {2020},
       number = {2},
          eid = {012},
        pages = {012},
          doi = {10.1088/1475-7516/2020/02/012},
archivePrefix = {arXiv},
       eprint = {1812.06986},
 primaryClass = {astro-ph.CO},
       adsurl = {https://ui.adsabs.harvard.edu/abs/2020JCAP...02..012H}
}

@ARTICLE{Haidar2022,
       author = {{Haidar}, Houda and {Habouzit}, M{\'e}lanie and {Volonteri}, Marta and {Mezcua}, Mar and {Greene}, Jenny and {Neumayer}, Nadine and {Angl{\'e}s-Alc{\'a}zar}, Daniel and {Martin-Navarro}, Ignacio and {Hoyer}, Nils and {Dubois}, Yohan and {Dav{\'e}}, Romeel},
        title = "{The black hole population in low-mass galaxies in large-scale cosmological simulations}",
      journal = {\mnras},
         year = 2022,
        month = aug,
       volume = {514},
       number = {4},
        pages = {4912-4931},
          doi = {10.1093/mnras/stac1659},
archivePrefix = {arXiv},
       eprint = {2201.09888},
 primaryClass = {astro-ph.GA},
       adsurl = {https://ui.adsabs.harvard.edu/abs/2022MNRAS.514.4912H}
}

@ARTICLE{Hayashi2020,
       author = {{Hayashi}, Kohei and {Chiba}, Masashi and {Ishiyama}, Tomoaki},
        title = "{Diversity of Dark Matter Density Profiles in the Galactic Dwarf Spheroidal Satellites}",
      journal = {\apj},
         year = 2020,
        month = nov,
       volume = {904},
       number = {1},
          eid = {45},
        pages = {45},
          doi = {10.3847/1538-4357/abbe0a},
archivePrefix = {arXiv},
       eprint = {2007.13780},
 primaryClass = {astro-ph.GA},
       adsurl = {https://ui.adsabs.harvard.edu/abs/2020ApJ...904...45H}
}

@ARTICLE{Hopkins2018,
       author = {{Hopkins}, Philip F. and {Wetzel}, Andrew and {Kere{\v{s}}}, Du{\v{s}}an and {Faucher-Gigu{\`e}re}, Claude-Andr{\'e} and {Quataert}, Eliot and {Boylan-Kolchin}, Michael and {Murray}, Norman and {Hayward}, Christopher C. and {Garrison-Kimmel}, Shea and {Hummels}, Cameron and {Feldmann}, Robert and {Torrey}, Paul and {Ma}, Xiangcheng and {Angl{\'e}s-Alc{\'a}zar}, Daniel and {Su}, Kung-Yi and {Orr}, Matthew and {Schmitz}, Denise and {Escala}, Ivanna and {Sanderson}, Robyn and {Grudi{\'c}}, Michael Y. and {Hafen}, Zachary and {Kim}, Ji-Hoon and {Fitts}, Alex and {Bullock}, James S. and {Wheeler}, Coral and {Chan}, T.~K. and {Elbert}, Oliver D. and {Narayanan}, Desika},
        title = "{FIRE-2 simulations: physics versus numerics in galaxy formation}",
      journal = {\mnras},
         year = 2018,
        month = oct,
       volume = {480},
       number = {1},
        pages = {800-863},
          doi = {10.1093/mnras/sty1690},
archivePrefix = {arXiv},
       eprint = {1702.06148},
 primaryClass = {astro-ph.GA},
       adsurl = {https://ui.adsabs.harvard.edu/abs/2018MNRAS.480..800H}
}

@ARTICLE{Hu2024,
       author = {{Hu}, Xiao-Song and {Zhu}, Ben-Yang and {Liu}, Tian-Ci and {Liang}, Yun-Feng},
        title = "{Constraints on the annihilation of heavy dark matter in dwarf spheroidal galaxies with gamma-ray observations}",
      journal = {\prd},
         year = 2024,
        month = mar,
       volume = {109},
       number = {6},
          eid = {063036},
        pages = {063036},
          doi = {10.1103/PhysRevD.109.063036},
archivePrefix = {arXiv},
       eprint = {2309.06151},
 primaryClass = {astro-ph.HE},
       adsurl = {https://ui.adsabs.harvard.edu/abs/2024PhRvD.109f3036H}
}

@ARTICLE{Hui2017,
   author = {{Hui}, L. and {Ostriker}, J.~P. and {Tremaine}, S. and {Witten}, E.
	},
    title = "{Ultralight scalars as cosmological dark matter}",
  journal = {\prd},
archivePrefix = "arXiv",
   eprint = {1610.08297},
     year = 2017,
    month = feb,
   volume = 95,
   number = 4,
      eid = {043541},
    pages = {043541},
      doi = {10.1103/PhysRevD.95.043541},
   adsurl = {http://adsabs.harvard.edu/abs/2017PhRvD..95d3541H}
}

@ARTICLE{Inayoshi2022,
       author = {{Inayoshi}, Kohei and {Nakatani}, Riouhei and {Toyouchi}, Daisuke and {Hosokawa}, Takashi and {Kuiper}, Rolf and {Onoue}, Masafusa},
        title = "{Rapid Growth of Seed Black Holes during Early Bulge Formation}",
      journal = {\apj},
         year = 2022,
        month = mar,
       volume = {927},
       number = {2},
          eid = {237},
        pages = {237},
          doi = {10.3847/1538-4357/ac4751},
archivePrefix = {arXiv},
       eprint = {2110.10693},
 primaryClass = {astro-ph.GA},
       adsurl = {https://ui.adsabs.harvard.edu/abs/2022ApJ...927..237I}
}

@ARTICLE{IrwinHatzidimitriou1995,
   author = {{Irwin}, M. and {Hatzidimitriou}, D.},
    title = "{Structural parameters for the Galactic dwarf spheroidals}",
  journal = {\mnras},
     year = 1995,
    month = dec,
   volume = 277,
    pages = {1354-1378},
      doi = {10.1093/mnras/277.4.1354},
   adsurl = {http://adsabs.harvard.edu/abs/1995MNRAS.277.1354I}
}

@ARTICLE{Jardel2012,
   author = {{Jardel}, J.~R. and {Gebhardt}, K.},
    title = "{The Dark Matter Density Profile of the Fornax Dwarf}",
  journal = {\apj},
archivePrefix = "arXiv",
   eprint = {1112.0319},
     year = 2012,
    month = feb,
   volume = 746,
      eid = {89},
    pages = {89},
      doi = {10.1088/0004-637X/746/1/89},
   adsurl = {http://adsabs.harvard.edu/abs/2012ApJ...746...89J}
}

@ARTICLE{Jensen2024,
       author = {{Jensen}, Jaclyn and {Hayes}, Christian R. and {Sestito}, Federico and {McConnachie}, Alan W. and {Waller}, Fletcher and {Smith}, Simon E.~T. and {Navarro}, Julio and {Venn}, Kim A.},
        title = "{Small-scale stellar haloes: detecting low surface brightness features in the outskirts of Milky Way dwarf satellites}",
      journal = {\mnras},
         year = 2024,
        month = jan,
       volume = {527},
       number = {2},
        pages = {4209-4233},
          doi = {10.1093/mnras/stad3322},
archivePrefix = {arXiv},
       eprint = {2308.07394},
 primaryClass = {astro-ph.GA},
       adsurl = {https://ui.adsabs.harvard.edu/abs/2024MNRAS.527.4209J}
}

@ARTICLE{Gaia,
       author = {{Gaia Collaboration} and {Brown}, A.~G.~A. and {Vallenari}, A. and {Prusti}, T. and {de Bruijne}, J.~H.~J. and {Babusiaux}, C. and {Biermann}, M. and {Creevey}, O.~L. and {Evans}, D.~W. and {Eyer}, L. and {Hutton}, A. and {Jansen}, F. and {Jordi}, C. and {Klioner}, S.~A. and {Lammers}, U. and {Lindegren}, L. and {Luri}, X. and {Mignard}, F. and {Panem}, C. and {Pourbaix}, D. and {Randich}, S. and {Sartoretti}, P. and {Soubiran}, C. and {Walton}, N.~A. and {Arenou}, F. and {Bailer-Jones}, C.~A.~L. and {Bastian}, U. and {Cropper}, M. and {Drimmel}, R. and {Katz}, D. and {Lattanzi}, M.~G. and {van Leeuwen}, F. and {Bakker}, J. and {Cacciari}, C. and {Casta{\~n}eda}, J. and {De Angeli}, F. and {Ducourant}, C. and {Fabricius}, C. and {Fouesneau}, M. and {Fr{\'e}mat}, Y. and {Guerra}, R. and {Guerrier}, A. and {Guiraud}, J. and {Jean-Antoine Piccolo}, A. and {Masana}, E. and {Messineo}, R. and {Mowlavi}, N. and {Nicolas}, C. and {Nienartowicz}, K. and {Pailler}, F. and {Panuzzo}, P. and {Riclet}, F. and {Roux}, W. and {Seabroke}, G.~M. and {Sordo}, R. and {Tanga}, P. and {Th{\'e}venin}, F. and {Gracia-Abril}, G. and {Portell}, J. and {Teyssier}, D. and {Altmann}, M. and {Andrae}, R. and {Bellas-Velidis}, I. and {Benson}, K. and {Berthier}, J. and {Blomme}, R. and {Brugaletta}, E. and {Burgess}, P.~W. and {Busso}, G. and {Carry}, B. and {Cellino}, A. and {Cheek}, N. and {Clementini}, G. and {Damerdji}, Y. and {Davidson}, M. and {Delchambre}, L. and {Dell'Oro}, A. and {Fern{\'a}ndez-Hern{\'a}ndez}, J. and {Galluccio}, L. and {Garc{\'\i}a-Lario}, P. and {Garcia-Reinaldos}, M. and {Gonz{\'a}lez-N{\'u}{\~n}ez}, J. and {Gosset}, E. and {Haigron}, R. and {Halbwachs}, J. -L. and {Hambly}, N.~C. and {Harrison}, D.~L. and {Hatzidimitriou}, D. and {Heiter}, U. and {Hern{\'a}ndez}, J. and {Hestroffer}, D. and {Hodgkin}, S.~T. and {Holl}, B. and {Jan{\ss}en}, K. and {Jevardat de Fombelle}, G. and {Jordan}, S. and {Krone-Martins}, A. and {Lanzafame}, A.~C. and {L{\"o}ffler}, W. and {Lorca}, A. and {Manteiga}, M. and {Marchal}, O. and {Marrese}, P.~M. and {Moitinho}, A. and {Mora}, A. and {Muinonen}, K. and {Osborne}, P. and {Pancino}, E. and {Pauwels}, T. and {Petit}, J. -M. and {Recio-Blanco}, A. and {Richards}, P.~J. and {Riello}, M. and {Rimoldini}, L. and {Robin}, A.~C. and {Roegiers}, T. and {Rybizki}, J. and {Sarro}, L.~M. and {Siopis}, C. and {Smith}, M. and {Sozzetti}, A. and {Ulla}, A. and {Utrilla}, E. and {van Leeuwen}, M. and {van Reeven}, W. and {Abbas}, U. and {Abreu Aramburu}, A. and {Accart}, S. and {Aerts}, C. and {Aguado}, J.~J. and {Ajaj}, M. and {Altavilla}, G. and {{\'A}lvarez}, M.~A. and {{\'A}lvarez Cid-Fuentes}, J. and {Alves}, J. and {Anderson}, R.~I. and {Anglada Varela}, E. and {Antoja}, T. and {Audard}, M. and {Baines}, D. and {Baker}, S.~G. and {Balaguer-N{\'u}{\~n}ez}, L. and {Balbinot}, E. and {Balog}, Z. and {Barache}, C. and {Barbato}, D. and {Barros}, M. and {Barstow}, M.~A. and {Bartolom{\'e}}, S. and {Bassilana}, J. -L. and {Bauchet}, N. and {Baudesson-Stella}, A. and {Becciani}, U. and {Bellazzini}, M. and {Bernet}, M. and {Bertone}, S. and {Bianchi}, L. and {Blanco-Cuaresma}, S. and {Boch}, T. and {Bombrun}, A. and {Bossini}, D. and {Bouquillon}, S. and {Bragaglia}, A. and {Bramante}, L. and {Breedt}, E. and {Bressan}, A. and {Brouillet}, N. and {Bucciarelli}, B. and {Burlacu}, A. and {Busonero}, D. and {Butkevich}, A.~G. and {Buzzi}, R. and {Caffau}, E. and {Cancelliere}, R. and {C{\'a}novas}, H. and {Cantat-Gaudin}, T. and {Carballo}, R. and {Carlucci}, T. and {Carnerero}, M.~I. and {Carrasco}, J.~M. and {Casamiquela}, L. and {Castellani}, M. and {Castro-Ginard}, A. and {Castro Sampol}, P. and {Chaoul}, L. and {Charlot}, P. and {Chemin}, L. and {Chiavassa}, A. and {Cioni}, M. -R.~L. and {Comoretto}, G. and {Cooper}, W.~J. and {Cornez}, T. and {Cowell}, S. and {Crifo}, F. and {Crosta}, M. and {Crowley}, C. and {Dafonte}, C. and {Dapergolas}, A. and {David}, M. and {David}, P. and {de Laverny}, P. and {De Luise}, F. and {De March}, R. and {De Ridder}, J. and {de Souza}, R. and {de Teodoro}, P. and {de Torres}, A. and {del Peloso}, E.~F. and {del Pozo}, E. and {Delbo}, M. and {Delgado}, A. and {Delgado}, H.~E. and {Delisle}, J. -B. and {Di Matteo}, P. and {Diakite}, S. and {Diener}, C. and {Distefano}, E. and {Dolding}, C. and {Eappachen}, D. and {Edvardsson}, B. and {Enke}, H. and {Esquej}, P. and {Fabre}, C. and {Fabrizio}, M. and {Faigler}, S. and {Fedorets}, G. and {Fernique}, P. and {Fienga}, A. and {Figueras}, F. and {Fouron}, C. and {Fragkoudi}, F. and {Fraile}, E. and {Franke}, F. and {Gai}, M. and {Garabato}, D. and {Garcia-Gutierrez}, A. and {Garc{\'\i}a-Torres}, M. and {Garofalo}, A. and {Gavras}, P. and {Gerlach}, E. and {Geyer}, R. and {Giacobbe}, P. and {Gilmore}, G. and {Girona}, S. and {Giuffrida}, G. and {Gomel}, R. and {Gomez}, A. and {Gonzalez-Santamaria}, I. and {Gonz{\'a}lez-Vidal}, J.~J. and {Granvik}, M. and {Guti{\'e}rrez-S{\'a}nchez}, R. and {Guy}, L.~P. and {Hauser}, M. and {Haywood}, M. and {Helmi}, A. and {Hidalgo}, S.~L. and {Hilger}, T. and {H{\l}adczuk}, N. and {Hobbs}, D. and {Holland}, G. and {Huckle}, H.~E. and {Jasniewicz}, G. and {Jonker}, P.~G. and {Juaristi Campillo}, J. and {Julbe}, F. and {Karbevska}, L. and {Kervella}, P. and {Khanna}, S. and {Kochoska}, A. and {Kontizas}, M. and {Kordopatis}, G. and {Korn}, A.~J. and {Kostrzewa-Rutkowska}, Z. and {Kruszy{\'n}ska}, K. and {Lambert}, S. and {Lanza}, A.~F. and {Lasne}, Y. and {Le Campion}, J. -F. and {Le Fustec}, Y. and {Lebreton}, Y. and {Lebzelter}, T. and {Leccia}, S. and {Leclerc}, N. and {Lecoeur-Taibi}, I. and {Liao}, S. and {Licata}, E. and {Lindstr{\o}m}, E.~P. and {Lister}, T.~A. and {Livanou}, E. and {Lobel}, A. and {Madrero Pardo}, P. and {Managau}, S. and {Mann}, R.~G. and {Marchant}, J.~M. and {Marconi}, M. and {Marcos Santos}, M.~M.~S. and {Marinoni}, S. and {Marocco}, F. and {Marshall}, D.~J. and {Martin Polo}, L. and {Mart{\'\i}n-Fleitas}, J.~M. and {Masip}, A. and {Massari}, D. and {Mastrobuono-Battisti}, A. and {Mazeh}, T. and {McMillan}, P.~J. and {Messina}, S. and {Michalik}, D. and {Millar}, N.~R. and {Mints}, A. and {Molina}, D. and {Molinaro}, R. and {Moln{\'a}r}, L. and {Montegriffo}, P. and {Mor}, R. and {Morbidelli}, R. and {Morel}, T. and {Morris}, D. and {Mulone}, A.~F. and {Munoz}, D. and {Muraveva}, T. and {Murphy}, C.~P. and {Musella}, I. and {Noval}, L. and {Ord{\'e}novic}, C. and {Orr{\`u}}, G. and {Osinde}, J. and {Pagani}, C. and {Pagano}, I. and {Palaversa}, L. and {Palicio}, P.~A. and {Panahi}, A. and {Pawlak}, M. and {Pe{\~n}alosa Esteller}, X. and {Penttil{\"a}}, A. and {Piersimoni}, A.~M. and {Pineau}, F. -X. and {Plachy}, E. and {Plum}, G. and {Poggio}, E. and {Poretti}, E. and {Poujoulet}, E. and {Pr{\v{s}}a}, A. and {Pulone}, L. and {Racero}, E. and {Ragaini}, S. and {Rainer}, M. and {Raiteri}, C.~M. and {Rambaux}, N. and {Ramos}, P. and {Ramos-Lerate}, M. and {Re Fiorentin}, P. and {Regibo}, S. and {Reyl{\'e}}, C. and {Ripepi}, V. and {Riva}, A. and {Rixon}, G. and {Robichon}, N. and {Robin}, C. and {Roelens}, M. and {Rohrbasser}, L. and {Romero-G{\'o}mez}, M. and {Rowell}, N. and {Royer}, F. and {Rybicki}, K.~A. and {Sadowski}, G. and {Sagrist{\`a} Sell{\'e}s}, A. and {Sahlmann}, J. and {Salgado}, J. and {Salguero}, E. and {Samaras}, N. and {Sanchez Gimenez}, V. and {Sanna}, N. and {Santove{\~n}a}, R. and {Sarasso}, M. and {Schultheis}, M. and {Sciacca}, E. and {Segol}, M. and {Segovia}, J.~C. and {S{\'e}gransan}, D. and {Semeux}, D. and {Shahaf}, S. and {Siddiqui}, H.~I. and {Siebert}, A. and {Siltala}, L. and {Slezak}, E. and {Smart}, R.~L. and {Solano}, E. and {Solitro}, F. and {Souami}, D. and {Souchay}, J. and {Spagna}, A. and {Spoto}, F. and {Steele}, I.~A. and {Steidelm{\"u}ller}, H. and {Stephenson}, C.~A. and {S{\"u}veges}, M. and {Szabados}, L. and {Szegedi-Elek}, E. and {Taris}, F. and {Tauran}, G. and {Taylor}, M.~B. and {Teixeira}, R. and {Thuillot}, W. and {Tonello}, N. and {Torra}, F. and {Torra}, J. and {Turon}, C. and {Unger}, N. and {Vaillant}, M. and {van Dillen}, E. and {Vanel}, O. and {Vecchiato}, A. and {Viala}, Y. and {Vicente}, D. and {Voutsinas}, S. and {Weiler}, M. and {Wevers}, T. and {Wyrzykowski}, {\L}. and {Yoldas}, A. and {Yvard}, P. and {Zhao}, H. and {Zorec}, J. and {Zucker}, S. and {Zurbach}, C. and {Zwitter}, T.},
        title = "{Gaia Early Data Release 3. Summary of the contents and survey properties}",
      journal = {\aap},
         year = 2021,
        month = may,
       volume = {649},
          eid = {A1},
        pages = {A1},
          doi = {10.1051/0004-6361/202039657},
archivePrefix = {arXiv},
       eprint = {2012.01533},
 primaryClass = {astro-ph.GA},
       adsurl = {https://ui.adsabs.harvard.edu/abs/2021A&A...649A...1G}
}

@ARTICLE{Kaplinghat2016,
       author = {{Kaplinghat}, Manoj and {Tulin}, Sean and {Yu}, Hai-Bo},
        title = "{Dark Matter Halos as Particle Colliders: Unified Solution to Small-Scale Structure Puzzles from Dwarfs to Clusters}",
      journal = {\prl},
         year = 2016,
        month = jan,
       volume = {116},
       number = {4},
          eid = {041302},
        pages = {041302},
          doi = {10.1103/PhysRevLett.116.041302},
archivePrefix = {arXiv},
       eprint = {1508.03339},
 primaryClass = {astro-ph.CO},
       adsurl = {https://ui.adsabs.harvard.edu/abs/2016PhRvL.116d1302K}
}

@ARTICLE{Kormendy2013,
       author = {{Kormendy}, John and {Ho}, Luis C.},
        title = "{Coevolution (Or Not) of Supermassive Black Holes and Host Galaxies}",
      journal = {\araa},
         year = 2013,
        month = aug,
       volume = {51},
       number = {1},
        pages = {511-653},
          doi = {10.1146/annurev-astro-082708-101811},
archivePrefix = {arXiv},
       eprint = {1304.7762},
 primaryClass = {astro-ph.CO},
       adsurl = {https://ui.adsabs.harvard.edu/abs/2013ARA&A..51..511K}
}

@ARTICLE{Gherghinescu2024,
       author = {{Gherghinescu}, Paula and {Das}, Payel and {Grand}, Robert J.~J. and {Orkney}, Matthew D.~A.},
        title = "{Action-based dynamical models of M31-like galaxies}",
      journal = {\mnras},
         year = 2024,
        month = oct,
       volume = {533},
       number = {4},
        pages = {4393-4409},
          doi = {10.1093/mnras/stae1960},
archivePrefix = {arXiv},
       eprint = {2307.09963},
 primaryClass = {astro-ph.GA},
       adsurl = {https://ui.adsabs.harvard.edu/abs/2024MNRAS.533.4393G}
}

@ARTICLE{Gherghinescu2026,
       author = {{Gherghinescu}, Paula and {Vasiliev}, Eugene and {Das}, Payel and {Read}, Justin},
        title = "{How much can we learn from resolved stellar kinematics of galactic haloes using action-based dynamical models?}",
      journal = {\mnras},
         year = 2026,
        month = mar,
       volume = {546},
       number = {4},
          eid = {stag254},
        pages = {stag254},
          doi = {10.1093/mnras/stag254},
archivePrefix = {arXiv},
       eprint = {2511.10445},
 primaryClass = {astro-ph.GA},
       adsurl = {https://ui.adsabs.harvard.edu/abs/2026MNRAS.546ag254G}
}

@ARTICLE{Kirby2013,
       author = {{Kirby}, Evan N. and {Cohen}, Judith G. and {Guhathakurta}, Puragra and {Cheng}, Lucy and {Bullock}, James S. and {Gallazzi}, Anna},
        title = "{The Universal Stellar Mass-Stellar Metallicity Relation for Dwarf Galaxies}",
      journal = {\apj},
         year = 2013,
        month = dec,
       volume = {779},
       number = {2},
          eid = {102},
        pages = {102},
          doi = {10.1088/0004-637X/779/2/102},
archivePrefix = {arXiv},
       eprint = {1310.0814},
 primaryClass = {astro-ph.GA},
       adsurl = {https://ui.adsabs.harvard.edu/abs/2013ApJ...779..102K}
}

@ARTICLE{Kleyna2003,
   author = {{Kleyna}, J.~T. and {Wilkinson}, M.~I. and {Gilmore}, G. and 
	{Evans}, N.~W.},
    title = "{A Dynamical Fossil in the Ursa Minor Dwarf Spheroidal Galaxy}",
  journal = {\apjl},
   eprint = {astro-ph/0304093},
     year = 2003,
    month = may,
   volume = 588,
    pages = {L21-L24},
      doi = {10.1086/375522},
   adsurl = {http://adsabs.harvard.edu/abs/2003ApJ...588L..21K}
}

@ARTICLE{Kokorev2024,
       author = {{Kokorev}, Vasily and {Caputi}, Karina I. and {Greene}, Jenny E. and {Dayal}, Pratika and {Trebitsch}, Maxime and {Cutler}, Sam E. and {Fujimoto}, Seiji and {Labb{\'e}}, Ivo and {Miller}, Tim B. and {Iani}, Edoardo and {Navarro-Carrera}, Rafael and {Rinaldi}, Pierluigi},
        title = "{A Census of Photometrically Selected Little Red Dots at 4 < z < 9 in JWST Blank Fields}",
      journal = {\apj},
         year = 2024,
        month = jun,
       volume = {968},
       number = {1},
          eid = {38},
        pages = {38},
          doi = {10.3847/1538-4357/ad4265},
archivePrefix = {arXiv},
       eprint = {2401.09981},
 primaryClass = {astro-ph.GA},
       adsurl = {https://ui.adsabs.harvard.edu/abs/2024ApJ...968...38K}
}

@ARTICLE{Lindegren2018,
       author = {{Lindegren}, L. and {Hern{\'a}ndez}, J. and {Bombrun}, A. and {Klioner}, S. and {Bastian}, U. and {Ramos-Lerate}, M. and {de Torres}, A. and {Steidelm{\"u}ller}, H. and {Stephenson}, C. and {Hobbs}, D. and {Lammers}, U. and {Biermann}, M. and {Geyer}, R. and {Hilger}, T. and {Michalik}, D. and {Stampa}, U. and {McMillan}, P.~J. and {Casta{\~n}eda}, J. and {Clotet}, M. and {Comoretto}, G. and {Davidson}, M. and {Fabricius}, C. and {Gracia}, G. and {Hambly}, N.~C. and {Hutton}, A. and {Mora}, A. and {Portell}, J. and {van Leeuwen}, F. and {Abbas}, U. and {Abreu}, A. and {Altmann}, M. and {Andrei}, A. and {Anglada}, E. and {Balaguer-N{\'u}{\~n}ez}, L. and {Barache}, C. and {Becciani}, U. and {Bertone}, S. and {Bianchi}, L. and {Bouquillon}, S. and {Bourda}, G. and {Br{\"u}semeister}, T. and {Bucciarelli}, B. and {Busonero}, D. and {Buzzi}, R. and {Cancelliere}, R. and {Carlucci}, T. and {Charlot}, P. and {Cheek}, N. and {Crosta}, M. and {Crowley}, C. and {de Bruijne}, J. and {de Felice}, F. and {Drimmel}, R. and {Esquej}, P. and {Fienga}, A. and {Fraile}, E. and {Gai}, M. and {Garralda}, N. and {Gonz{\'a}lez-Vidal}, J.~J. and {Guerra}, R. and {Hauser}, M. and {Hofmann}, W. and {Holl}, B. and {Jordan}, S. and {Lattanzi}, M.~G. and {Lenhardt}, H. and {Liao}, S. and {Licata}, E. and {Lister}, T. and {L{\"o}ffler}, W. and {Marchant}, J. and {Martin-Fleitas}, J. -M. and {Messineo}, R. and {Mignard}, F. and {Morbidelli}, R. and {Poggio}, E. and {Riva}, A. and {Rowell}, N. and {Salguero}, E. and {Sarasso}, M. and {Sciacca}, E. and {Siddiqui}, H. and {Smart}, R.~L. and {Spagna}, A. and {Steele}, I. and {Taris}, F. and {Torra}, J. and {van Elteren}, A. and {van Reeven}, W. and {Vecchiato}, A.},
        title = "{Gaia Data Release 2. The astrometric solution}",
      journal = {\aap},
         year = 2018,
        month = aug,
       volume = {616},
          eid = {A2},
        pages = {A2},
          doi = {10.1051/0004-6361/201832727},
archivePrefix = {arXiv},
       eprint = {1804.09366},
 primaryClass = {astro-ph.IM},
       adsurl = {https://ui.adsabs.harvard.edu/abs/2018A&A...616A...2L}
}

@ARTICLE{Lodato2006,
       author = {{Lodato}, Giuseppe and {Natarajan}, Priyamvada},
        title = "{Supermassive black hole formation during the assembly of pre-galactic discs}",
      journal = {\mnras},
         year = 2006,
        month = oct,
       volume = {371},
       number = {4},
        pages = {1813-1823},
          doi = {10.1111/j.1365-2966.2006.10801.x},
archivePrefix = {arXiv},
       eprint = {astro-ph/0606159},
 primaryClass = {astro-ph},
       adsurl = {https://ui.adsabs.harvard.edu/abs/2006MNRAS.371.1813L}
}

@ARTICLE{Lodato2007,
       author = {{Lodato}, Giuseppe and {Natarajan}, Priyamvada},
        title = "{The mass function of high-redshift seed black holes}",
      journal = {\mnras},
         year = 2007,
        month = may,
       volume = {377},
       number = {1},
        pages = {L64-L68},
          doi = {10.1111/j.1745-3933.2007.00304.x},
archivePrefix = {arXiv},
       eprint = {astro-ph/0702340},
 primaryClass = {astro-ph},
       adsurl = {https://ui.adsabs.harvard.edu/abs/2007MNRAS.377L..64L}
}

@ARTICLE{Lokas2002,
       author = {{{\L}okas}, Ewa L.},
        title = "{Dark matter distribution in dwarf spheroidal galaxies}",
      journal = {\mnras},
         year = 2002,
        month = jul,
       volume = {333},
       number = {3},
        pages = {697-708},
          doi = {10.1046/j.1365-8711.2002.05457.x},
archivePrefix = {arXiv},
       eprint = {astro-ph/0112023},
 primaryClass = {astro-ph},
       adsurl = {https://ui.adsabs.harvard.edu/abs/2002MNRAS.333..697L}
}

@ARTICLE{Lokas2005,
       author = {{{\L}okas}, Ewa L. and {Mamon}, Gary A. and {Prada}, Francisco},
        title = "{Dark matter distribution in the Draco dwarf from velocity moments}",
      journal = {\mnras},
         year = 2005,
        month = nov,
       volume = {363},
       number = {3},
        pages = {918-928},
          doi = {10.1111/j.1365-2966.2005.09497.x},
archivePrefix = {arXiv},
       eprint = {astro-ph/0411694},
 primaryClass = {astro-ph},
       adsurl = {https://ui.adsabs.harvard.edu/abs/2005MNRAS.363..918L}
}

@ARTICLE{Lora2009,
       author = {{Lora}, V. and {S{\'a}nchez-Salcedo}, F.~J. and {Raga}, A.~C. and {Esquivel}, A.},
        title = "{An Upper Limit on the Mass of the Black Hole in Ursa Minor Dwarf Galaxy}",
      journal = {\apjl},
         year = 2009,
        month = jul,
       volume = {699},
       number = {2},
        pages = {L113-L117},
          doi = {10.1088/0004-637X/699/2/L113},
archivePrefix = {arXiv},
       eprint = {0906.0951},
 primaryClass = {astro-ph.CO},
       adsurl = {https://ui.adsabs.harvard.edu/abs/2009ApJ...699L.113L}
}

@ARTICLE{Manni2015,
       author = {{Manni}, L. and {Nucita}, A.~A. and {De Paolis}, F. and {Testa}, V. and {Ingrosso}, G.},
        title = "{A XMM-Newton observation of a sample of four close dwarf spheroidal galaxies}",
      journal = {\mnras},
         year = 2015,
        month = aug,
       volume = {451},
       number = {3},
        pages = {2735-2749},
          doi = {10.1093/mnras/stv1009},
archivePrefix = {arXiv},
       eprint = {1509.01076},
 primaryClass = {astro-ph.GA},
       adsurl = {https://ui.adsabs.harvard.edu/abs/2015MNRAS.451.2735M}
}

@ARTICLE{MadauRees2001,
       author = {{Madau}, Piero and {Rees}, Martin J.},
        title = "{Massive Black Holes as Population III Remnants}",
      journal = {\apjl},
         year = 2001,
        month = apr,
       volume = {551},
       number = {1},
        pages = {L27-L30},
          doi = {10.1086/319848},
archivePrefix = {arXiv},
       eprint = {astro-ph/0101223},
 primaryClass = {astro-ph},
       adsurl = {https://ui.adsabs.harvard.edu/abs/2001ApJ...551L..27M}
}

@ARTICLE{Mayer2015,
       author = {{Mayer}, Lucio and {Fiacconi}, Davide and {Bonoli}, Silvia and {Quinn}, Thomas and {Ro{\v{s}}kar}, Rok and {Shen}, Sijing and {Wadsley}, James},
        title = "{Direct Formation of Supermassive Black Holes in Metal-enriched Gas at the Heart of High-redshift Galaxy Mergers}",
      journal = {\apj},
         year = 2015,
        month = sep,
       volume = {810},
       number = {1},
          eid = {51},
        pages = {51},
          doi = {10.1088/0004-637X/810/1/51},
archivePrefix = {arXiv},
       eprint = {1411.5683},
 primaryClass = {astro-ph.GA},
       adsurl = {https://ui.adsabs.harvard.edu/abs/2015ApJ...810...51M}
}

@ARTICLE{MarelCioni2001,
       author = {{van der Marel}, Roeland P. and {Cioni}, Maria-Rosa L.},
        title = "{Magellanic Cloud Structure from Near-Infrared Surveys. I. The Viewing Angles of the Large Magellanic Cloud}",
      journal = {\aj},
         year = 2001,
        month = oct,
       volume = {122},
       number = {4},
        pages = {1807-1826},
          doi = {10.1086/323099},
archivePrefix = {arXiv},
       eprint = {astro-ph/0105339},
 primaryClass = {astro-ph},
       adsurl = {https://ui.adsabs.harvard.edu/abs/2001AJ....122.1807V}
}

@ARTICLE{Marigo2008,
       author = {{Marigo}, P. and {Girardi}, L. and {Bressan}, A. and {Groenewegen}, M.~A.~T. and {Silva}, L. and {Granato}, G.~L.},
        title = "{Evolution of asymptotic giant branch stars. II. Optical to far-infrared isochrones with improved TP-AGB models}",
      journal = {\aap},
         year = 2008,
        month = may,
       volume = {482},
       number = {3},
        pages = {883-905},
          doi = {10.1051/0004-6361:20078467},
archivePrefix = {arXiv},
       eprint = {0711.4922},
 primaryClass = {astro-ph},
       adsurl = {https://ui.adsabs.harvard.edu/abs/2008A&A...482..883M}
}

@ARTICLE{Martin2008,
       author = {{Martin}, Nicolas F. and {de Jong}, Jelte T.~A. and {Rix}, Hans-Walter},
        title = "{A Comprehensive Maximum Likelihood Analysis of the Structural Properties of Faint Milky Way Satellites}",
      journal = {\apj},
         year = 2008,
        month = sep,
       volume = {684},
       number = {2},
        pages = {1075-1092},
          doi = {10.1086/590336},
archivePrefix = {arXiv},
       eprint = {0805.2945},
 primaryClass = {astro-ph},
       adsurl = {https://ui.adsabs.harvard.edu/abs/2008ApJ...684.1075M}
}

@ARTICLE{Martinez2001,
       author = {{Mart{\'\i}nez-Delgado}, D. and {Alonso-Garc{\'\i}a}, J. and {Aparicio}, A. and {G{\'o}mez-Flechoso}, M.~A.},
        title = "{A Tidal Extension in the Ursa Minor Dwarf Spheroidal Galaxy}",
      journal = {\apjl},
         year = 2001,
        month = mar,
       volume = {549},
       number = {1},
        pages = {L63-L66},
          doi = {10.1086/319150},
archivePrefix = {arXiv},
       eprint = {astro-ph/0101456},
 primaryClass = {astro-ph},
       adsurl = {https://ui.adsabs.harvard.edu/abs/2001ApJ...549L..63M}
}

@ARTICLE{Martinez2023,
       author = {{Mart{\'\i}nez-Garc{\'\i}a}, Alberto Manuel and {del Pino}, Andr{\'e}s and {Aparicio}, Antonio},
        title = "{Tidally induced velocity gradients in the Milky Way dwarf spheroidal satellites}",
      journal = {\mnras},
         year = 2023,
        month = jan,
       volume = {518},
       number = {2},
        pages = {3083-3094},
          doi = {10.1093/mnras/stac3305},
archivePrefix = {arXiv},
       eprint = {2206.06339},
 primaryClass = {astro-ph.GA},
       adsurl = {https://ui.adsabs.harvard.edu/abs/2023MNRAS.518.3083M}
}

@ARTICLE{Martnez2021,
       author = {{Mart{\'\i}nez-Garc{\'\i}a}, Alberto Manuel and {del Pino}, Andr{\'e}s and {Aparicio}, Antonio and {van der Marel}, Roeland P. and {Watkins}, Laura L.},
        title = "{Internal rotation of Milky Way dwarf spheroidal satellites with Gaia Early Data Release 3}",
      journal = {\mnras},
         year = 2021,
        month = aug,
       volume = {505},
       number = {4},
        pages = {5884-5895},
          doi = {10.1093/mnras/stab1568},
archivePrefix = {arXiv},
       eprint = {2104.00662},
 primaryClass = {astro-ph.GA},
       adsurl = {https://ui.adsabs.harvard.edu/abs/2021MNRAS.505.5884M}
}

@ARTICLE{Martin2016,
       author = {{Martin}, Nicolas F. and {Ibata}, Rodrigo A. and {Lewis}, Geraint F. and {McConnachie}, Alan and {Babul}, Arif and {Bate}, Nicholas F. and {Bernard}, Edouard and {Chapman}, Scott C. and {Collins}, Michelle M.~L. and {Conn}, Anthony R. and {Crnojevi{\'c}}, Denija and {Fardal}, Mark A. and {Ferguson}, Annette M.~N. and {Irwin}, Michael and {Mackey}, A. Dougal and {McMonigal}, Brendan and {Navarro}, Julio F. and {Rich}, R. Michael},
        title = "{The PAndAS View of the Andromeda Satellite System. II. Detailed Properties of 23 M31 Dwarf Spheroidal Galaxies}",
      journal = {\apj},
         year = 2016,
        month = dec,
       volume = {833},
       number = {2},
          eid = {167},
        pages = {167},
          doi = {10.3847/1538-4357/833/2/167},
archivePrefix = {arXiv},
       eprint = {1610.01158},
 primaryClass = {astro-ph.GA},
       adsurl = {https://ui.adsabs.harvard.edu/abs/2016ApJ...833..167M}
}

@ARTICLE{Mateo1998,
       author = {{Mateo}, Mario L.},
        title = "{Dwarf Galaxies of the Local Group}",
      journal = {Annual Review of Astronomy and Astrophysics},
         year = 1998,
        month = Jan,
       volume = {36},
        pages = {435-506},
          doi = {10.1146/annurev.astro.36.1.435},
archivePrefix = {arXiv},
       eprint = {astro-ph/9810070},
 primaryClass = {astro-ph},
       adsurl = {https://ui.adsabs.harvard.edu/#abs/1998ARA&A..36..435M}
}

@ARTICLE{McConnachie2012,
   author = {{McConnachie}, A.~W.},
    title = "{The Observed Properties of Dwarf Galaxies in and around the Local Group}",
  journal = {\aj},
archivePrefix = "arXiv",
   eprint = {1204.1562},
     year = 2012,
    month = jul,
   volume = 144,
      eid = {4},
    pages = {4},
      doi = {10.1088/0004-6256/144/1/4},
   adsurl = {https://ui.adsabs.harvard.edu/abs/2012AJ....144....4M}
}

@ARTICLE{McConnachie2020,
       author = {{McConnachie}, Alan W. and {Venn}, Kim A.},
        title = "{Revised and New Proper Motions for Confirmed and Candidate Milky Way Dwarf Galaxies}",
      journal = {\aj},
         year = 2020,
        month = sep,
       volume = {160},
       number = {3},
          eid = {124},
        pages = {124},
          doi = {10.3847/1538-3881/aba4ab},
archivePrefix = {arXiv},
       eprint = {2007.05011},
 primaryClass = {astro-ph.GA},
       adsurl = {https://ui.adsabs.harvard.edu/abs/2020AJ....160..124M}
}

@ARTICLE{McConnell2013,
       author = {{McConnell}, Nicholas J. and {Ma}, Chung-Pei},
        title = "{Revisiting the Scaling Relations of Black Hole Masses and Host Galaxy Properties}",
      journal = {\apj},
         year = 2013,
        month = feb,
       volume = {764},
       number = {2},
          eid = {184},
        pages = {184},
          doi = {10.1088/0004-637X/764/2/184},
archivePrefix = {arXiv},
       eprint = {1211.2816},
 primaryClass = {astro-ph.CO},
       adsurl = {https://ui.adsabs.harvard.edu/abs/2013ApJ...764..184M}
}

@ARTICLE{Merritt1985,
   author = {{Merritt}, D.},
    title = "{Distribution functions for spherical galaxies}",
  journal = {\mnras},
     year = 1985,
    month = jun,
   volume = 214,
    pages = {25P-28P},
      doi = {10.1093/mnras/214.1.25P},
   adsurl = {https://ui.adsabs.harvard.edu/abs/1985MNRAS.214P..25M}
}

@ARTICLE{Merritt2001,
       author = {{Merritt}, David and {Ferrarese}, Laura},
        title = "{The M$_{{\textbullet}}$-{\ensuremath{\sigma}} Relation for Supermassive Black Holes}",
      journal = {\apj},
         year = 2001,
        month = jan,
       volume = {547},
       number = {1},
        pages = {140-145},
          doi = {10.1086/318372},
archivePrefix = {arXiv},
       eprint = {astro-ph/0008310},
 primaryClass = {astro-ph},
       adsurl = {https://ui.adsabs.harvard.edu/abs/2001ApJ...547..140M}
}

@ARTICLE{Merrifield1990,
       author = {{Merrifield}, Michael R. and {Kent}, Stephen M.},
        title = "{Fourth Moments and the Dynamics of Spherical Systems}",
      journal = {\aj},
         year = 1990,
        month = may,
       volume = {99},
        pages = {1548},
          doi = {10.1086/115438},
       adsurl = {https://ui.adsabs.harvard.edu/abs/1990AJ.....99.1548M}
}

@ARTICLE{Miller2015,
       author = {{Miller}, Brendan P. and {Gallo}, Elena and {Greene}, Jenny E. and {Kelly}, Brandon C. and {Treu}, Tommaso and {Woo}, Jong-Hak and {Baldassare}, Vivienne},
        title = "{X-Ray Constraints on the Local Supermassive Black Hole Occupation Fraction}",
      journal = {\apj},
         year = 2015,
        month = jan,
       volume = {799},
       number = {1},
          eid = {98},
        pages = {98},
          doi = {10.1088/0004-637X/799/1/98},
archivePrefix = {arXiv},
       eprint = {1403.4246},
 primaryClass = {astro-ph.GA},
       adsurl = {https://ui.adsabs.harvard.edu/abs/2015ApJ...799...98M}
}

@ARTICLE{Muni2025,
       author = {{Muni}, Claudia and {Pontzen}, Andrew and {Read}, Justin I. and {Agertz}, Oscar and {Rey}, Martin P. and {Taylor}, Ethan and {Kim}, Stacy Y. and {Gray}, Emily I.},
        title = "{EDGE: dark matter core creation depends on the timing of star formation}",
      journal = {\mnras},
         year = 2025,
        month = jan,
       volume = {536},
       number = {1},
        pages = {314-323},
          doi = {10.1093/mnras/stae2748},
archivePrefix = {arXiv},
       eprint = {2407.14579},
 primaryClass = {astro-ph.GA},
       adsurl = {https://ui.adsabs.harvard.edu/abs/2025MNRAS.536..314M}
}

@ARTICLE{Munoz2018,
       author = {{Mu{\~n}oz}, Ricardo R. and {C{\^o}t{\'e}}, Patrick and {Santana}, Felipe A. and {Geha}, Marla and {Simon}, Joshua D. and {Oyarz{\'u}n}, Grecco A. and {Stetson}, Peter B. and {Djorgovski}, S.~G.},
        title = "{A MegaCam Survey of Outer Halo Satellites. III. Photometric and Structural Parameters}",
      journal = {\apj},
         year = 2018,
        month = jun,
       volume = {860},
       number = {1},
          eid = {66},
        pages = {66},
          doi = {10.3847/1538-4357/aac16b},
archivePrefix = {arXiv},
       eprint = {1806.06891},
 primaryClass = {astro-ph.GA},
       adsurl = {https://ui.adsabs.harvard.edu/abs/2018ApJ...860...66M}
}

@ARTICLE{Natarajan2021,
       author = {{Natarajan}, Priyamvada},
        title = "{A new channel to form IMBHs throughout cosmic time}",
      journal = {\mnras},
         year = 2021,
        month = feb,
       volume = {501},
       number = {1},
        pages = {1413-1425},
          doi = {10.1093/mnras/staa3724},
archivePrefix = {arXiv},
       eprint = {2009.09156},
 primaryClass = {astro-ph.GA},
       adsurl = {https://ui.adsabs.harvard.edu/abs/2021MNRAS.501.1413N}
}

@ARTICLE{Navarro1997,
       author = {{Navarro}, Julio F. and {Frenk}, Carlos S. and {White}, Simon D.~M.},
        title = "{A Universal Density Profile from Hierarchical Clustering}",
      journal = {\apj},
         year = 1997,
        month = dec,
       volume = {490},
       number = {2},
        pages = {493-508},
          doi = {10.1086/304888},
archivePrefix = {arXiv},
       eprint = {astro-ph/9611107},
 primaryClass = {astro-ph},
       adsurl = {https://ui.adsabs.harvard.edu/abs/1997ApJ...490..493N}
}

@ARTICLE{NavarroFrenkWhite1996,
   author = {{Navarro}, J.~F. and {Frenk}, C.~S. and {White}, S.~D.~M.},
    title = "{The Structure of Cold Dark Matter Halos}",
  journal = {\apj},
   eprint = {astro-ph/9508025},
     year = 1996,
    month = may,
   volume = 462,
    pages = {563},
      doi = {10.1086/177173},
   adsurl = {http://adsabs.harvard.edu/abs/1996ApJ...462..563N}
}

@ARTICLE{Nelson2014,
       author = {{Nelson}, Benjamin and {Ford}, Eric B. and {Payne}, Matthew J.},
        title = "{RUN DMC: An Efficient, Parallel Code for Analyzing Radial Velocity Observations Using N-body Integrations and Differential Evolution Markov Chain Monte Carlo}",
      journal = {\apjs},
         year = 2014,
        month = jan,
       volume = {210},
       number = {1},
          eid = {11},
        pages = {11},
          doi = {10.1088/0067-0049/210/1/11},
archivePrefix = {arXiv},
       eprint = {1311.5229},
 primaryClass = {astro-ph.EP},
       adsurl = {https://ui.adsabs.harvard.edu/abs/2014ApJS..210...11N}
}

@ARTICLE{Nguyen2018,
       author = {{Nguyen}, Dieu D. and {Seth}, Anil C. and {Neumayer}, Nadine and {Kamann}, Sebastian and {Voggel}, Karina T. and {Cappellari}, Michele and {Picotti}, Arianna and {Nguyen}, Phuong M. and {B{\"o}ker}, Torsten and {Debattista}, Victor and {Caldwell}, Nelson and {McDermid}, Richard and {Bastian}, Nathan and {Ahn}, Christopher C. and {Pechetti}, Renuka},
        title = "{Nearby Early-type Galactic Nuclei at High Resolution: Dynamical Black Hole and Nuclear Star Cluster Mass Measurements}",
      journal = {\apj},
         year = 2018,
        month = may,
       volume = {858},
       number = {2},
          eid = {118},
        pages = {118},
          doi = {10.3847/1538-4357/aabe28},
archivePrefix = {arXiv},
       eprint = {1711.04314},
 primaryClass = {astro-ph.GA},
       adsurl = {https://ui.adsabs.harvard.edu/abs/2018ApJ...858..118N}
}

@ARTICLE{Nguyen2019,
       author = {{Nguyen}, Dieu D. and {Seth}, Anil C. and {Neumayer}, Nadine and {Iguchi}, Satoru and {Cappellari}, Michelle and {Strader}, Jay and {Chomiuk}, Laura and {Tremou}, Evangelia and {Pacucci}, Fabio and {Nakanishi}, Kouichiro and {Bahramian}, Arash and {Nguyen}, Phuong M. and {den Brok}, Mark and {Ahn}, Christopher C. and {Voggel}, Karina T. and {Kacharov}, Nikolay and {Tsukui}, Takafumi and {Ly}, Cuc K. and {Dumont}, Antoine and {Pechetti}, Renuka},
        title = "{Improved Dynamical Constraints on the Masses of the Central Black Holes in Nearby Low-mass Early-type Galactic Nuclei and the First Black Hole Determination for NGC 205}",
      journal = {\apj},
         year = 2019,
        month = feb,
       volume = {872},
       number = {1},
          eid = {104},
        pages = {104},
          doi = {10.3847/1538-4357/aafe7a},
archivePrefix = {arXiv},
       eprint = {1901.05496},
 primaryClass = {astro-ph.GA},
       adsurl = {https://ui.adsabs.harvard.edu/abs/2019ApJ...872..104N}
}

@ARTICLE{NipotiBinney2015,
   author = {{Nipoti}, C. and {Binney}, J.},
    title = "{Early flattening of dark matter cusps in dwarf spheroidal galaxies}",
  journal = {\mnras},
archivePrefix = "arXiv",
   eprint = {1410.6169},
     year = 2015,
    month = jan,
   volume = 446,
    pages = {1820-1828},
      doi = {10.1093/mnras/stu2217},
   adsurl = {http://adsabs.harvard.edu/abs/2015MNRAS.446.1820N}
}

@ARTICLE{Nucita2013,
       author = {{Nucita}, A.~A. and {De Paolis}, F. and {Manni}, L. and {Ingrosso}, G.},
        title = "{Hint for a faint intermediate mass black hole in the Ursa Minor dwarf galaxy}",
      journal = {\na},
         year = 2013,
        month = oct,
       volume = {23},
        pages = {107-112},
          doi = {10.1016/j.newast.2013.03.003},
       adsurl = {https://ui.adsabs.harvard.edu/abs/2013NewA...23..107N}
}

@ARTICLE{Onorbe2015,
       author = {{O{\~n}orbe}, Jose and {Boylan-Kolchin}, Michael and {Bullock}, James S. and {Hopkins}, Philip F. and {Kere{\v{s}}}, Du{\v{s}}an and {Faucher-Gigu{\`e}re}, Claude-Andr{\'e} and {Quataert}, Eliot and {Murray}, Norman},
        title = "{Forged in FIRE: cusps, cores and baryons in low-mass dwarf galaxies}",
      journal = {\mnras},
         year = 2015,
        month = dec,
       volume = {454},
       number = {2},
        pages = {2092-2106},
          doi = {10.1093/mnras/stv2072},
archivePrefix = {arXiv},
       eprint = {1502.02036},
 primaryClass = {astro-ph.GA},
       adsurl = {https://ui.adsabs.harvard.edu/abs/2015MNRAS.454.2092O}
}

@ARTICLE{Oh2011,
       author = {{Oh}, Se-Heon and {Brook}, Chris and {Governato}, Fabio and {Brinks}, Elias and {Mayer}, Lucio and {de Blok}, W.~J.~G. and {Brooks}, Alyson and {Walter}, Fabian},
        title = "{The Central Slope of Dark Matter Cores in Dwarf Galaxies: Simulations versus THINGS}",
      journal = {\aj},
         year = 2011,
        month = jul,
       volume = {142},
       number = {1},
          eid = {24},
        pages = {24},
          doi = {10.1088/0004-6256/142/1/24},
archivePrefix = {arXiv},
       eprint = {1011.2777},
 primaryClass = {astro-ph.CO},
       adsurl = {https://ui.adsabs.harvard.edu/abs/2011AJ....142...24O}
}

@ARTICLE{Oh2015,
       author = {{Oh}, Se-Heon and {Hunter}, Deidre A. and {Brinks}, Elias and
         {Elmegreen}, Bruce G. and {Schruba}, Andreas and {Walter}, Fabian and
         {Rupen}, Michael P. and {Young}, Lisa M. and {Simpson}, Caroline E. and
         {Johnson}, Megan C. and {Herrmann}, Kimberly A. and
         {Ficut-Vicas}, Dana and {Cigan}, Phil and {Heesen}, Volker and
         {Ashley}, Trisha and {Zhang}, Hong-Xin},
        title = "{High-resolution Mass Models of Dwarf Galaxies from LITTLE THINGS}",
      journal = {\aj},
         year = 2015,
        month = jun,
       volume = {149},
       number = {6},
          eid = {180},
        pages = {180},
          doi = {10.1088/0004-6256/149/6/180},
archivePrefix = {arXiv},
       eprint = {1502.01281},
 primaryClass = {astro-ph.GA},
       adsurl = {https://ui.adsabs.harvard.edu/abs/2015AJ....149..180O}
}

@ARTICLE{Orkney2021,
       author = {{Orkney}, Matthew D.~A. and {Read}, Justin I. and {Rey}, Martin P. and {Nasim}, Imran and {Pontzen}, Andrew and {Agertz}, Oscar and {Kim}, Stacy Y. and {Delorme}, Maxime and {Dehnen}, Walter},
        title = "{EDGE: two routes to dark matter core formation in ultra-faint dwarfs}",
      journal = {\mnras},
         year = 2021,
        month = jul,
       volume = {504},
       number = {3},
        pages = {3509-3522},
          doi = {10.1093/mnras/stab1066},
archivePrefix = {arXiv},
       eprint = {2101.02688},
 primaryClass = {astro-ph.GA},
       adsurl = {https://ui.adsabs.harvard.edu/abs/2021MNRAS.504.3509O}
}

@ARTICLE{Osipkov1979,
   author = {{Osipkov}, L.~P.},
    title = "{Spherical systems of gravitating bodies with an ellipsoidal velocity distribution}",
  journal = {Soviet Astronomy Letters},
     year = 1979,
   volume = 5,
    pages = {42-44},
   adsurl = {https://ui.adsabs.harvard.edu/abs/1979SvAL....5...42O}
}

@ARTICLE{Pace2014,
       author = {{Pace}, Andrew B. and {Martinez}, Gregory D. and {Kaplinghat}, Manoj and {Mu{\~n}oz}, Ricardo R.},
        title = "{Evidence for substructure in Ursa Minor dwarf spheroidal galaxy using a Bayesian object detection method}",
      journal = {\mnras},
         year = 2014,
        month = aug,
       volume = {442},
       number = {2},
        pages = {1718-1730},
          doi = {10.1093/mnras/stu938},
archivePrefix = {arXiv},
       eprint = {1208.4146},
 primaryClass = {astro-ph.GA},
       adsurl = {https://ui.adsabs.harvard.edu/abs/2014MNRAS.442.1718P}
}

@ARTICLE{Pace2020,
       author = {{Pace}, Andrew B. and {Kaplinghat}, Manoj and {Kirby}, Evan and {Simon}, Joshua D. and {Tollerud}, Erik and {Mu{\~n}oz}, Ricardo R. and {C{\^o}t{\'e}}, Patrick and {Djorgovski}, S.~G. and {Geha}, Marla},
        title = "{Multiple chemodynamic stellar populations of the Ursa Minor dwarf spheroidal galaxy}",
      journal = {\mnras},
         year = 2020,
        month = jul,
       volume = {495},
       number = {3},
        pages = {3022-3040},
          doi = {10.1093/mnras/staa1419},
archivePrefix = {arXiv},
       eprint = {2002.09503},
 primaryClass = {astro-ph.GA},
       adsurl = {https://ui.adsabs.harvard.edu/abs/2020MNRAS.495.3022P}
}

@ARTICLE{Pascale2018,
       author = {{Pascale}, Raffaele and {Posti}, Lorenzo and {Nipoti}, Carlo and
         {Binney}, James},
        title = "{Action-based dynamical models of dwarf spheroidal galaxies: application to Fornax}",
      journal = {\mnras},
         year = "2018",
        month = "Oct",
       volume = {480},
        pages = {927-946},
          doi = {10.1093/mnras/sty1860},
archivePrefix = {arXiv},
       eprint = {1802.02606},
 primaryClass = {astro-ph.GA},
       adsurl = {https://ui.adsabs.harvard.edu/\#abs/2018MNRAS.480..927P}
}

@ARTICLE{Pascale2019,
       author = {{Pascale}, Raffaele and {Binney}, James and {Nipoti}, Carlo and
         {Posti}, Lorenzo},
        title = "{Action-based models for dwarf spheroidal galaxies and globular clusters}",
      journal = {\mnras},
         year = "2019",
        month = "Sep",
       volume = {488},
       number = {2},
        pages = {2423-2439},
          doi = {10.1093/mnras/stz1617},
archivePrefix = {arXiv},
       eprint = {1904.08447},
 primaryClass = {astro-ph.GA},
       adsurl = {https://ui.adsabs.harvard.edu/abs/2019MNRAS.488.2423P}
}

@ARTICLE{Pascale2024a,
       author = {{Pascale}, R. and {Nipoti}, C. and {Calura}, F. and {Della Croce}, A.},
        title = "{The central black hole in the dwarf spheroidal galaxy Leo I: Not supermassive, at most an intermediate-mass candidate}",
      journal = {\aap},
         year = 2024,
        month = apr,
       volume = {684},
          eid = {L19},
        pages = {L19},
          doi = {10.1051/0004-6361/202449620},
archivePrefix = {arXiv},
       eprint = {2403.14784},
 primaryClass = {astro-ph.GA},
       adsurl = {https://ui.adsabs.harvard.edu/abs/2024A&A...684L..19P}
}

@ARTICLE{Pascale2025,
       author = {{Pascale}, R. and {Nipoti}, C. and {Calura}, F. and {Della Croce}, A.},
        title = "{Leo I: The classical dwarf spheroidal galaxy with the highest dark matter density}",
      journal = {\aap},
         year = 2025,
        month = aug,
       volume = {700},
          eid = {A77},
        pages = {A77},
          doi = {10.1051/0004-6361/202555004},
archivePrefix = {arXiv},
       eprint = {2506.13847},
 primaryClass = {astro-ph.GA},
       adsurl = {https://ui.adsabs.harvard.edu/abs/2025A&A...700A..77P}
}

@ARTICLE{Pontzen2012,
   author = {{Pontzen}, A. and {Governato}, F.},
    title = "{How supernova feedback turns dark matter cusps into cores}",
  journal = {\mnras},
archivePrefix = "arXiv",
   eprint = {1106.0499},
     year = 2012,
    month = apr,
   volume = 421,
    pages = {3464-3471},
      doi = {10.1111/j.1365-2966.2012.20571.x},
   adsurl = {http://adsabs.harvard.edu/abs/2012MNRAS.421.3464P}
}

@ARTICLE{PortegiesZwart2004,
       author = {{Portegies Zwart}, Simon F. and {Baumgardt}, Holger and {Hut}, Piet and {Makino}, Junichiro and {McMillan}, Stephen L.~W.},
        title = "{Formation of massive black holes through runaway collisions in dense young star clusters}",
      journal = {\nat},
         year = 2004,
        month = apr,
       volume = {428},
       number = {6984},
        pages = {724-726},
          doi = {10.1038/nature02448},
archivePrefix = {arXiv},
       eprint = {astro-ph/0402622},
 primaryClass = {astro-ph},
       adsurl = {https://ui.adsabs.harvard.edu/abs/2004Natur.428..724P}
}

@ARTICLE{Posti2015,
   author = {{Posti}, L. and {Binney}, J. and {Nipoti}, C. and {Ciotti}, L.
	},
    title = "{Action-based distribution functions for spheroidal galaxy components}",
  journal = {\mnras},
archivePrefix = "arXiv",
   eprint = {1411.7897},
     year = 2015,
    month = mar,
   volume = 447,
    pages = {3060-3068},
      doi = {10.1093/mnras/stu2608},
   adsurl = {http://adsabs.harvard.edu/abs/2015MNRAS.447.3060P}
}

@INPROCEEDINGS{Pryor93,
   author = {{Pryor}, C. and {Meylan}, G.},
    title = "{Velocity Dispersions for Galactic Globular Clusters}",
booktitle = {Structure and Dynamics of Globular Clusters},
     year = 1993,
   series = {Astronomical Society of the Pacific Conference Series},
   volume = 50,
   editor = {{Djorgovski}, S.~G. and {Meylan}, G.},
    month = jan,
    pages = {357},
   adsurl = {http://adsabs.harvard.edu/abs/1993ASPC...50..357P}
}

@ARTICLE{Quinlan1995,
   author = {{Quinlan}, G.~D. and {Hernquist}, L. and {Sigurdsson}, S.},
    title = "{Models of Galaxies with Central Black Holes: Adiabatic Growth in Spherical Galaxies}",
  journal = {\apj},
   eprint = {astro-ph/9407005},
     year = 1995,
    month = feb,
   volume = 440,
    pages = {554},
      doi = {10.1086/175295},
   adsurl = {http://adsabs.harvard.edu/abs/1995ApJ...440..554Q}
}

@ARTICLE{Read2017,
   author = {{Read}, J.~I. and {Iorio}, G. and {Agertz}, O. and {Fraternali}, F.
	},
    title = "{The stellar mass-halo mass relation of isolated field dwarfs: a critical test of {$\Lambda$}CDM at the edge of galaxy formation}",
  journal = {\mnras},
archivePrefix = "arXiv",
   eprint = {1607.03127},
     year = 2017,
    month = may,
   volume = 467,
    pages = {2019-2038},
      doi = {10.1093/mnras/stx147},
   adsurl = {http://adsabs.harvard.edu/abs/2017MNRAS.467.2019R}
}

@ARTICLE{Read2017b,
       author = {{Read}, J.~I. and {Steger}, P.},
        title = "{How to break the density-anisotropy degeneracy in spherical stellar systems}",
      journal = {\mnras},
         year = 2017,
        month = nov,
       volume = {471},
       number = {4},
        pages = {4541-4558},
          doi = {10.1093/mnras/stx1798},
archivePrefix = {arXiv},
       eprint = {1701.04833},
 primaryClass = {astro-ph.GA},
       adsurl = {https://ui.adsabs.harvard.edu/abs/2017MNRAS.471.4541R}
}

@ARTICLE{Read2021,
       author = {{Read}, J.~I. and {Mamon}, G.~A. and {Vasiliev}, E. and {Watkins}, L.~L. and {Walker}, M.~G. and {Pe{\~n}arrubia}, J. and {Wilkinson}, M. and {Dehnen}, W. and {Das}, P.},
        title = "{Breaking beta: a comparison of mass modelling methods for spherical systems}",
      journal = {\mnras},
         year = 2021,
        month = feb,
       volume = {501},
       number = {1},
        pages = {978-993},
          doi = {10.1093/mnras/staa3663},
archivePrefix = {arXiv},
       eprint = {2011.09493},
 primaryClass = {astro-ph.GA},
       adsurl = {https://ui.adsabs.harvard.edu/abs/2021MNRAS.501..978R}
}

@ARTICLE{Read2018,
       author = {{Read}, J.~I. and {Walker}, M.~G. and {Steger}, P.},
        title = "{The case for a cold dark matter cusp in Draco}",
      journal = {\mnras},
         year = 2018,
        month = nov,
       volume = {481},
       number = {1},
        pages = {860-877},
          doi = {10.1093/mnras/sty2286},
archivePrefix = {arXiv},
       eprint = {1805.06934},
 primaryClass = {astro-ph.GA},
       adsurl = {https://ui.adsabs.harvard.edu/abs/2018MNRAS.481..860R}
}

@ARTICLE{Read2019,
   author = {{Read}, J.~I. and {Walker}, M.~G. and {Steger}, P.},
    title = "{Dark matter heats up in dwarf galaxies}",
  journal = {\mnras},
archivePrefix = "arXiv",
   eprint = {1808.06634},
     year = 2019,
    month = mar,
   volume = 484,
    pages = {1401-1420},
      doi = {10.1093/mnras/sty3404},
   adsurl = {http://adsabs.harvard.edu/abs/2019MNRAS.484.1401R}
}

@ARTICLE{Relatores2019,
       author = {{Relatores}, Nicole C. and {Newman}, Andrew B. and {Simon}, Joshua D. and {Ellis}, Richard S. and {Truong}, Phuongmai and {Blitz}, Leo and {Bolatto}, Alberto and {Martin}, Christopher and {Matuszewski}, Matt and {Morrissey}, Patrick and {Neill}, James D.},
        title = "{The Dark Matter Distributions in Low-mass Disk Galaxies. II. The Inner Density Profiles}",
      journal = {\apj},
         year = 2019,
        month = dec,
       volume = {887},
       number = {1},
          eid = {94},
        pages = {94},
          doi = {10.3847/1538-4357/ab5305},
archivePrefix = {arXiv},
       eprint = {1911.05836},
 primaryClass = {astro-ph.GA},
       adsurl = {https://ui.adsabs.harvard.edu/abs/2019ApJ...887...94R}
}

@ARTICLE{Richardson2013,
       author = {{Richardson}, Thomas and {Fairbairn}, Malcolm},
        title = "{Analytical solutions to the mass-anisotropy degeneracy with higher order Jeans analysis: a general method}",
      journal = {\mnras},
         year = 2013,
        month = jul,
       volume = {432},
       number = {4},
        pages = {3361-3380},
          doi = {10.1093/mnras/stt686},
archivePrefix = {arXiv},
       eprint = {1207.1709},
 primaryClass = {astro-ph.CO},
       adsurl = {https://ui.adsabs.harvard.edu/abs/2013MNRAS.432.3361R}
}

@ARTICLE{Riello2021,
       author = {{Riello}, M. and {De Angeli}, F. and {Evans}, D.~W. and {Montegriffo}, P. and {Carrasco}, J.~M. and {Busso}, G. and {Palaversa}, L. and {Burgess}, P.~W. and {Diener}, C. and {Davidson}, M. and {Rowell}, N. and {Fabricius}, C. and {Jordi}, C. and {Bellazzini}, M. and {Pancino}, E. and {Harrison}, D.~L. and {Cacciari}, C. and {van Leeuwen}, F. and {Hambly}, N.~C. and {Hodgkin}, S.~T. and {Osborne}, P.~J. and {Altavilla}, G. and {Barstow}, M.~A. and {Brown}, A.~G.~A. and {Castellani}, M. and {Cowell}, S. and {De Luise}, F. and {Gilmore}, G. and {Giuffrida}, G. and {Hidalgo}, S. and {Holland}, G. and {Marinoni}, S. and {Pagani}, C. and {Piersimoni}, A.~M. and {Pulone}, L. and {Ragaini}, S. and {Rainer}, M. and {Richards}, P.~J. and {Sanna}, N. and {Walton}, N.~A. and {Weiler}, M. and {Yoldas}, A.},
        title = "{Gaia Early Data Release 3. Photometric content and validation}",
      journal = {\aap},
         year = 2021,
        month = may,
       volume = {649},
          eid = {A3},
        pages = {A3},
          doi = {10.1051/0004-6361/202039587},
archivePrefix = {arXiv},
       eprint = {2012.01916},
 primaryClass = {astro-ph.IM},
       adsurl = {https://ui.adsabs.harvard.edu/abs/2021A&A...649A...3R}
}

@ARTICLE{Saglia2016,
       author = {{Saglia}, R.~P. and {Opitsch}, M. and {Erwin}, P. and {Thomas}, J. and {Beifiori}, A. and {Fabricius}, M. and {Mazzalay}, X. and {Nowak}, N. and {Rusli}, S.~P. and {Bender}, R.},
        title = "{The SINFONI Black Hole Survey: The Black Hole Fundamental Plane Revisited and the Paths of (Co)evolution of Supermassive Black Holes and Bulges}",
      journal = {\apj},
         year = 2016,
        month = feb,
       volume = {818},
       number = {1},
          eid = {47},
        pages = {47},
          doi = {10.3847/0004-637X/818/1/47},
archivePrefix = {arXiv},
       eprint = {1601.00974},
 primaryClass = {astro-ph.GA},
       adsurl = {https://ui.adsabs.harvard.edu/abs/2016ApJ...818...47S}
}

@ARTICLE{Sato2025,
       author = {{Sato}, Kyosuke S. and {Komiyama}, Yutaka and {Okamoto}, Sakurako and {Yagi}, Masafumi and {Ogami}, Itsuki and {Tanaka}, Mikito and {Arimoto}, Nobuo and {Chiba}, Masashi and {Kirby}, Evan N. and {Wyse}, Rosemary F.~G. and {Mori}, Rintaro},
        title = "{The star formation and chemical evolution histories of the Ursa Minor dwarf spheroidal galaxy}",
      journal = {\pasj},
         year = 2025,
        month = dec,
       volume = {77},
       number = {6},
        pages = {1259-1277},
          doi = {10.1093/pasj/psaf106},
archivePrefix = {arXiv},
       eprint = {2505.13161},
 primaryClass = {astro-ph.GA},
       adsurl = {https://ui.adsabs.harvard.edu/abs/2025PASJ...77.1259S}
}

@ARTICLE{Schlegel1998,
       author = {{Schlegel}, David J. and {Finkbeiner}, Douglas P. and {Davis}, Marc},
        title = "{Maps of Dust Infrared Emission for Use in Estimation of Reddening and Cosmic Microwave Background Radiation Foregrounds}",
      journal = {\apj},
         year = 1998,
        month = jun,
       volume = {500},
       number = {2},
        pages = {525-553},
          doi = {10.1086/305772},
archivePrefix = {arXiv},
       eprint = {astro-ph/9710327},
 primaryClass = {astro-ph},
       adsurl = {https://ui.adsabs.harvard.edu/abs/1998ApJ...500..525S}
}

@ARTICLE{Schwarzschild1979,
   author = {{Schwarzschild}, M.},
    title = "{A numerical model for a triaxial stellar system in dynamical equilibrium}",
  journal = {\apj},
     year = 1979,
    month = aug,
   volume = 232,
    pages = {236-247},
      doi = {10.1086/157282},
   adsurl = {http://adsabs.harvard.edu/abs/1979ApJ...232..236S}
}

@ARTICLE{Sestito2019,
       author = {{Sestito}, Federico and {Longeard}, Nicolas and {Martin}, Nicolas F. and {Starkenburg}, Else and {Fouesneau}, Morgan and {Gonz{\'a}lez Hern{\'a}ndez}, Jonay I. and {Arentsen}, Anke and {Ibata}, Rodrigo and {Aguado}, David S. and {Carlberg}, Raymond G. and {Jablonka}, Pascale and {Navarro}, Julio F. and {Tolstoy}, Eline and {Venn}, Kim A.},
        title = "{Tracing the formation of the Milky Way through ultra metal-poor stars}",
      journal = {\mnras},
         year = 2019,
        month = apr,
       volume = {484},
       number = {2},
        pages = {2166-2180},
          doi = {10.1093/mnras/stz043},
archivePrefix = {arXiv},
       eprint = {1811.03099},
 primaryClass = {astro-ph.GA},
       adsurl = {https://ui.adsabs.harvard.edu/abs/2019MNRAS.484.2166S}
}

@ARTICLE{Sestito2023,
       author = {{Sestito}, Federico and {Zaremba}, Daria and {Venn}, Kim A. and {D'Aoust}, Lina and {Hayes}, Christian and {Jensen}, Jaclyn and {Navarro}, Julio F. and {Jablonka}, Pascale and {Fern{\'a}ndez-Alvar}, Emma and {Glover}, Jennifer and {McConnachie}, Alan W. and {Chen{\'e}}, Andr{\'e}-Nicolas},
        title = "{The extended 'stellar halo' of the Ursa Minor dwarf galaxy}",
      journal = {\mnras},
         year = 2023,
        month = oct,
       volume = {525},
       number = {2},
        pages = {2875-2890},
          doi = {10.1093/mnras/stad2427},
archivePrefix = {arXiv},
       eprint = {2301.13214},
 primaryClass = {astro-ph.GA},
       adsurl = {https://ui.adsabs.harvard.edu/abs/2023MNRAS.525.2875S}
}

@ARTICLE{Sales2022,
       author = {{Sales}, Laura V. and {Wetzel}, Andrew and {Fattahi}, Azadeh},
        title = "{Baryonic solutions and challenges for cosmological models of dwarf galaxies}",
      journal = {Nature Astronomy},
         year = 2022,
        month = jun,
       volume = {6},
        pages = {897-910},
          doi = {10.1038/s41550-022-01689-w},
archivePrefix = {arXiv},
       eprint = {2206.05295},
 primaryClass = {astro-ph.GA},
       adsurl = {https://ui.adsabs.harvard.edu/abs/2022NatAs...6..897S}
}

@ARTICLE{Smith2018,
       author = {{Smith}, Britton D. and {Regan}, John A. and {Downes}, Turlough P. and {Norman}, Michael L. and {O'Shea}, Brian W. and {Wise}, John H.},
        title = "{The growth of black holes from Population III remnants in the Renaissance simulations}",
      journal = {\mnras},
         year = 2018,
        month = nov,
       volume = {480},
       number = {3},
        pages = {3762-3773},
          doi = {10.1093/mnras/sty2103},
archivePrefix = {arXiv},
       eprint = {1804.06477},
 primaryClass = {astro-ph.GA},
       adsurl = {https://ui.adsabs.harvard.edu/abs/2018MNRAS.480.3762S}
}

@ARTICLE{Smith2023,
       author = {{Smith}, Simon E.~T. and {Jensen}, Jaclyn and {Roediger}, Joel and {Sestito}, Federico and {Hayes}, Christian R. and {McConnachie}, Alan W. and {Cuillandre}, Jean-Charles and {Gwyn}, Stephen and {Magnier}, Eugene and {Chambers}, Ken and {Hammer}, Francois and {Hudson}, Mike J. and {Martin}, Nicolas and {Navarro}, Julio and {Scott}, Douglas},
        title = "{Discovery of a New Local Group Dwarf Galaxy Candidate in UNIONS: Bo{\"o}tes V}",
      journal = {\aj},
         year = 2023,
        month = aug,
       volume = {166},
       number = {2},
          eid = {76},
        pages = {76},
          doi = {10.3847/1538-3881/acdd77},
archivePrefix = {arXiv},
       eprint = {2209.08242},
 primaryClass = {astro-ph.GA},
       adsurl = {https://ui.adsabs.harvard.edu/abs/2023AJ....166...76S}
}

@ARTICLE{Spencer2018,
       author = {{Spencer}, Meghin E. and {Mateo}, Mario and {Olszewski}, Edward W. and {Walker}, Matthew G. and {McConnachie}, Alan W. and {Kirby}, Evan N.},
        title = "{The Binary Fraction of Stars in Dwarf Galaxies: The Cases of Draco and Ursa Minor}",
      journal = {\aj},
         year = 2018,
        month = dec,
       volume = {156},
       number = {6},
          eid = {257},
        pages = {257},
          doi = {10.3847/1538-3881/aae3e4},
archivePrefix = {arXiv},
       eprint = {1811.06597},
 primaryClass = {astro-ph.GA},
       adsurl = {https://ui.adsabs.harvard.edu/abs/2018AJ....156..257S}
}

@ARTICLE{terBraak2008,
    title = "Differential Evolution Markov Chain with snooker updater and fewer chains",
    author = "{ter Braak}, C.J.F. and J.A. Vrugt",
    year = "2008",
    doi = "10.1007/s11222-008-9104-9",
    volume = "18",
    pages = "435-446",
    journal = "Statistics and Computing",
    publisher = "Springer Verlag",
    number = "4",
}

@ARTICLE{Tollet2016,
   author = {{Tollet}, E. and {Macci{\`o}}, A.~V. and {Dutton}, A.~A. and 
	{Stinson}, G.~S. and {Wang}, L. and {Penzo}, C. and {Gutcke}, T.~A. and 
	{Buck}, T. and {Kang}, X. and {Brook}, C. and {Di Cintio}, A. and 
	{Keller}, B.~W. and {Wadsley}, J.},
    title = "{NIHAO - IV: core creation and destruction in dark matter density profiles across cosmic time}",
  journal = {\mnras},
archivePrefix = "arXiv",
   eprint = {1507.03590},
     year = 2016,
    month = mar,
   volume = 456,
    pages = {3542-3552},
      doi = {10.1093/mnras/stv2856},
   adsurl = {http://adsabs.harvard.edu/abs/2016MNRAS.456.3542T}
}

@ARTICLE{Tolstoy2004,
   author = {{Tolstoy}, E. and {Irwin}, M.~J. and {Helmi}, A. and {Battaglia}, G. and 
	{Jablonka}, P. and {Hill}, V. and {Venn}, K.~A. and {Shetrone}, M.~D. and 
	{Letarte}, B. and {Cole}, A.~A. and {Primas}, F. and {Francois}, P. and 
	{Arimoto}, N. and {Sadakane}, K. and {Kaufer}, A. and {Szeifert}, T. and 
	{Abel}, T.},
    title = "{Two Distinct Ancient Components in the Sculptor Dwarf Spheroidal Galaxy: First Results from the Dwarf Abundances and Radial Velocities Team}",
  journal = {\apjl},
   eprint = {astro-ph/0411029},
     year = 2004,
    month = dec,
   volume = 617,
    pages = {L119-L122},
      doi = {10.1086/427388},
   adsurl = {https://ui.adsabs.harvard.edu/abs/2004ApJ...617L.119T}
}

@ARTICLE{Tolstoy2009,
       author = {{Tolstoy}, Eline and {Hill}, Vanessa and {Tosi}, Monica},
        title = "{Star-Formation Histories, Abundances, and Kinematics of Dwarf Galaxies in the Local Group}",
      journal = {\araa},
         year = 2009,
        month = sep,
       volume = {47},
       number = {1},
        pages = {371-425},
          doi = {10.1146/annurev-astro-082708-101650},
archivePrefix = {arXiv},
       eprint = {0904.4505},
 primaryClass = {astro-ph.CO},
       adsurl = {https://ui.adsabs.harvard.edu/abs/2009ARA&A..47..371T}
}

@ARTICLE{Tolstoy2023,
       author = {{Tolstoy}, Eline and {Sk{\'u}lad{\'o}ttir}, {\'A}sa and {Battaglia}, Giuseppina and {Brown}, Anthony G.~A. and {Massari}, Davide and {Irwin}, Michael J. and {Starkenburg}, Else and {Salvadori}, Stefania and {Hill}, Vanessa and {Jablonka}, Pascale and {Salaris}, Maurizio and {van Essen}, Thom and {Olsthoorn}, Carla and {Helmi}, Amina and {Pritchard}, John},
        title = "{A 3D view of dwarf galaxies with Gaia and VLT/FLAMES. I. The Sculptor dwarf spheroidal}",
      journal = {\aap},
         year = 2023,
        month = jul,
       volume = {675},
          eid = {A49},
        pages = {A49},
          doi = {10.1051/0004-6361/202245717},
archivePrefix = {arXiv},
       eprint = {2304.11980},
 primaryClass = {astro-ph.GA},
       adsurl = {https://ui.adsabs.harvard.edu/abs/2023A&A...675A..49T}
}

@ARTICLE{Tolstoy2025,
       author = {{Tolstoy}, Eline and {Battaglia}, Giuseppina and {Arroyo-Polonio}, Jos{\'e} Mar{\'\i}a and {Brown}, Anthony G.~A. and {van Essen}, Thom and {Massari}, Davide and {Sk{\'u}lad{\'o}ttir}, {\'A}sa and {Irwin}, Michael J. and {Taibi}, Salvatore and {Pritchard}, John},
        title = "{A 3D view of dwarf galaxies with Gaia and VLT/FLAMES: II. The Sextans dwarf spheroidal}",
      journal = {\aap},
         year = 2025,
        month = jun,
       volume = {698},
          eid = {A53},
        pages = {A53},
          doi = {10.1051/0004-6361/202554176},
archivePrefix = {arXiv},
       eprint = {2504.02787},
 primaryClass = {astro-ph.GA},
       adsurl = {https://ui.adsabs.harvard.edu/abs/2025A&A...698A..53T}
}

@ARTICLE{Tremaine2002,
       author = {{Tremaine}, Scott and {Gebhardt}, Karl and {Bender}, Ralf and {Bower}, Gary and {Dressler}, Alan and {Faber}, S.~M. and {Filippenko}, Alexei V. and {Green}, Richard and {Grillmair}, Carl and {Ho}, Luis C. and {Kormendy}, John and {Lauer}, Tod R. and {Magorrian}, John and {Pinkney}, Jason and {Richstone}, Douglas},
        title = "{The Slope of the Black Hole Mass versus Velocity Dispersion Correlation}",
      journal = {\apj},
         year = 2002,
        month = aug,
       volume = {574},
       number = {2},
        pages = {740-753},
          doi = {10.1086/341002},
archivePrefix = {arXiv},
       eprint = {astro-ph/0203468},
 primaryClass = {astro-ph},
       adsurl = {https://ui.adsabs.harvard.edu/abs/2002ApJ...574..740T}
}

@ARTICLE{vandenBosch2016,
       author = {{van den Bosch}, Remco C.~E.},
        title = "{Unification of the fundamental plane and Super Massive Black Hole Masses}",
      journal = {\apj},
         year = 2016,
        month = nov,
       volume = {831},
       number = {2},
          eid = {134},
        pages = {134},
          doi = {10.3847/0004-637X/831/2/134},
archivePrefix = {arXiv},
       eprint = {1606.01246},
 primaryClass = {astro-ph.GA},
       adsurl = {https://ui.adsabs.harvard.edu/abs/2016ApJ...831..134V}
}

@ARTICLE{vanWassenhove2010,
       author = {{van Wassenhove}, S. and {Volonteri}, M. and {Walker}, M.~G. and {Gair}, J.~R.},
        title = "{Massive black holes lurking in Milky Way satellites}",
      journal = {\mnras},
         year = 2010,
        month = oct,
       volume = {408},
       number = {2},
        pages = {1139-1146},
          doi = {10.1111/j.1365-2966.2010.17189.x},
archivePrefix = {arXiv},
       eprint = {1001.5451},
 primaryClass = {astro-ph.CO},
       adsurl = {https://ui.adsabs.harvard.edu/abs/2010MNRAS.408.1139V}
}

@ARTICLE{Vasiliev2019,
       author = {{Vasiliev}, Eugene},
        title = "{AGAMA: action-based galaxy modelling architecture}",
      journal = {\mnras},
         year = "2019",
        month = "Jan",
       volume = {482},
        pages = {1525-1544},
          doi = {10.1093/mnras/sty2672},
archivePrefix = {arXiv},
       eprint = {1802.08239},
 primaryClass = {astro-ph.GA},
       adsurl = {https://ui.adsabs.harvard.edu/\#abs/2019MNRAS.482.1525V}
}

@ARTICLE{Vasiliev2026,
       author = {{Vasiliev}, Eugene and {Feldmeier-Krause}, Anja and {Sormani}, Mattia C.},
        title = "{Distribution function-based modelling of discrete kinematic datasets, in application to the Milky Way nuclear star cluster}",
      journal = {\apj},
         year = 2026,
       volume = {1002},
        pages = {71},
          doi = {10.3847/1538-4357/ae5a30},
archivePrefix = {arXiv},
       eprint = {2603.29502},
 primaryClass = {astro-ph.GA},
       adsurl = {https://ui.adsabs.harvard.edu/abs/2026arXiv260329502V}
}

@ARTICLE{Volonteri2003,
       author = {{Volonteri}, Marta and {Haardt}, Francesco and {Madau}, Piero},
        title = "{The Assembly and Merging History of Supermassive Black Holes in Hierarchical Models of Galaxy Formation}",
      journal = {\apj},
         year = 2003,
        month = jan,
       volume = {582},
       number = {2},
        pages = {559-573},
          doi = {10.1086/344675},
archivePrefix = {arXiv},
       eprint = {astro-ph/0207276},
 primaryClass = {astro-ph},
       adsurl = {https://ui.adsabs.harvard.edu/abs/2003ApJ...582..559V}
}

@ARTICLE{Volonteri2008,
       author = {{Volonteri}, Marta and {Lodato}, Giuseppe and {Natarajan}, Priyamvada},
        title = "{The evolution of massive black hole seeds}",
      journal = {\mnras},
         year = 2008,
        month = jan,
       volume = {383},
       number = {3},
        pages = {1079-1088},
          doi = {10.1111/j.1365-2966.2007.12589.x},
archivePrefix = {arXiv},
       eprint = {0709.0529},
 primaryClass = {astro-ph},
       adsurl = {https://ui.adsabs.harvard.edu/abs/2008MNRAS.383.1079V}
}

@ARTICLE{Volonteri2009,
       author = {{Volonteri}, Marta and {Natarajan}, Priyamvada},
        title = "{Journey to the M$_{BH}$-{\ensuremath{\sigma}} relation: the fate of low-mass black holes in the Universe}",
      journal = {\mnras},
         year = 2009,
        month = dec,
       volume = {400},
       number = {4},
        pages = {1911-1918},
          doi = {10.1111/j.1365-2966.2009.15577.x},
archivePrefix = {arXiv},
       eprint = {0903.2262},
 primaryClass = {astro-ph.CO},
       adsurl = {https://ui.adsabs.harvard.edu/abs/2009MNRAS.400.1911V}
}

@ARTICLE{Volonteri2016,
       author = {{Volonteri}, M. and {Dubois}, Y. and {Pichon}, C. and {Devriendt}, J.},
        title = "{The cosmic evolution of massive black holes in the Horizon-AGN simulation}",
      journal = {\mnras},
         year = 2016,
        month = aug,
       volume = {460},
       number = {3},
        pages = {2979-2996},
          doi = {10.1093/mnras/stw1123},
archivePrefix = {arXiv},
       eprint = {1602.01941},
 primaryClass = {astro-ph.GA},
       adsurl = {https://ui.adsabs.harvard.edu/abs/2016MNRAS.460.2979V}
}

@ARTICLE{Vitral2024,
       author = {{Vitral}, Eduardo and {van der Marel}, Roeland P. and {Sohn}, Sangmo Tony and {Libralato}, Mattia and {del Pino}, Andr{\'e}s and {Watkins}, Laura L. and {Bellini}, Andrea and {Walker}, Matthew G. and {Besla}, Gurtina and {Pawlowski}, Marcel S. and {Mamon}, Gary A.},
        title = "{HSTPROMO Internal Proper-motion Kinematics of Dwarf Spheroidal Galaxies. I. Velocity Anisotropy and Dark Matter Cusp Slope of Draco}",
      journal = {\apj},
         year = 2024,
        month = jul,
       volume = {970},
       number = {1},
          eid = {1},
        pages = {1},
          doi = {10.3847/1538-4357/ad571c},
archivePrefix = {arXiv},
       eprint = {2407.07769},
 primaryClass = {astro-ph.GA},
       adsurl = {https://ui.adsabs.harvard.edu/abs/2024ApJ...970....1V}
}

@ARTICLE{Vitral2026,
       author = {{Vitral}, Eduardo and {van der Marel}, Roeland P. and {Sohn}, Sangmo Tony and {Pe{\~n}arrubia}, Jorge and {Patel}, Ekta and {Watkins}, Laura L. and {Libralato}, Mattia and {McKinnon}, Kevin A. and {Bellini}, Andrea and {del Pino}, Andr{\'e}s and {Bennet}, Paul},
        title = "{HSTPROMO Internal Proper-motion Kinematics of Dwarf Spheroidal Galaxies. II. Velocity Anisotropy and Dark Matter Cusp Slope of Sculptor}",
      journal = {\apj},
         year = 2026,
        month = feb,
       volume = {998},
       number = {2},
          eid = {206},
        pages = {206},
          doi = {10.3847/1538-4357/ae1f8a},
archivePrefix = {arXiv},
       eprint = {2508.20711},
 primaryClass = {astro-ph.GA},
       adsurl = {https://ui.adsabs.harvard.edu/abs/2026ApJ...998..206V}
}

@ARTICLE{Walker2008,
       author = {{Walker}, Matthew G. and {Mateo}, Mario and {Olszewski}, Edward W.},
        title = "{Systemic Proper Motions of Milky Way Satellites from Stellar Redshifts: The Carina, Fornax, Sculptor, and Sextans Dwarf Spheroidals}",
      journal = {\apjl},
         year = 2008,
        month = dec,
       volume = {688},
       number = {2},
        pages = {L75},
          doi = {10.1086/595586},
archivePrefix = {arXiv},
       eprint = {0810.1511},
 primaryClass = {astro-ph},
       adsurl = {https://ui.adsabs.harvard.edu/abs/2008ApJ...688L..75W}
}

@ARTICLE{Walker2009a,
   author = {{Walker}, M.~G. and {Mateo}, M. and {Olszewski}, E.~W.},
    title = "{Stellar Velocities in the Carina, Fornax, Sculptor, and Sextans dSph Galaxies: Data From the Magellan/MMFS Survey}",
  journal = {\aj},
archivePrefix = "arXiv",
   eprint = {0811.0118},
     year = 2009,
    month = feb,
   volume = 137,
    pages = {3100-3108},
      doi = {10.1088/0004-6256/137/2/3100},
   adsurl = {http://adsabs.harvard.edu/abs/2009AJ....137.3100W}
}

@ARTICLE{Walker2009b,
   author = {{Walker}, M.~G. and {Mateo}, M. and {Olszewski}, E.~W. and {Pe{\~n}arrubia}, J. and 
	{Wyn Evans}, N. and {Gilmore}, G.},
    title = "{A Universal Mass Profile for Dwarf Spheroidal Galaxies?}",
  journal = {\apj},
archivePrefix = "arXiv",
   eprint = {0906.0341},
 primaryClass = "astro-ph.CO",
     year = 2009,
    month = oct,
   volume = 704,
    pages = {1274-1287},
      doi = {10.1088/0004-637X/704/2/1274},
   adsurl = {http://adsabs.harvard.edu/abs/2009ApJ...704.1274W}
}

@ARTICLE{Walker2015,
       author = {{Walker}, Matthew G. and {Olszewski}, Edward W. and {Mateo}, Mario},
        title = "{Bayesian analysis of resolved stellar spectra: application to MMT/Hectochelle observations of the Draco dwarf spheroidal}",
      journal = {\mnras},
         year = 2015,
        month = apr,
       volume = {448},
       number = {3},
        pages = {2717-2732},
          doi = {10.1093/mnras/stv099},
archivePrefix = {arXiv},
       eprint = {1503.02589},
 primaryClass = {astro-ph.GA},
       adsurl = {https://ui.adsabs.harvard.edu/abs/2015MNRAS.448.2717W}
}

@ARTICLE{Walker2023,
       author = {{Walker}, Matthew G. and {Caldwell}, Nelson and {Mateo}, Mario and {Olszewski}, Edward W. and {Pace}, Andrew B. and {Bailey}, John I. and {Koposov}, Sergey E. and {Roederer}, Ian U.},
        title = "{Magellan/M2FS and MMT/Hectochelle Spectroscopy of Dwarf Galaxies and Faint Star Clusters within the Galactic Halo}",
      journal = {\apjs},
         year = 2023,
        month = sep,
       volume = {268},
       number = {1},
          eid = {19},
        pages = {19},
          doi = {10.3847/1538-4365/acdd79},
archivePrefix = {arXiv},
       eprint = {2312.12738},
 primaryClass = {astro-ph.GA},
       adsurl = {https://ui.adsabs.harvard.edu/abs/2023ApJS..268...19W}
}

@ARTICLE{WalkerPen2011,
   author = {{Walker}, M.~G. and {Pe{\~n}arrubia}, J.},
    title = "{A Method for Measuring (Slopes of) the Mass Profiles of Dwarf Spheroidal Galaxies}",
  journal = {\apj},
archivePrefix = "arXiv",
   eprint = {1108.2404},
     year = 2011,
    month = nov,
   volume = 742,
      eid = {20},
    pages = {20},
      doi = {10.1088/0004-637X/742/1/20},
   adsurl = {http://adsabs.harvard.edu/abs/2011ApJ...742...20W}
}

@ARTICLE{Watkins2013,
       author = {{Watkins}, Laura L. and {van de Ven}, Glenn and {den Brok}, Mark and {van den Bosch}, Remco C.~E.},
        title = "{Discrete dynamical models of {\ensuremath{\omega}} Centauri}",
      journal = {\mnras},
         year = 2013,
        month = dec,
       volume = {436},
       number = {3},
        pages = {2598-2615},
          doi = {10.1093/mnras/stt1756},
archivePrefix = {arXiv},
       eprint = {1308.4789},
 primaryClass = {astro-ph.GA},
       adsurl = {https://ui.adsabs.harvard.edu/abs/2013MNRAS.436.2598W}
}

@ARTICLE{Wardana2025,
       author = {{Wardana}, Dafa and {Chiba}, Masashi and {Hayashi}, Kohei},
        title = "{Disentangling {\ensuremath{\gamma}} {\ensuremath{-}} {\ensuremath{\beta}}: The Fourth-order Velocity Moments Based on Spherical Jeans Analysis}",
      journal = {\apj},
         year = 2025,
        month = apr,
       volume = {982},
       number = {2},
          eid = {167},
        pages = {167},
          doi = {10.3847/1538-4357/adb8e4},
archivePrefix = {arXiv},
       eprint = {2404.12671},
 primaryClass = {astro-ph.GA},
       adsurl = {https://ui.adsabs.harvard.edu/abs/2025ApJ...982..167W}
}

@ARTICLE{Weisz2014,
       author = {{Weisz}, Daniel R. and {Dolphin}, Andrew E. and {Skillman}, Evan D. and
         {Holtzman}, Jon and {Gilbert}, Karoline M. and
         {Dalcanton}, Julianne J. and {Williams}, Benjamin F.},
        title = "{The Star Formation Histories of Local Group Dwarf Galaxies. I. Hubble Space Telescope/Wide Field Planetary Camera 2 Observations}",
      journal = {\apj},
         year = "2014",
        month = "Jul",
       volume = {789},
          eid = {147},
        pages = {147},
          doi = {10.1088/0004-637X/789/2/147},
archivePrefix = {arXiv},
       eprint = {1404.7144},
 primaryClass = {astro-ph.GA},
       adsurl = {https://ui.adsabs.harvard.edu/abs/2014ApJ...789..147W}
}

@ARTICLE{Yang2025,
       author = {{Yang}, Hao and {Wang}, Wenting and {Zhu}, Ling and {Li}, Ting S. and {Koposov}, Sergey E. and {Han}, Jiaxin and {Li}, Songting and {Shi}, Rui and {Valluri}, Monica and {Riley}, Alexander H. and {Dey}, Arjun and {Rockosi}, Constance and {Palau}, Carles G. and {Aguilar}, Jessica Nicole and {Ahlen}, Steven and {Brooks}, David and {Claybaugh}, Todd and {Cooper}, Andrew and {de la Macorra}, Axel and {Doel}, Peter and {Ferraro}, Simone and {Forero-Romero}, Jaime E. and {Gazta{\~n}aga}, Enrique and {Gontcho A Gontcho}, Satya and {Gonzalez Morales}, Alma Xochitl and {Gutierrez}, Gaston and {Guy}, Julien and {Honscheid}, Klaus and {Ishak}, Mustapha and {Joyce}, Dick and {Kehoe}, Robert and {Kisner}, Theodore and {Kizhuprakkat}, Namitha and {Kremin}, Anthony and {Lahav}, Ofer and {Landriau}, Martin and {Le Guillou}, Laurent and {Medina}, Gustavo E. and {Meisner}, Aaron and {Miquel}, Ramon and {Palanque-Delabrouille}, Nathalie and {Prada}, Francisco and {P{\'e}rez-R{\`a}fols}, Ignasi and {Rossi}, Graziano and {Sanchez}, Eusebio and {Schlegel}, David and {Schubnell}, Michael and {Silber}, Joseph Harry and {Sprayberry}, David and {Tarl{\'e}}, Gregory and {Weaver}, Benjamin Alan and {Zhou}, Rongpu and {Zou}, Hu},
        title = "{The Dark Matter Content of Milky Way Dwarf Spheroidal Galaxies: Draco, Sextans, and Ursa Minor}",
      journal = {\apj},
         year = 2025,
        month = nov,
       volume = {993},
       number = {2},
          eid = {249},
        pages = {249},
          doi = {10.3847/1538-4357/ae07ce},
archivePrefix = {arXiv},
       eprint = {2507.02284},
 primaryClass = {astro-ph.GA},
       adsurl = {https://ui.adsabs.harvard.edu/abs/2025ApJ...993..249Y}
}

@ARTICLE{Young1980,
       author = {{Young}, P.},
        title = "{Numerical models of star clusters with a central black hole. I - Adiabatic models.}",
      journal = {\apj},
         year = 1980,
        month = dec,
       volume = {242},
        pages = {1232-1237},
          doi = {10.1086/158553},
       adsurl = {https://ui.adsabs.harvard.edu/abs/1980ApJ...242.1232Y}
}

@ARTICLE{Zhu2016,
   author = {{Zhu}, L. and {van de Ven}, G. and {Watkins}, L.~L. and {Posti}, L.
	},
    title = "{A discrete chemo-dynamical model of the dwarf spheroidal galaxy Sculptor: mass profile, velocity anisotropy and internal rotation}",
  journal = {\mnras},
archivePrefix = "arXiv",
   eprint = {1608.08239},
     year = 2016,
    month = nov,
   volume = 463,
    pages = {1117-1135},
      doi = {10.1093/mnras/stw2081},
   adsurl = {http://adsabs.harvard.edu/abs/2016MNRAS.463.1117Z}
}

\appendix

\section{Bayesian inference}
\label{app:A}

In this paper we make use of Bayesian inference in several independent applications, including the construction of photometric and kinematic membership samples and the dynamical modelling based on DFs.  Although the specific form of the likelihood, $\LL$, and the choice of the priors, $\Pi$, vary between these applications, they all share a common methodology. In this Appendix we therefore outline the general framework adopted throughout the paper, while the different definitions of likelihood and priors is provided in the relevant Sections. 

In general, the posterior probability distribution, $p$, of the model parameters $\btheta$, given the data $\DD$, follows from the Bayes' theorem
\begin{equation}
    p\,(\btheta \mid \DD) \propto \LL(\DD \mid \btheta)\, \Pi(\btheta),
\end{equation}
i.e. the posterior distribution is directly proportional to the likelihood function. In each application, to explore the posterior distributions we rely on Markov Chain Monte Carlo (MCMC) techniques, implemented with the affine-invariant ensemble sampler \texttt{emcee} \citep{ForemanMackey2013}. For the analyses in Sections~\ref{appBsub:memb} and \ref{app:Bkin}, we use a number of walkers equal to 6 times the number of free parameters, while for the dynamical models we adopt 12 times the number of free parameters. For the dynamical models we prefer more walkers, even for the same number of free parameters, as this typically improves convergence. The sampling strategy combines the Stretch Move \citep{Goodman2010}, the differential evolution move \citep{Nelson2014} and the snooker update \citep{terBraak2008}, which provide robust performance across a wide range of posterior shapes.

Each MCMC run is post-processed to ensure that the retained samples are independent and representative of the converged posterior. In particular: i) we remove an initial burn-in phase long enough to ensure model convergence; ii) we perform an additional convergence assessment by inspecting the behaviour of the remaining walkers, verifying that they all converge toward the same region of parameter space and reach consistent values of the log-posterior, and excluding cases in which a subset of walkers remains trapped in local maxima; iii) we apply a final thinning of the chains, typically corresponding to 0.25-0.5 times the median autocorrelation time, computed using the remaining walkers. We prefer more aggressive thinning for higher-dimensional problems. For dynamical models runs, which typically exhibit slow convergence, the MCMC exploration is performed in two stages: an initial run is used to approach convergence, after which we reinitialise the walkers within the converged region and perform a new run to efficiently sample the posterior distribution.

From the cleaned posterior distributions, we adopt the median of each distribution as the reference value for the model parameters and derived quantities. Credible intervals are computed using the 16th and 84th percentiles for the 1$\sigma$ interval, and the 5th and 95th percentiles for the 2$\sigma$ interval. For completeness, we also consider highest posterior density (HPD) intervals, defined as the smallest intervals enclosing a given fraction of the posterior probability. We report 1$\sigma$ credible regions only for parameters whose 95\% HPD interval is fully contained within the prior range, i.e. such that neither of its bounds coincides with a prior boundary. If one bound of the 95\% HPD interval reaches the prior limit, we instead quote the parameter value as a lower or upper limit given by the HPD. This approach provides more meaningful constraints for parameters whose posteriors are limited by the prior, for which percentile-based intervals would be misleading. We adopt this convention in Tables~\ref{tab:modparams} and ~\ref{tab:dm} and throughout the text. In the supplementary material tables, instead, we uniformly report credible intervals based solely on percentiles, without applying the HPD-based criterion.

\section{Construction of the catalogues}
\label{sec:data}

In this Appendix, we describe the Gaia-based photometric and astrometric catalogue (hereafter, the Gaia catalogue) and the spectroscopic datasets from which the radial-velocity and metallicity samples adopted for the dynamical modelling of Draco and Ursa Minor are constructed. Sections~\ref{subsec:gaia} and~\ref{subsec:kin} present the parent catalogues and describe the selection flags and quality criteria adopted. The procedures used to identify high-probability member stars and to construct the final samples are postponed to Appendix~\ref{app:B}.

\subsection{Gaia photometric catalogue}
\label{subsec:gaia}

To construct a Gaia-based dataset of member stars for Draco and Ursa Minor, we first selected sources from the Gaia DR3 catalogue within circular regions centred on the coordinates reported by \citet{Munoz2018}: $\alpha = 260.0684^\circ$ and $\delta = 57.9185^\circ$ for Draco, $\alpha = 227.2420^\circ$ and $\delta = 67.2221^\circ$ for Ursa Minor. The initial selections extend out $1.25^\circ$ for Draco and $2.5^\circ$ for Ursa Minor. As discussed in detail in Appendix~\ref{app:B}, stars located within the inner $1.0^\circ$ ($2.0^\circ$) are used to construct the bona fide member dataset of Draco (Ursa Minor), while stars at larger projected distances are retained only to characterise the foreground contamination. These angular distances have been chosen based on literature estimates of the spatial extent of the systems, and correspond to approximately $6.5$ times the projected half-mass radius of both galaxies \citep{Munoz2018}, thereby ensuring that the vast majority of stellar body of each galaxy is sampled \footnote{Recent works indicate that a small number of member stars in Ursa Minor may extend to very large radii, up to $\sim$11 times the half-light radius \citep{Sestito2023,Jensen2024}.}.

We retained only stars with a full five-parameter astrometric solution and high-quality astrometry, requiring \texttt{astrometric\_params\_solved} > 31 and a renormalised unit weight error \texttt{ruwe} < 1.4 \citep{Lindegren2018, Battaglia2022}. Also, to ensure reliable photometric measurements, we kept stars with non-null $\bprp$ colour, and applied a quality filter on the normalised flux-excess factor, requiring $|C^*/\sigma_C|<3$, according to the colour-dependent prescription of $C^*$ provided by \citet[][see their Table 2]{Riello2021}, and their $\Gmag$ band magnitude dependent estimate of $\sigma_C$ (see their equation 18). Large absolute values of $|C^*/\sigma_C|$ may suggest the presence of crowding effects or contamination by (extended) non-stellar sources. The apparent $\Gmag$ magnitudes for sources with six-parameter solutions have then been corrected on the fly using the query provided in Appendix A of \cite{Gaia2021}. We also retained the $\Gmag$ magnitude, the $\col$ colour, and their associated uncertainties, since these quantities are required to identify member stars according to the procedure described in Appendix~\ref{app:B}. 

The standard errors associated with the Gaia $\Gmag$-band magnitude and $\col$ and $\bprp$ colours were estimated by propagating the uncertainties in the measured fluxes, according to the following expressions:
\begin{equation}
    \begin{split}
   & \Delta G^2  = \left(\frac{2.5}{\ln 10}\frac{\Delta\FG}{\FG} \right)^2 + \Go^2, \,\,\,\,  \Go = 0.002755,\\
   & \Delta \GBP^2  = \left( \frac{2.5}{\ln 10} \frac{\Delta\FGBP}{\FGBP} \right)^2 + \GBPo^2, \,\,\,\,  \GBPo = 0.002790, \\
   & \Delta\GRP^2  = \left( \frac{2.5}{\ln 10} \frac{\Delta\FGRP}{\FGRP} \right)^2 + \GRPo^2,  \,\,\,\, \GRPo = 0.003779,\\
    \end{split}
\end{equation}
and     
\begin{equation}
    \begin{split}
    & \Delta(\col)^2  = \Delta G^2+\Delta \GRP^2, \\
    & \Delta(\bprp)^2 = \Delta \GBP^2+\Delta \GRP^2.
    \end{split}
\end{equation}
In the equations above, $\FG$, $\FGBP$, and $\FGRP$ represent the mean fluxes in the $\Gmag$, $\GBP$, and $\GRP$ bands, respectively, while $\Delta\FG$, $\Delta\FGBP$, and $\Delta\FGRP$ denote their corresponding uncertainties. The constants $\Go$, $\GBPo$, $\GRPo$ represent the zero-point uncertainties associated with each band\footnote{\url{https://www.cosmos.esa.int/web/gaia/edr3-passbands}}.

We adopted $\col$ rather than $\bprp$ because the $\GBP$ flux of faint sources is known to be biased \citep{Gaia}. Mitigating this bias through quality cuts on $\GBP$ would introduce undesirable selection effects. Using $\col$ therefore provides a more robust choice, particularly at low magnitudes. For Draco, we verified that adopting either colour definition does not produce any significant difference in our final results.

The apparent $\Gmag$ band magnitude and $\col$ colour of each star have been corrected for Galactic extinction. The $\ebmv$ values were obtained using the reddening maps of \citet{Schlegel1998}, accessed through the \texttt{dustmaps} Python package \citep{Green2018}. The maps have been interpolated at the star position. Afterwards, we use the \cite{Marigo2008} coefficients for the Gaia filters \citep[see][]{Evans2018} and apply the following formulae to compute the de-reddened $\Gmago$ band magnitude and $\colo$ colour \citep[see][]{Sestito2019}:
\begin{equation}\begin{split}
    &\Gmago = \Gmag - 2.664\,\ebmv, \\
    & \colo = (\col) - 0.643\,\ebmv. 
\end{split}\end{equation}
We do not consider errors on the estimates of the $\ebmv$, thus $\Delta\colo=\Delta(\col)$ and $\Delta\Gmago=\Delta\Gmag$. 

Finally, we limited the sample to stars with $\Gmago$-band magnitudes brighter than 20.5. To avoid contamination from known AGNs, we further cross-matched our samples with the Gaia catalogue of AGN candidates classified as such based on time-series photometry. 

The stars selected in these steps are used to construct a catalogue of high-confidence Draco and Ursa Minor members. The methodology underlying this selection is described in detail in the following Appendix~\ref{app:B}. Briefly, the method combines information on the position on the plane of the sky, proper motions (PMs), parallaxes, and CMD to identify stars with a high probability of membership. 

\subsection{Spectroscopic catalogues}
\label{subsec:kin}


Several spectroscopic catalogues, acquired with different facilities, have been published over the years, with the number of identified member stars increasing from a few dozens to a few hundreds \citep[see, e.g., Table~1 in][]{Spencer2018}. Combining datasets from different facilities is certainly possible, provided that the measurements can be accurately mapped onto a common scale and that their uncertainties are well characterised. However, when studying systems with intrinsically low velocity dispersions, minimising additional sources of noise becomes particularly advantageous.

For this reason, rather than merging heterogeneous spectroscopic samples, we adopt a different strategy. We analyse two independent kinematic catalogues separately, and compare the chemo-dynamical inferences obtained from each of them. This approach allows us to assess the stability of our results against the choice of spectroscopic dataset, while avoiding potential systematic effects introduced by merging spectroscopic catalogues from different instruments and reduction pipelines. 

The first dataset is drawn from the homogeneous multi-epoch survey presented by \cite{Walker2015} and \citetalias{Walker2023}. The second is taken from the recently published Keck/DEIMOS Stellar Archive of \citetalias{Geha2026}, which provides a large, uniformly reduced compilation of radial velocities and Ca II Triplet-based
metallicities for MW satellites. The details of the construction and selection procedures for the two catalogues are described in the following subsections.

\subsubsection{Walker et al. (2015, 2023)}
\label{subsec:wal}


The first chemo-kinematic dataset adopted here is constructed from the \cite{Walker2015} and \citetalias{Walker2023} catalogues, who provide multi-epoch, medium-resolution spectra for thousands of targets observed with the M2FS and Hectochelle spectrographs. These data include line-of-sight velocities, atmospheric parameters, chemical abundances, and signal-to-noise (S/N) estimates derived through template-fitting procedures. The catalogues contain repeated observations for several thousand objects across all target systems, with the Draco and Ursa Minor subsamples including both single- and multi-epoch measurements, making a rigorous treatment of quality selection and multi-epoch combination essential for producing clean and robust samples. The construction of catalogues used in this work begins with a set of base quality cuts applied to individual spectral measurements. We exclude all sources flagged by Gaia~DR3 as photometric variables (\texttt{f\_gaiavar = VARIABLE}), RR~Lyrae (\texttt{RRL = 1}), or AGN (\texttt{AGN = 1}), since their variability or non-stellar nature can compromise velocity determinations. Measurements marked as $\chi^2$ outliers (\texttt{f\_Chi2 = 1}) are removed because their spectra show large residuals with respect to the template fits. We also discard stars flagged as carbon-rich (\texttt{f\_C = 1}), whose spectral features are not well represented by the template grid and typically correspond to foreground contaminants. Finally, measurements indicating significant radial-velocity variability (\texttt{f\_Vlosvar = 1}) are excluded to avoid binary-induced motions\footnote{In any case, we note that undetected binaries do not introduce significant bias in dynamical models of dSphs \citep{Arroyo2026}.}. Together, these six criteria define a baseline set of stable, well-behaved spectra suitable for identifying Draco and Ursa Minor kinematic members.

After establishing these base samples, we apply additional filters targeting the quality of the velocity and metallicity determinations. For line-of-sight velocities, we require raw velocity uncertainties not exceeding $5\kms$ (\texttt{e\_Vlosraw $\leq$ 5}) and constrain the shape of the posterior distribution by imposing $|\texttt{skVlosraw}| \leq 1$ and $|\texttt{ktVlosraw}| \leq 1$. These conditions select measurements with approximately symmetric, well determined line-of-sight velocities probability denisty distributions. Analogous criteria are applied to metallicities: we require $|\texttt{skFeHraw}| \leq 1$, $|\texttt{ktFeHraw}| \leq 1$, and a formal uncertainty $\texttt{e\_FeHraw} < 1$~dex. 


Stars with multiple spectroscopic observations are combined as follows. For a star $j$ with $K$ measurements of a scalar quantity, $s_{ij}$, and associated uncertainty, $\Delta s_{ij}$ and $i=1,...,K$, we compute the inverse-variance--weighted mean
\begin{equation}
    s_j = \frac{\sum_{i=1}^{K} s_{ij} \, (\Delta s_{ij})^{-2}}{\sum_{i=1}^{K} (\Delta s_{ij})^{-2}}, \qquad
    \Delta s_j = \left(\sum_{i=1}^{K} (\Delta s_{ij})^{-2}\right)^{-1/2}.
\end{equation}
This procedure is applied to both the line-of-sight velocities ($s=\vlos$) and metallicities ($s=\feh$). The resulting catalogues therefore contain one velocity and one metallicity measurement per star, with uncertainties propagated from the individual measurements.

\subsubsection{Geha et al. (2026)}
\label{subsec:geha}

The second chemo-kinematic dataset used is drawn from the Keck/DEIMOS Stellar Archive presented by \citetalias{Geha2026}. This catalogue provides homogeneously reduced spectra obtained with the DEIMOS spectrograph on the Keck II telescope, including radial velocities and CaT-based metallicities for more than 22000 stars across 78 MW dwarf galaxies and GCs. The DEIMOS reduction is performed using a uniform PypeIt-based pipeline, and radial velocities are derived through forward modelling of stellar and telluric templates.  Membership probabilities are provided for each star based on a combination of spatial, photometric, and kinematic information.

For Draco and Ursa Minor, we adopt the membership probabilities provided in the DEIMOS catalogue and, following the criteria recommended by the authors, retain stars with membership probability $>0.5$ for the dynamical analysis. For these stars, we use the published line-of-sight velocities and CaT-based metallicities. Since the DEIMOS catalogue provides one measurement per star (after its internal quality control), no additional multi-epoch combination is required.

\section{Identifying samples of members}
\label{app:B}

To identify high-probability member stars of the target systems using the photometric and kinematic catalogues described in Sections~\ref{subsec:gaia} and~\ref{subsec:wal}, we adopt a parametric classification framework based on mixture modelling. In this approach, the observed stellar population is interpreted as a realization of a composite statistical model comprising multiple components, each representing a distinct population. The parameters of these components are inferred by maximizing the likelihood of the observed data under the assumed model. The classification procedure is applied independently to the two catalogues, yielding separate member datasets, which are subsequently cross-matched to construct the final samples. The datasets are treated separately because they have different selection functions, with the spectroscopic sample based on discrete, non-uniform pointings of the target. A joint analysis would require modelling its selection function relative to the Gaia sample, which is beyond the scope of this work. 


In the following, we provide a brief overview of the methodology, focusing on the aspects that are common to its application to the two catalogues. In both applications, the model consists of two components: one representing the stellar population of the target dSph and the other accounting for foreground contamination. Given a set of stars within a specified region of the sky, we assume that the probability of observing any individual star, denoted $\PP$, is
\begin{equation}\label{for:tmodel}
\PP(\bxi\,|\,\btheta) = w\PPdsph(\bxi\,|\,\btheta) + (1 - w)\PPC(\bxi\,|\,\btheta).
\end{equation}
Here, $\PPdsph$ and $\PPC$ are the dSph and contaminants components, respectively, both normalised to unity, with $0<w<1$ and $(1-w)$ denoting the relative contribution of the two components. The vector $\btheta$ represents the set of free parameters defining the model, while $\bxi$ encodes the variables on which $\PP$ depends. Depending on the specific application and, thus, the input data, $\bxi$ may include sky positions, PMs, parallax, magnitude and colour in specific bands, or line-of-sight velocity and metallicity. The specific forms of $\PPdsph$ and $\PPC$ for the Gaia catalogue is described in Section~\ref{app:Bgaia}, for the spectroscopic kinematic catalogue is presented in Section~\ref{app:Bkin}. 

Given the model~(\ref{for:tmodel}), in all applications the likelihood function is defined as
\begin{equation}\label{for:loglgaia}
    \LL(\btheta,\DD) = \prod_{j=1}^{\Nsamp} (\PP \ast \EE)\bigl({\DD}_j \mid \btheta\bigr),
\end{equation}
where $\Nsamp$ is the total number of stars in the sample, $\DD$ denotes the full set of observed quantities, and ${\DD}_j$ represents the measurements associated with the $j$-th star. The symbol $\ast$ indicates convolution with the error function $\EE$.
We adopt uniform priors for all model parameters. The posterior is sampled using a MCMC approach. Further details on the sampling strategy, the computation of confidence intervals, and the implementation are provided in Appendix~\ref{app:A}.

The mixture–model framework allows to evaluate self-consistently membership probabilities for individual stars. For each application, membership probabilities of stars are computed by evaluating the relative contribution of each component to the total likelihood. In particular, the probability that the $j$-th star belongs to the target dSph population is defined as
\begin{equation}
    \Pmembj = \frac{w(\PPdsph \ast \EE)\bigl({\DD}_j\bigr)}{(\PP \ast \EE)\bigl({\DD}_j\bigr)}.
\end{equation}
Membership probabilities are evaluated over the posterior distribution of the model parameters, so the membership probability of each star is associated with a probability distribution. We adopt the median of this distribution as the representative membership value. In a  mixture model the expected number of members stars of a component is given by $\Nsamp w$. Rather than adopting a fixed threshold on membership probability, we define the final member sample by ranking stars according to their membership probabilities and retaining the $\Nsamp w|_{16-{\rm th}}$ highest-probability objects, where $w|_{16-{\rm th}}$ is the 16-th percentile of the $w$ distribution after the fit (i.e. the lower limit of its $1\sigma$ credible interval). 

After applying the method to the photometric and spectroscopic catalogues, we obtain two datasets of member stars. To build a unified sample, we retain all stars from the photometric catalogue and include only those spectroscopic stars that have a counterpart in the photometric sample, identified through a positional cross-match. The final dataset therefore consists of $N$ stars with measured positions and a subsample of $\Mvlos$ stars with kinematic information. Stars with line-of-sight velocities but lacking a Gaia counterpart are excluded, since the Gaia sample offers a complete and uniform spatial coverage of the systems. Table \ref{tab:samples} summarises the main properties of the samples, as the number of stars in each catalogue.


Similar classification frameworks, or closely related variants of this methodology, have been widely employed in the literature. These approaches have proven to be a robust and flexible tool for membership assignment and/or to characterise the properties of a population. Specific implementations and modelling choices often depend on the available datasets and their quality. Variations include, for example, the assumption of spherical symmetry, the use of multiple stellar components to describe the dwarf galaxy population, the inclusion or omission of colour–magnitude information, and the incorporation of additional observables or likelihood terms. Representative applications and extensions of these methods can be found in several studies \citep[see, e.g., ][]{Martin2008,WalkerPen2011,Martin2016,Pace2020,McConnachie2020,Battaglia2022,Smith2023,Arroyo2024}.

\subsection{Gaia catalogue}
\label{app:Bgaia}

In the application of the mixture model to the Gaia catalogue, the model equation~(\ref{for:tmodel}) is defined in terms of the variable vector
$\bxi \equiv (x, y, \mura, \mudec, \rhoad, \varpi, \Gmago, \colo)$.
The corresponding input data include, for the $j$-th star, the measured values of these quantities, together with their associated observational uncertainties where available, and are given by $ {\DD}_j \equiv \{ \xj, \yj, \muraj, \Delta\muraj, \mudecj, \Delta\mudecj, \rhoadj, \varpij, \Delta\varpij, \Gmagoj, \coloj, \Delta\Gmagoj, \Delta\coloj \},$ with $j$ that runs across stars in the sample.

We assume that each of the two probability components, $\PPdsph$ and $\PPC$, can be factorised into distinct terms, each describing the contribution of a specific set of observables: the spatial position on the plane of the sky, the PM, the parallax, and the location in the CMD. Under this assumption, we can rewrite the terms in equation~(\ref{for:tmodel})
\begin{equation}\label{for:factors}
\begin{split}
     \PPi(\bxi) & = \PPxyi(x,y) \\
    & \PPmuprli(\mura,\mudec,\rhoad,\varpi) \PPcmdi(\Gmago,\colo).
\end{split}
\end{equation}
In the above equation, when $i$=dSph we refer to the dSph (Draco or Ursa Minor), while when $i$=C to the contaminants. 

In the following subsections, we describe the functional forms adopted for each of these components.

\subsubsection{Spatial distributions}
\label{appBsub:spatial}

The spatial probability term representing stars in the dSph is
\begin{equation}\label{for:mixsp}
    \PPxydsph(x,y) = \frac{S(x,y)\Sigma(x,y)}{\int S(x,y)\Sigma(x,y)\dd x\dd y},
\end{equation}
where $\Sigma$ denotes the projected spatial distribution of the model component, and $S$ represents the spatial selection function, which accounts for the fact that stars may have been observed over a limited area of the sky. The denominator in equation~(\ref{for:mixsp}) ensures that the spatial probability density $\PPxydsph$ is properly normalised to unity. 
We adopt the flattened Exponential model
\begin{equation}\label{for:exp}
    \Sigma(x,y) = \frac{k^2e^{-k\bigl(\frac{m}{\Reff}\bigl)}}{2\pi\Reff^2(1-e)}, {\rm with}\, k=1.678,
\end{equation}
to describe the dSph spatial distribution. Given the non-negligible angular extent of a dSph, we applied the transformation given in equation (4) of \cite{MarelCioni2001} to ensure an accurate projection onto the plane of the sky. Thus, the coordinates $(x,y)$ are defined on the plane tangent to the celestial sphere at a the centres from \citet{Munoz2018}. In equation (\ref{for:exp}), $m$ is the elliptical radius
\begin{equation}\begin{split}\label{for:m}
    m^2 \equiv \, & \biggl[\frac{(x-\xo)\cos\phi - (y-\yo)\sin\phi)}{(1-e)}\biggr]^2 + \\ & [(x-\xo)\sin\phi + (y-\yo)\cos\phi]^2,
\end{split}\end{equation}
where $(\xo, \yo)$ represents a possible offset of the system’s centre with respect to the reference value, and the ellipticity is defined as $e \equiv 1 - b/a$, where $a$ and $b$ are the semi-major and semi-minor axes of the isodensity contours, respectively. The quantity $\Reff$ denotes the effective (or half-counts) radius, defined as the elliptical radius enclosing half of the total star counts. The angle $\phi$ is the PA of the major axis, measured from north through east.

Since the region on the tangent plane where stars have been selected is limited to a circle of maximum radius, $\Rmax$, here we assume $S(x,y)=1$, while is zero otherwise. We assume that the density of the foreground is constant on this region, thus 
\begin{equation}
    \PPxyBG(x,y) = \dfrac{1}{\pi\Rmax^2}. 
\end{equation}
We adopt $\Rmax=1\deg$ for Draco and $\Rmax=2\deg$ for Ursa Minor.

\subsubsection{PM and parallax distributions}
\label{appBsub:pms}

We model the joint probability distribution of PMs and parallaxes of stars in the dSph as a trivariate Gaussian, combining a bivariate Gaussian in PM and a univariate Gaussian in parallax. The foreground is modelled similarly, as a weighted sum of two such trivariate Gaussians, each with its own parameters, and with a relative contribution $\wiii$ between the two. Thus, called 
\begin{equation}\label{for:gmuprli}
    \GGmuprli(\bmu) = \frac{1}{2\pi |\bSigmai|^{1/2}} 
\exp\left[ 
-\frac{1}{2} (\bmu - \mui)^\mathrm{T} (\bSigmai)^{-1} (\bmu - \mui) 
\right]
\end{equation}
any of these Guassians, with $i$=dSph for the dwarf, and $i$=C1, C2 for the contamination components, then
\begin{equation}\begin{split}
   & \PPmuprldsph(\bmu) = \GGmuprldsph(\bmu)\\
   & \PPmuprlBG(\bmu)  = \wiii\,\GGmuprlbgi(\bmu) + (1-\wiii)\,\GGmuprlbgii(\bmu).
\end{split}\end{equation}
For clarity of notation, we have defined $\bmu \equiv (\mura, \mudec,\varpi)$, while, in equation~(\ref{for:gmuprli}), $\mui\equiv(\murai,\mudeci,\prli)$ is the centre of the $i$-th Gaussian and
\begin{equation}
\bSigmai =
\begin{pmatrix}
(\smurai)^2 & \rhoadi\smurai\smudeci & 0\\
\rhoadi\smurai\smudeci & (\smudeci)^2 & 0 \\
0 & 0 & (\sprli)^2 \\
\end{pmatrix}
\end{equation}
its corresponding covariance matrix. In the above equations, $\murai$, $\mudeci$ are the mean PM of the $i$-th component, $\smurai$ and $\smudeci$ the corresponding dispersions, $\rhoadi$ their correlation, while $\prli$ and $\sprli$ the mean parallax and parallax dispersion.

\begin{figure*}[h!]
    \centering
    \includegraphics[width=1\hsize]{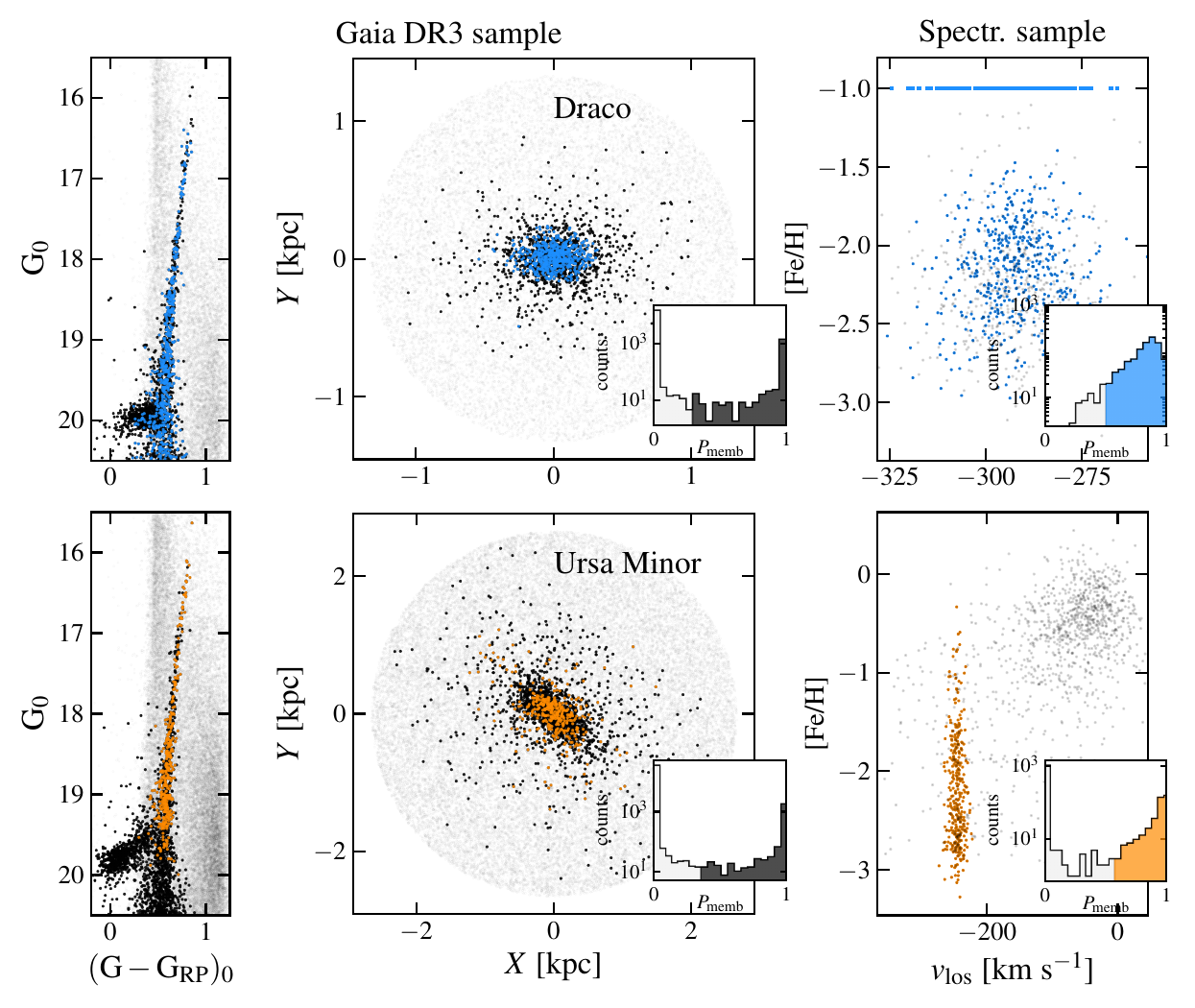}
    \caption{Examples of datasets adopted in our analysis. The top panels refer to Draco, while the bottom panels refer to Ursa Minor. Left panels: ($\Gmago$, $\colo$) CMDs. Grey points show the full Gaia-based catalogues, while black points indicate the $N$ stars classified as members according to the methodology of Appendix \ref{app:B}. Coloured points (blue for Draco, orange for Ursa Minor) highlight the subsample of $\Mvlos$ stars with available spectroscopic measurements. Middle panels: spatial distribution of the selected stars on the plane of the sky; the colour-code is the same as in the left panels. Right panels: distribution of stars in the line-of-sight velocity–metallicity plane. For Draco we show the sample based on the G26, while for Ursa Minor the sample based on W23. Here, the light points show all the stars with measured kinematics, while the coloured points indicate the subsample of stars classified as members. For Draco, the points at [Fe/H]=-1 only indicate stars with measured line-of-sight velocity but without metallicity. The small insets in the middle and right panels display the membership probability distributions resulting from applying the methods of Appendices ~\ref{sec:data} and ~\ref{app:B}, with the shaded coloured regions highlighting stars classified as members. Details on the construction of the catalogues are provided in in Appendices ~\ref{sec:data} and ~\ref{app:B}. Additional properties of the adopted samples are summarised in Table~\ref{tab:samples}.}
    \label{fig:datasets}
    \vspace{-0.3cm}
\end{figure*}

\subsubsection{CMDs}
\label{appBsub:cmds}

The CMD components, $\PPcmddsph$ and $\PPcmdBG$, are derived directly from the data following a methodology similar to that of \citet{McConnachie2020} and \citet{Battaglia2022}. The foreground CMD is constructed using stars at $1^\circ < R < 1.25^\circ$ for Draco and $2^\circ < R < 2.5^\circ$ for Ursa Minor. These regions lie at projected distances exceeding 6.5 times the half-mass radius and are therefore dominated by foreground stars.  Stars in these regions are mapped in the CMD space, and their distribution is smoothed in colour $\colo$ and magnitude $\Gmago$ according to the individual photometric uncertainties, $\Delta\colo$ and $\Delta\Gmago$, yielding a probabilistic CMD that explicitly accounts for measurement errors; to ensure a smooth representation, the adopted photometric uncertainties in magnitude are doubled.

The CMD of the dSph is constructed in a similar manner, using stars selected within an elliptical area centred on each system, and with semi-major axis equal to half of the galaxy half mass radius. We adopt reference values for the centre, PA and half-mass radius from the exponential fits of \citet{Munoz2018}. To further reduce contamination, we retain only stars with $-0.5 < \colo < 1.65$ for Draco and $-0.75 < \colo < 1.35$ for Ursa Minor. Despite these selections, the resulting dwarf CMD is still a superposition of galaxy and foreground stars, denoted $\PPcmdsum$. The dSph CMD component is therefore obtained as
\begin{equation}
    \PPcmddsph = \PPcmdsum - \lambda \PPcmdBG ,
\end{equation}
where $\PPcmdsum$ and $\PPcmdBG$ are rescaled to have the same total area prior to subtraction. Then both $\PPcmddsph$ and $\PPcmdBG$ are normalised to unit integral; any negative values of $\PPcmddsph$ resulting from the subtraction are set to zero. The scaling parameter $\lambda$ is treated as a free nuisance parameter in the analysis, allowing the model to absorb residual contamination arising from imperfect sampling, selection effects, and Poisson noise.

\subsubsection{Deriving the models parameters}
\label{appBsub:memb}

The model includes a total of 29 free parameters. All free parameters, together with the adopted priors, the values inferred from the fit, and the corresponding $1\sigma$ and $2\sigma$ credible regions, are listed in a table provided as supplementary material.


We account for observational uncertainties only in PMs and parallaxes, including their mutual correlations. Accordingly, $\EE$ is modelled as a three-dimensional Gaussian, whose covariance matrix has diagonal elements given by the squared uncertainties on PMs and parallax, and a single off-diagonal term $\rhoadj$ describing the correlation between $\muraj$ and $\mudecj$. Further details in the MCMC strategy are given in Appendix~\ref{app:A}.

\subsection{Selecting members from \citetalias{Walker2023} catalogue}
\label{app:Bkin}

When using the spectroscopic catalogues described in Section~\ref{app:Bkin}, we apply an additional pre-filtering step to the \citetalias{Walker2023} dataset, which contains both likely members and foreground contaminants. In this case, kinematic members are identified using the probabilistic method described in Section~\ref{app:B}. By contrast, stars from the DEIMOS catalogue of \citetalias{Geha2026} are already selected as likely members.

When applying the method to the spectroscopic \citetalias{Walker2023} catalogues of Section~\ref{subsec:wal}, we assume that the model depends on the variable vector $\bxi\equiv\{\vlos,\feh\}$, and that the input data consist of the corresponding measured quantities together with their associated uncertainties. For the $j$-th star, the data vector is therefore $\{\DD\}_j = \{\vlosj, \Delta\vlosj, \fehj, \Delta\fehj\}$.

In this application, the term describing the dSph population in the line-of-sight velocity–metallicity plane is modelled as a two-dimensional, uncorrelated Gaussian distribution in velocity and metallicity. To account for the complexity of the foreground distribution in this plane, we model it as a weighted sum of two two-dimensional, uncorrelated Gaussian components. Denoting any of these Gaussian components by $\NNi$, we write
\begin{equation}\begin{split}
    \PPdsph(\bxi) = & \GGdSph(\bxi), \\ \PPC(\bxi) = & \wiii\GGBGi(\bxi) + (1-\wiii)\GGBGii(\bxi), \\
    \end{split}\end{equation}
where $i$=dSph refers to the target dwarf galaxy, and $i$=C1, C2 to the two foreground components, while $\wiii$ weights the relative contribution of $\GGBGi$. Each Gaussian component is characterised by four free parameters: the mean line-of-sight velocity, $\vlosi$, and the mean metallicity of that component, $\fehi$, together with their corresponding intrinsic dispersions, $\sigmalosi$ and $\sigmafehi$.

The model includes 14 free parameters. All free parameters, priors, credible regions, are provided as supplementary material. In this application, the error function $\EE$ is modelled as a bi-variate, uncorrelated, Gaussian, with covariance matrix with diagonal elements given by the squared
uncertainties on line-of-sight velocity and metallicity.

\subsection{The final samples}
\label{app:cfinal}
At this stage, for each galaxy we have two independent spectroscopic datasets. Before combining the spectroscopic information with the Gaia catalogues, we apply the same kinematic corrections to both samples. Following the prescription given in the appendix of \cite{Walker2008}, we correct the line-of-sight velocities for the position-dependent velocity gradient induced by the large angular extent of the systems. Both datasets are shifted to their respective systemic velocities\footnote{For the \citetalias{Walker2023} samples we adopt $\vsys = -291.7\kms$ for Draco and $\vsys = -245.7\kms$ for Ursa Minor. For the \citetalias{Geha2026} catalogue, we instead adopt $\vsys = -292.6\kms$ for Draco and $\vsys = -246.1\kms$ for Ursa Minor, consistent with the systemic velocities reported in that work.}, while the projection corrections are computed using identical systemic PMs and distances for the two catalogues. For Draco we adopt\footnote{The adopted PMs are those derived in Appendix~\ref{app:Bgaia} and listed with the model parameters in the supplementary material. They are consistent with the literature \citep[e.g.][]{Battaglia2022}.} $(\mura,\mudec)=(0.042,-0.188)\mas,\yr^{-1}$, and for Ursa Minor $(\mura,\mudec)=(-0.131,0.070)\mas,\yr^{-1}$, with systemic PMs corrected for the Gaia zero-point offsets estimated by \cite{Battaglia2022}. The distance is assumed to be $76\kpc$ for both galaxies \citep{Munoz2018}.

We then combine the spectroscopic and astrometric information through a spatial cross-match with the Gaia catalogue, retaining only stars with a Gaia counterpart. For the \citetalias{Walker2023} samples, nearly all spectroscopic members have a Gaia match: we exclude only 18 stars in Draco and 7 in Ursa Minor. The situation is different for the DEIMOS catalogue, which extends to magnitudes fainter than the Gaia limit. In this case, we exclude 480 out of 1190, and 351 out of 876 stars in Draco and Ursa Minor, respectively. Although this choice reduces the size of the DEIMOS-based samples, we adopt a uniform Gaia-matched selection for both datasets. This avoids the need to introduce an explicit selection function to describe the different magnitude depths of the two spectroscopic catalogues and allows a more direct comparison between them. Note, however, that even with these cuts the samples contain a very large number of spectroscopic tracers, larger than in the catalogues based on \citetalias{Walker2023} data. Finally, we note that in the \citetalias{Walker2023}-based samples all stars with measured kinematics also have metallicity estimates. In contrast, in the DEIMOS catalogue only a fraction of the stars with measured $\vlos$ have reliable $\feh$ determinations, and this difference is accounted for in the chemo-dynamical analysis. The main properties of the resulting samples are summarised in Table~\ref{tab:samples}, while Fig.~\ref{fig:datasets} provides a schematic representation of the structure of the catalogues. For clarity, in the remainder of this paper we refer to the samples as W23 and G26 for the \citetalias{Walker2023} and \citetalias{Geha2026} based catalogues, respectively.

\begin{table}[h!]
\centering
\caption{Number of stars in each dataset.}\label{tab:samples}
\tiny
\renewcommand{\arraystretch}{1.}
\begin{tabular}{llccc}
\toprule
\toprule
Galaxy & Dataset & $N$ & $\Mvlos$ & $\Mfeh$ \\
\midrule
\multirow{2}{*}{Draco}
  & W23 & \multirow{2}{*}{1610} & 375 & 375 \\
  & G26 &                       & 710 &    514  \\
\addlinespace[0.5em]
\multirow{2}{*}{Ursa Minor}
  & W23 & \multirow{2}{*}{2137} & 361 & 361 \\
  & G26 &                       & 525 &    416  \\
\bottomrule
\bottomrule
\end{tabular}
\tablefoot{$N$ is the total number of stars. $\Mvlos$ is the subsample of stars with a measured line-of-sight velocity, $\Mfeh$ is the subsample of $\Mvlos$ with an associated metallicity measurement ($N > \Mvlos > \Mfeh$). W23 and G26 label datasets based on \citetalias{Walker2023} and \citetalias{Geha2026} catalogues, respectively. Details on the data reduction are provided in Appendices~\ref{sec:data} and~\ref{app:B}.}
\end{table}

\vspace{-0.5cm}

\begin{table*}
\centering
\tiny
\caption{Structural, kinematic, and chemical parameters of the stellar populations in Draco and Ursa Minor inferred from the two-component, flattened models of Section~\ref{sec:mod}.}
\label{tab:modparams}
\setlength{\tabcolsep}{4.25pt}
\begin{tabular}{lrrrrrrrrrrrr}
\toprule
\toprule
\addlinespace[6pt]
 & \multicolumn{6}{c}{ {\large Draco} } & \multicolumn{6}{c}{{\large Ursa Minor}} \\
 \addlinespace[3pt]
\cmidrule(lr){2-7} \cmidrule(lr){8-13}
\cmidrule(lr){2-7} \cmidrule(lr){8-13}
\cmidrule(lr){2-7} \cmidrule(lr){8-13}
\cmidrule(lr){2-7} \cmidrule(lr){8-13}
 & \multicolumn{2}{c}{MR} & \multicolumn{2}{c}{MP} & \multicolumn{2}{c}{Total}
 & \multicolumn{2}{c}{MR} & \multicolumn{2}{c}{MP} & \multicolumn{2}{c}{Total} \\
\cmidrule(lr){2-3} \cmidrule(lr){4-5} \cmidrule(lr){6-7}
\cmidrule(lr){8-9} \cmidrule(lr){10-11} \cmidrule(lr){12-13}
\cmidrule(lr){2-3} \cmidrule(lr){4-5} \cmidrule(lr){6-7}
\cmidrule(lr){8-9} \cmidrule(lr){10-11} \cmidrule(lr){12-13}
\cmidrule(lr){2-3} \cmidrule(lr){4-5} \cmidrule(lr){6-7}
\cmidrule(lr){8-9} \cmidrule(lr){10-11} \cmidrule(lr){12-13}
Param 
& G26 & W23 & G26 & W23 & G26 & W23
& G26 & W23 & G26 & W23 & G26 & W23 \\
\midrule
\addlinespace[6pt]
$\MM$ [dex]           
& $-1.98_{-0.06}^{+0.05}$ & $-1.70_{-0.09}^{+0.07}$ & $-2.37_{-0.08}^{+0.07}$ & $-2.29_{-0.16}^{+0.12}$ & -- & --
& $-2.18_{-0.03}^{+0.03}$ & $-1.88_{-0.09}^{+0.07}$ 
& $-2.52_{-0.11}^{+0.11}$ & $-2.58_{-0.09}^{+0.09}$ & -- &  -- \\
\addlinespace[6pt]
$\sigma$ [dex] 
& $0.13_{-0.06}^{+0.05}$ & $0.20_{-0.05}^{+0.05}$ & $0.18_{-0.05}^{+0.05}$ & $0.38_{-0.07}^{+0.07}$ & -- & --
& $0.21_{-0.03}^{+0.03}$ & $0.27_{-0.07}^{+0.09}$ & $0.29_{-0.11}^{+0.11}$ &  $0.29_{-0.07}^{+0.07}$ & -- &  --\\
\addlinespace[6pt]
$\Reff$ [pc] 
& $140_{-17}^{+19}$ & $159_{-18}^{+19}$ & $299_{-31}^{+46}$ & $343_{-49}^{+57}$ & $211_{-9}^{+9}$ & $212_{-9}^{+9}$ 
& $374_{-22}^{+21}$ & $352_{-21}^{+23}$ & $1034_{-123}^{+141}$ & $882_{-77}^{+84}$ & $463_{-18}^{+20}$ & $455_{-18}^{+20}$ \\
\addlinespace[6pt]
$\eeff$ 
& $0.23_{-0.11}^{+0.09}$ & $0.27_{-0.09}^{+0.07}$ & $0.33_{-0.07}^{+0.06}$ & $0.32_{-0.10}^{+0.08}$ & $0.28_{-0.05}^{+0.04}$ & $0.28_{-0.05}^{+0.05}$ 
& $0.61_{-0.03}^{+0.03}$ & $0.62_{-0.04}^{+0.03}$ & $0.29_{-0.11}^{+0.09}$ & $0.32_{-0.08}^{+0.08}$ & $0.56_{-0.03}^{+0.03}$ & $0.55_{-0.03}^{+0.03}$ \\
\addlinespace[6pt]
$\betaz$ 
& $0.43_{-0.35}^{+0.24}$ &  $0.45_{-0.34}^{+0.21}$ & $0.83_{-0.17}^{+0.08}$ & $0.73_{-0.20}^{+0.12}$ & $0.71_{-0.14}^{+0.10}$ & $0.59_{-0.16}^{+0.13}$
& $0.92_{-0.04}^{+0.03}$ & $0.89_{-0.05}^{+0.04}$ & $0.40_{-0.37}^{+0.28}$ & $0.73_{-0.17}^{+0.10}$ & $0.84_{-0.05}^{+0.04}$ &  $0.80_{0.06}^{+0.04}$\\
\addlinespace[6pt]
$\frac{\sloso}{[\kms]}$
& $8.5_{-1.5}^{+1.5}$ & $7.0_{-1.4}^{+1.6}$ & $14.5_{-1.8}^{+1.7}$ & $12.2_{-2.6}^{+2.6}$ & $11.0_{-0.9}^{+1.0}$ & $8.4_{-1.1}^{+1.2}$
& $7.9_{-0.9}^{+0.9}$ & $6.4_{-0.7}^{+0.8}$ & $14.8_{-2.9}^{+2.6}$ & $12.5_{-2.2}^{+1.8}$ & $8.2_{-0.8}^{+0.8}$ & $7.1_{-0.7}^{+0.7}$ \\
\addlinespace[6pt]
$\frac{|\Delta\vrotmax|}{[\kms]}$
& $0.66^{+0.04}_{-0.04}$ & $0.75^{+0.04}_{-0.04}$ & $1.40^{+0.06}_{-0.06}$ & $1.66^{+0.07}_{-0.07}$ & $1.00^{+0.05}_{-0.05}$ & $1.14^{+0.06}_{-0.06}$ & $0.29^{+0.04}_{-0.04}$ & $0.60^{+0.04}_{-0.04}$ & $2.87^{+0.09}_{-0.09}$ & $3.01^{+0.07}_{-0.07}$ & $0.56^{+0.06}_{-0.06}$ & $1.57^{+0.06}_{-0.06}$ \\
\addlinespace[6pt]
$w$ & $0.41_{-0.13}^{+0.13}$ & $0.55_{-0.13}^{+0.13}$ & $0.59_{-0.13}^{+0.13}$ & $0.45_{-0.13}^{+0.13}$ & -- & -- & $0.68_{-0.07}^{+0.06}$ & $0.58_{-0.05}^{+0.05}$ & $0.32_{-0.06}^{+0.07}$ & $0.42_{-0.05}^{+0.05}$ & -- & --\\ \addlinespace[6pt]
$N$ & $660^{+209}_{-209}$ & $886^{+209}_{-209}$ & $950^{+209}_{-209}$ & $724^{+209}_{-209}$ & -- & -- & $1453^{+128}_{-150}$ & $1239^{+107}_{-107}$ & $684^{+150}_{-128}$ & $898^{+107}_{-107}$ & -- & --  \\
\addlinespace[6pt]
$\Mvlos$ & 403 & 173 & 307 & 202 & -- & -- & 454 & 231 & 71 & 130 & -- & -- \\
\addlinespace[6pt]
\bottomrule
\bottomrule
\end{tabular}
\tablefoot{Mean metallicities, $\MM$, metallicity dispersion, $\sigma$, the total fraction of MR and MP stars, $w$, the total number of MR and MP stars, and the median number of MR and MP stars with available kinematics are provided for the MR and MP only. Projected half-mass radii, $\Reff$, effective ellipticity, $\eeff$, anisotropy, $\betaz$, central velocity dispersion, $\sloso$, and maximum mean line-of-sight velocity offset, $|\Delta\vrotmax|$, are given for the MR, MP population and the overall stellar system. The effective ellipticity is defined as $\eeff \equiv 1-\qeff$, where $\qeff$ is the mass-weighted axis ratio inferred from the projected surface-density profiles (Eq.~\ref{for:qeff}).}
\end{table*}

\begin{table*}
\centering
\tiny
\caption{DM halo parameters inferred for Draco and Ursa Minor from the modelling described in Section~\ref{sec:mod}. Results are shown for the G26 and W23 datasets. The reported quantities refer to the different modelling configurations adopted: flattened two-component models (F2), flattened single-component models (F1), and spherical two-component models (S2).}\label{tab:dm}
\setlength{\tabcolsep}{3pt}
\begin{tabular}{l
>{\columncolor{gray!15}}r >{\columncolor{gray!15}}r
r r r r
>{\columncolor{gray!15}}r >{\columncolor{gray!15}}r
r r r r}
\toprule
\toprule
\addlinespace[6pt]
 & \multicolumn{6}{c}{ {\normalsize Draco} } & \multicolumn{6}{c}{{\normalsize Ursa Minor}} \\
 \addlinespace[3pt]
\cmidrule(lr){2-7} \cmidrule(lr){8-13}
 & \multicolumn{2}{c}{F2} & \multicolumn{2}{c}{F1} & \multicolumn{2}{c}{S2}
 & \multicolumn{2}{c}{F2} & \multicolumn{2}{c}{F1} & \multicolumn{2}{c}{S2} \\
\cmidrule(lr){2-3} \cmidrule(lr){4-5} \cmidrule(lr){6-7}
\cmidrule(lr){8-9} \cmidrule(lr){10-11} \cmidrule(lr){12-13}
Param 
& G26 & W23 & G26 & W23 & G26 & W23 & G26 & W23 & G26 & W23 & G26 & W23  \\
\midrule

$\gamma$ 
& $0.98_{-0.26}^{+0.28}$ & $0.82_{-0.31}^{+0.29}$ & $0.90_{-0.29}^{+0.27}$ & $0.75_{-0.36}^{+0.32}$ & $1.17_{-0.31}^{+0.25}$ & $0.95_{-0.32}^{+0.37}$ & $0.37_{-0.24}^{+0.31}$ & $0.74_{-0.30}^{+0.22}$ & $0.48_{-0.25}^{+0.23}$ & $0.55_{-0.28}^{+0.26}$ & $0.83_{-0.37}^{+0.32}$ & $1.13_{-0.32}^{+0.27}$ \\[6pt]

$\beta$ & $>3.41$ & $5.07_{-1.27}^{+1.34}$ & $4.95_{-1.20}^{+1.35}$ & $4.90_{-1.20}^{+1.45}$ & $4.97_{-1.18}^{+1.37}$ & $>3.65$ & $>3.60$ & $>3.74$ & $4.76_{-1.06}^{+1.42}$ & $4.68_{-1.05}^{+1.50}$ & $4.94_{-1.14}^{+1.37}$ & $4.77_{-1.08}^{+1.44}$  \\[6pt]

$\alpha$ & $>1.18$ & $>1.29$ & $>1.06$ & $>1.16$ & $>1.19$ & $>1.13$ & $>1.90$ & $3.25_{-1.38}^{+1.54}$ & $>1.41$ & $>1.51$ & $>1.53$ & $>1.01$  \\[6pt]

$\rs$ $[\kpc$] & $>1.10$ & $>1.27$ & $>0.97$ & $>1.19$ & $>0.97$ & $>1.10$ & $>2.00$ & $>1.82$ & $>2.55$ & $>2.41$ & $>1.73$ & $>1.38$ \\[6pt]

$\frac{\rhodmcl}{[10^8\Msun\kpc^{-3}]}$ 
& $2.10_{-0.31}^{+0.34}$ & $1.52_{-0.31}^{+0.35}$ & $1.96_{-0.32}^{+0.39}$ & $1.50_{-0.32}^{+0.37}$ & $2.54_{-0.30}^{+0.35}$ & $1.86_{-0.31}^{+0.31}$ & $0.43_{-0.11}^{+0.15}$ & $0.44_{-0.13}^{+0.14}$ & $0.33_{-0.93}^{+0.12}$ & $0.31_{-0.10}^{+0.13}$ & $0.95_{-0.27}^{+0.31}$ & $0.91_{-0.24}^{+0.27}$ \\[6pt]

$\frac{\Mdm(<\Reff)}{[10^6\Msun]}$ &
$8.96_{-1.49}^{+1.69}$ &
$6.30_{-1.40}^{+1.73}$ &
$8.20_{-1.48}^{+1.93}$ &
$6.18_{-1.40}^{+1.78}$ &
$11.2_{-1.6}^{+1.7}$ &
$7.90_{-1.53}^{+1.65}$ &
$12.7_{-2.0}^{+2.2}$ &
$10.2_{-1.8}^{+1.9}$ &
$8.99_{-1.62}^{+1.81}$ &
$8.02_{-1.70}^{+1.93}$ &
$20.6_{-3.4}^{+4.0}$ &
$16.2_{-2.6}^{+4.0}$ \\[6pt]

$\frac{\Mdm(<1.8\Reff)}{[10^7\Msun]}$ &
$2.81_{-0.34}^{+0.48}$ &
$2.21_{-0.31}^{+0.38}$ &
$2.73_{-0.33}^{+0.44}$ &
$2.24_{-0.32}^{+0.44}$ &
$3.19_{-0.38}^{+0.56}$ &
$2.53_{-0.31}^{+0.39}$ &
$5.82_{-0.94}^{+1.06}$ &
$3.74_{-0.51}^{+0.65}$ &
$3.88_{-0.50}^{+0.57}$ &
$3.32_{-0.47}^{+0.57}$ &
$7.20_{-1.45}^{+2.03}$ &
$4.66_{-0.63}^{+0.79}$ \\[6pt]

$\frac{\Mdm(<5\Reff)}{[10^8\Msun]}$ &
$1.96_{-0.54}^{+0.78}$ &
$1.86_{-0.53}^{+0.75}$ &
$2.00_{-0.52}^{+0.74}$ &
$2.01_{-0.63}^{+0.99}$ &
$1.88_{-0.55}^{+0.95}$ &
$1.81_{-0.59}^{+0.87}$ &
$7.18_{-2.66}^{+3.47}$ &
$2.94_{-0.75}^{+1.09}$ &
$4.38_{-1.01}^{+1.33}$ &
$3.51_{-0.86}^{+1.24}$ &
$5.34_{-2.13}^{+4.17}$ &
$2.40_{-0.62}^{+1.00}$ \\[6pt]

$\log\frac{J_{0.5^\circ}}{[\GeV^2\cm^{-5}]}$ 
& $18.60_{-0.17}^{+0.19}$ & $18.59_{-0.20}^{+0.23}$ & $18.60_{-0.17}^{+0.21}$ & $18.64_{-0.22}^{+0.26}$ & $18.56_{-0.18}^{+0.23}$ & $18.55_{-0.20}^{+0.24}$ 
& $18.51_{-0.18}^{+0.18}$ & $18.17_{-0.09}^{+0.12}$ & $18.87_{-0.14}^{+0.21}$ & $18.82_{-0.21}^{+0.33}$ & $18.44_{-0.17}^{+0.21}$ & $18.16_{-0.09}^{+0.11}$ \\[6pt]

$\log\frac{D_{0.5^\circ}}{[\GeV\cm^{-2}]}$ 
& $18.88_{-0.14}^{+0.21}$ & $18.77_{-0.18}^{+0.26}$ & $18.33_{-0.11}^{+0.12}$ & $18.23_{-0.11}^{+0.13}$ & $18.95_{-0.12}^{+0.20}$ & $18.80_{-0.15}^{+0.25}$ 
& $18.31_{-0.22}^{+0.24}$ & $17.83_{-0.12}^{+0.14}$ & $17.94_{-0.12}^{+0.15}$ & $17.79_{-0.14}^{+0.16}$ & $18.42_{-0.18}^{+0.24}$ & $18.10_{-0.13}^{+0.15}$ \\[6pt]

$\log\frac{\BH}{[\Msun]}$ & $<5.20$ & $<4.94$ & $<5.40$ & $<5.07$ & $<5.47$ & $<5.42$ & $<3.33$ & $<4.19$ & $<3.45$ & $<4.21$ & $<5.17$ & $<5.66$ \\

\bottomrule
\bottomrule
\end{tabular}
\tablefoot{Inner slope $\gamma$, transition parameter $\alpha$, outer slope $\beta$, scale radius $\rs$, DM density at $150\pc$, $\rhodmcl$, the DM mass within $\Reff$, $1.8\Reff$ and $5\Reff$, the astrophysical factors $\logJ$ and $\logD$, and the IMBH mass $\log\BH$. We report $1\sigma$ credible regions only for parameters whose 95\% HPD interval (see Appendix~\ref{app:A} for details) lies within the prior. If the 95\% HPD interval reaches or is limited by the prior boundary, we instead quote the corresponding lower or upper limit.}
\end{table*}

\section{Tables}
\label{app:D}
Tables summarizing the characteristic models quantities. Table~\ref{tab:modparams} reports the stellar component parameters for the different datasets, while Table~\ref{tab:dm} focuses on the DM properties computed using different modelling assumptions and datasets.

\clearpage

\section{Supplementary Material}
Tables given as supplementary material.

\begin{table*}[p]
    \renewcommand{\arraystretch}{1.5} 
    \centering
    \small
    \caption{Model parameters inferred from the mixture model described in Appendix~B and applied to the Gaia DR3 data.}
    \begin{tabular}{lrrr}
        \toprule
        \toprule
        Parameter & Prior & Draco & Ursa Minor \\
        \midrule
        $\xo$ [$\asec$] & [-50, 50] & $28.86_{-12.46\,(-24.60)}^{+12.59\,(+25.25)}$ & $-1.82_{-20.33\,(-40.84)}^{+20.45\,(+41.39)}$ \\
        $\yo$ [$\asec$] & [-50, 50] & $1.99_{-8.45\,(-16.88)}^{+8.33\,(+16.89)}$ & $21.77_{-18.65\,(-36.77)}^{+18.80\,(+38.85)}$ \\
        $\Reff$ [$\amin$] & [0, 25] & $10.04_{-0.25\,(-0.49)}^{+0.25\,(+0.52)}$ & $23.71_{-0.54\,(-1.08)}^{+0.55\,(+1.05)}$ \\
        $e$ & [0, 0.9] & $0.29_{-0.02\,(-0.05)}^{+0.02\,(+0.04)}$ & $0.39_{-0.02\,(-0.04)}^{+0.02\,(+0.04)}$ \\
        $\phi$ & [0, 180] & $88.22_{-2.54\,(-5.15)}^{+2.58\,(+5.24)}$ & $129.90_{-1.63\,(-3.26)}^{+1.59\,(+3.22)}$ \\
        $\muradsph$ [$\mas\yr^{-1}$] & [-7, 7] & $0.04_{-0.01\,(-0.01)}^{+0.01\,(+0.01)}$ & $-0.13_{-0.01\,(-0.01)}^{+0.01\,(+0.01)}$ \\
        $\mudecdsph$ [$\mas\yr^{-1}$] & [-7, 7] & $-0.19_{-0.01\,(-0.01)}^{+0.01\,(+0.01)}$ & $0.07_{-0.01\,(-0.01)}^{+0.01\,(+0.01)}$ \\
        $\smuradsph$ [$\mas\yr^{-1}$] & [0, 0.4] & $0.03_{-0.02\,(-0.03)}^{+0.01\,(+0.03)}$ & $0.07_{-0.01\,(-0.02)}^{+0.01\,(+0.02)}$ \\
        $\smudecdsph$ [$\mas\yr^{-1}$] & [0, 0.4] & $0.03_{-0.02\,(-0.03)}^{+0.02\,(+0.03)}$ & $0.05_{-0.01\,(-0.03)}^{+0.01\,(+0.02)}$ \\
        $\rhoaddsph$ & [0, 1] & $0.24_{-0.66\,(-1.13)}^{+0.48\,(+0.72)}$ & $-0.16_{-0.23\,(-0.47)}^{+0.25\,(+0.52)}$ \\
        $\prldsph$ [$\mas$] & [-4, 2] & $-0.022_{-0.005\,(-0.009)}^{+0.005\,(+0.009)}$ & $-0.014_{-0.004\,(-0.008)}^{+0.004\,(+0.008)}$ \\
        $\sprldsph$ [$\mas$] & [0, 2] & $0.02_{-0.01\,(-0.02)}^{+0.01\,(+0.02)}$ & $0.02_{-0.01\,(-0.02)}^{+0.01\,(+0.02)}$ \\
        $\murabgi$ [$\mas\yr^{-1}$] & [-30, 30] & $-5.59_{-0.25\,(-0.50)}^{+0.25\,(+0.50)}$ & $-11.22_{-0.24\,(-0.47)}^{+0.24\,(+0.47)}$ \\
        $\mudecbgi$ [$\mas\yr^{-1}$] & [-30, 30] & $-1.69_{-0.37\,(-0.73)}^{+0.36\,(+0.72)}$ & $-1.81_{-0.24\,(-0.48)}^{+0.25\,(+0.49)}$ \\
        $\smurabgi$ [$\mas\yr^{-1}$] & [0, 60] & $14.52_{-0.19\,(-0.36)}^{+0.19\,(+0.38)}$ & $20.47_{-0.17\,(-0.34)}^{+0.16\,(+0.33)}$ \\
        $\smudecbgi$ [$\mas\yr^{-1}$] & [0, 60] & $21.13_{-0.26\,(-0.53)}^{+0.27\,(+0.55)}$ & $21.52_{-0.18\,(-0.36)}^{+0.18\,(+0.37)}$ \\
        $\rhoadbgi$ & [-1, 1] & $-0.15_{-0.02\,(-0.03)}^{+0.02\,(+0.03)}$ & $-0.35_{-0.01\,(-0.02)}^{+0.01\,(+0.02)}$ \\
        $\prlbgi$ [$\mas$] & [0, 10] & $1.85_{-0.02\,(-0.04)}^{+0.02\,(+0.04)}$ & $2.08_{-0.01\,(-0.03)}^{+0.02\,(+0.03)}$ \\
        $\sprlbgi$ [$\mas$] & [0, 10] & $1.14_{-0.01\,(-0.03)}^{+0.01\,(+0.03)}$ & $1.22_{-0.01\,(-0.02)}^{+0.01\,(+0.02)}$ \\
        $\murabgii$ [$\mas\yr^{-1}$] & [-10, 10] & $-2.17_{-0.03\,(-0.07)}^{+0.03\,(+0.07)}$ & $-2.81_{-0.04\,(-0.07)}^{+0.04\,(+0.07)}$ \\
        $\mudecbgii$ [$\mas\yr^{-1}$] & [-10, 10] & $-2.80_{-0.06\,(-0.11)}^{+0.06\,(+0.11)}$ & $-2.60_{-0.04\,(-0.07)}^{+0.04\,(+0.08)}$ \\
        $\smurabgii$ [$\mas\yr^{-1}$] & [0, 20] & $3.57_{-0.03\,(-0.06)}^{+0.03\,(+0.06)}$ & $5.40_{-0.03\,(-0.06)}^{+0.03\,(+0.06)}$ \\
        $\smudecbgii$ [$\mas\yr^{-1}$] & [0, 20] & $5.79_{-0.05\,(-0.10)}^{+0.05\,(+0.10)}$ & $5.57_{-0.03\,(-0.07)}^{+0.03\,(+0.07)}$ \\
        $\rhoadbgii$ & [0, 1] & $-0.22_{-0.01\,(-0.02)}^{+0.01\,(+0.02)}$ & $-0.42_{-0.01\,(-0.01)}^{+0.01\,(+0.01)}$ \\
        $\prlbgii$ [$\mas$] & [0, 10] & $0.513_{-0.004\,(-0.008)}^{+0.004\,(+0.008)}$ & $0.609_{-0.003\,(-0.007)}^{+0.003\,(+0.007)}$ \\
        $\sprlbgii$ [$\mas$] & [0, 10] & $0.329_{-0.004\,(-0.008)}^{+0.004\,(+0.008)}$ & $0.402_{-0.003\,(-0.006)}^{+0.003\,(+0.006)}$ \\
        $\wiii$ & [0, 1] & $0.229_{-0.004\,(-0.009)}^{+0.004\,(+0.009)}$ & $0.244_{-0.003\,(-0.006)}^{+0.003\,(+0.006)}$ \\
        $\wii$ & [0, 1] & $0.095_{-0.002\,(-0.004)}^{+0.002\,(+0.005)}$ & $0.057_{-0.001\,(-0.002)}^{+0.001\,(+0.002)}$ \\
        $\lambda$ & [0.2, 2.5] & $1.01_{-0.11\,(-0.27)}^{+0.07\,(+0.11)}$ & $1.50_{-0.09\,(-0.14)}^{+0.11\,(+0.19)}$ \\
\bottomrule
\bottomrule
    \end{tabular}
    \tablefoot{The table reports the model parameters, the adopted priors, and the corresponding inferred estimates for Draco and Ursa Minor. Parameter constraints are given in the form $\mathrm{median}^{+1\sigma\,(+2\sigma)}_{-1\sigma\,(-2\sigma)}$, where the $1\sigma$ and $2\sigma$ uncertainties are computed as described in Appendix~A.}
    \label{tab:placeholder}
\end{table*}

\begin{table*}[p]
    \renewcommand{\arraystretch}{1.5} 
    \centering
    \small
    \caption{Model parameters inferred from the mixture model described in Appendix~B and applied to the spectroscopic sample of \citetalias{Walker2023}.}
    \begin{tabular}{lrrr}
        \toprule
        \toprule
        Parameter & Prior & Draco & Ursa Minor \\
        \midrule
        $\vlosdsph$ [$\kms$] & [-400, -200] & $-291.7_{-0.5\,(-1.0)}^{+0.5\,(+0.9)}$ & $-245.7_{-0.5\,(-0.9)}^{+0.5\,(+1.0)}$ \\
        $\svlosdsph$ [$\kms$] & [0, 30] & $9.1_{-0.4\,(-0.8)}^{+0.4\,(+0.8)}$ & $8.4_{-0.4\,(-0.8)}^{+0.4\,(+0.9)}$ \\
        $\vlosbgi$ [$\kms$] & [-250, -100] & $-178.9_{-5.4\,(-11.0)}^{+5.3\,(+10.5)}$ & $-176.5_{-7.6\,(-15.5)}^{+7.4\,(+14.8)}$ \\
        $\slosbgi$ [$\kms$] & [0, 200] & $91.6_{-2.9\,(-5.6)}^{+2.9\,(+6.1)}$ & $94.7_{-4.1\,(-8.1)}^{+4.4\,(+9.0)}$ \\
        $\vlosbgii$ [$\kms$] & [-100, -100] & $-51.4_{-1.7\,(-3.5)}^{+1.7\,(+3.5)}$ & $-50.8_{-1.9\,(-3.9)}^{+1.9\,(+3.8)}$ \\
        $\slosbgii$ [$\kms$] & [0, 100] & $38.5_{-1.3\,(-2.7)}^{+1.4\,(+2.8)}$ & $42.5_{-1.5\,(-2.9)}^{+1.5\,(+3.1)}$ \\
        $\fehdsph$ [dex] & [-3, -1] & $-1.96_{-0.03\,(-0.07)}^{+0.03\,(+0.07)}$ & $-2.13_{-0.04\,(-0.07)}^{+0.04\,(+0.07)}$ \\
        $\sfehdsph$ [dex] & [0, 1] & $0.45_{-0.03\,(-0.05)}^{+0.03\,(+0.06)}$ & $0.44_{-0.03\,(-0.06)}^{+0.03\,(+0.06)}$ \\
        $\fehbgi$ [dex]  & [-1.5, -0.5] & $-1.11_{-0.04\,(-0.08)}^{+0.04\,(+0.07)}$ & $-1.25_{-0.05\,(-0.10)}^{+0.05\,(+0.10)}$ \\
        $\sfehbgi$ [dex] & [0, 1] & $-0.36_{-0.01\,(-0.02)}^{+0.01\,(+0.02)}$ & $-0.42_{-0.01\,(-0.03)}^{+0.01\,(+0.02)}$ \\
        $\fehbgii$ [dex]  & [-0.5, 1] & $0.21_{-0.01\,(-0.02)}^{+0.01\,(+0.02)}$ & $0.24_{-0.01\,(-0.02)}^{+0.01\,(+0.02)}$ \\
        $\sfehbgii$ [dex] & [0, 1] & $0.21_{-0.01\,(-0.02)}^{+0.01\,(+0.03)}$ & $0.24_{-0.01\,(-0.03)}^{+0.01\,(+0.03)}$ \\
        $\wii$ & [0, 1] & $0.230_{-0.010\,(-0.020)}^{+0.010\,(+0.021)}$ & $0.261_{-0.012\,(-0.024)}^{+0.012\,(+0.025)}$ \\
        $\wiii$ & [0, 1] & $0.394_{-0.018\,(-0.038)}^{+0.019\,(+0.038)}$ & $0.303_{-0.021\,(-0.043)}^{+0.022\,(+0.045)}$ \\
\bottomrule
\bottomrule
    \end{tabular}
    \tablefoot{The table reports the model parameters, the adopted priors, and the corresponding inferred estimates for Draco and Ursa Minor. Parameter constraints are given in the form $\mathrm{median}^{+1\sigma\,(+2\sigma)}_{-1\sigma\,(-2\sigma)}$, where the $1\sigma$ and $2\sigma$ uncertainties are computed as described in Appendix~A.}
    \label{tab:placeholder}
\end{table*}

\begin{table*}[p]
    \renewcommand{\arraystretch}{1.5} 
    \centering
    \small
    \caption{Model parameters inferred from the action-based dynamical models applied to the G26 and W23 samples for both Draco and Ursa Minor.}
    \begin{tabular}{lrrrrr}
        \toprule
        \toprule
         &  & \multicolumn{2}{c}{Draco} & \multicolumn{2}{c}{Ursa Minor} \\
        \midrule
        Parameter & Prior & G26 & W23 & G26 & W23 \\
        \midrule
        $\log\JcstMR$ [$\kpc\kms$] & [-1.8, 0] & $-0.50_{-0.84\,(-1.23)}^{+0.53\,(+0.79)}$ & $-0.53_{-0.81\,(-1.20)}^{+0.49\,(+0.77)}$ & $-0.42_{-0.92\,(-1.32)}^{+0.55\,(+0.79)}$ & $-0.70_{-0.78\,(-1.05)}^{+0.74\,(+1.02)}$ \\
        $\log\JstMR$ [$\kpc\kms$] & [0, 3] & $0.45_{-0.18\,(-0.38)}^{+0.26\,(+0.62)}$ & $0.48_{-0.25\,(-0.49)}^{+0.35\,(+0.78)}$ & $0.51_{-0.35\,(-0.58)}^{+0.37\,(+0.69)}$ & $0.28_{-0.22\,(-0.39)}^{+0.27\,(+0.56)}$ \\
        $\GammastMR$ & [0, 2] & $0.87_{-0.59\,(-0.83)}^{+0.66\,(+1.05)}$ & $0.93_{-0.60\,(-0.88)}^{+0.70\,(+1.01)}$ & $0.72_{-0.50\,(-0.69)}^{+0.81\,(+1.21)}$ & $0.50_{-0.37\,(-0.48)}^{+0.72\,(+1.34)}$ \\
        $\BMR$ & [3, 20] & $12.75_{-5.24\,(-7.37)}^{+4.99\,(+6.97)}$ & $11.68_{-4.54\,(-6.54)}^{+5.59\,(+7.93)}$ & $10.52_{-4.34\,(-5.73)}^{+6.51\,(+9.12)}$ & $11.37_{-4.67\,(-6.15)}^{+5.36\,(+8.10)}$ \\
        $\etastMR$ & [0, 5] & $2.58_{-1.13\,(-1.76)}^{+1.36\,(+2.23)}$ & $1.98_{-0.92\,(-1.31)}^{+1.59\,(+2.72)}$ & $1.50_{-0.57\,(-0.86)}^{+1.45\,(+2.96)}$ & $2.69_{-1.10\,(-1.78)}^{+1.23\,(+2.15)}$ \\
        $\hrMR$ & [0.2, 3] & $1.12_{-0.59\,(-0.87)}^{+0.75\,(+1.26)}$ & $1.25_{-0.69\,(-1.00)}^{+0.70\,(+1.20)}$ & $0.79_{-0.43\,(-0.57)}^{+0.69\,(+1.32)}$ & $0.87_{-0.46\,(-0.65)}^{+0.74\,(+1.46)}$ \\
        $\hzMR$ & [0.2, 3] & $0.96_{-0.53\,(-0.72)}^{+0.69\,(+1.31)}$ & $0.98_{-0.52\,(-0.74)}^{+0.66\,(+1.33)}$ & $1.33_{-0.74\,(-1.09)}^{+0.97\,(+1.30)}$ & $1.19_{-0.68\,(-0.94)}^{+1.01\,(+1.43)}$ \\
        $\grMR$ & [0.2, 3] & $1.15_{-0.42\,(-0.76)}^{+0.40\,(+0.80)}$ & $1.16_{-0.43\,(-0.77)}^{+0.42\,(+0.90)}$ & $0.34_{-0.09\,(-0.13)}^{+0.17\,(+0.52)}$ & $0.47_{-0.16\,(-0.25)}^{+0.21\,(+0.55)}$ \\
        $\gzMR$ & [0.2, 3] & $1.30_{-0.37\,(-0.72)}^{+0.40\,(+0.78)}$ & $1.36_{-0.42\,(-0.79)}^{+0.41\,(+0.80)}$ & $2.43_{-0.19\,(-0.68)}^{+0.13\,(+0.22)}$ & $2.33_{-0.23\,(-0.70)}^{+0.17\,(+0.30)}$ \\
        $\MMMR$ [dex] & [-2.2, -1.5]* & $-1.98_{-0.06\,(-0.13)}^{+0.05\,(+0.09)}$ & $-1.70_{-0.08\,(-0.23)}^{+0.06\,(+0.13)}$ & $-2.18_{-0.03\,(-0.06)}^{+0.03\,(+0.06)}$ & $-1.88_{-0.09\,(-0.19)}^{+0.07\,(+0.13)}$ \\
        $\sigmaMR$ [dex] & [0, 1.5] & $0.13_{-0.06\,(-0.12)}^{+0.05\,(+0.09)}$ & $0.20_{-0.05\,(-0.10)}^{+0.08\,(+0.21)}$ & $0.21_{-0.03\,(-0.06)}^{+0.03\,(+0.06)}$ & $0.26_{-0.07\,(-0.11)}^{+0.09\,(+0.19)}$ \\
        $\log\JcstMP$ [$\kpc\kms$] & [-1.8, 0] & $-0.36_{-0.93\,(-1.36)}^{+0.72\,(+1.15)}$ & $-0.45_{-0.88\,(-1.29)}^{+0.81\,(+1.29)}$ & $-0.31_{-0.98\,(-1.42)}^{+1.15\,(+1.76)}$ & $-0.31_{-1.00\,(-1.41)}^{+0.97\,(+1.51)}$ \\
        $\log\JstMP$ [$\kpc\kms$] & [0, 3] & $0.92_{-0.23\,(-0.40)}^{+0.26\,(+0.62)}$ & $1.05_{-0.22\,(-0.50)}^{+0.21\,(+0.43)}$ & $1.61_{-0.13\,(-0.26)}^{+0.15\,(+0.35)}$ & $1.41_{-0.15\,(-0.30)}^{+0.13\,(+0.26)}$ \\
        $\GammastMP$ & [0, 2] & $0.89_{-0.57\,(-0.85)}^{+0.69\,(+1.04)}$ & $1.19_{-0.72\,(-1.11)}^{+0.57\,(+0.77)}$ & $1.07_{-0.70\,(-1.01)}^{+0.64\,(+0.89)}$ & $1.19_{-0.85\,(-1.14)}^{+0.62\,(+0.78)}$ \\
        $\BMP$ & [3, 20] & $13.10_{-3.84\,(-6.44)}^{+4.42\,(+6.53)}$ & $13.13_{-4.30\,(-6.77)}^{+4.58\,(+6.54)}$ & $13.51_{-3.62\,(-5.86)}^{+4.11\,(+6.18)}$ & $13.87_{-3.63\,(-6.13)}^{+3.90\,(+5.82)}$ \\
        $\etastMP$ & [0, 5] & $2.39_{-1.00\,(-1.47)}^{+1.49\,(+2.42)}$ & $2.35_{-1.02\,(-1.51)}^{+1.68\,(+2.49)}$ & $3.43_{-1.30\,(-2.15)}^{+1.08\,(+1.48)}$ & $3.55_{-1.17\,(-2.00)}^{+1.03\,(+1.40)}$ \\
        $\hrMP$ & [0.2, 3] & $0.82_{-0.46\,(-0.60)}^{+0.77\,(+1.41)}$ & $0.80_{-0.43\,(-0.57)}^{+0.73\,(+1.39)}$ & $0.71_{-0.40\,(-0.50)}^{+0.72\,(+1.56)}$ & $0.53_{-0.26\,(-0.32)}^{+0.65\,(+1.36)}$ \\
        $\hzMP$ & [0.2, 3] & $1.34_{-0.74\,(-1.08)}^{+0.80\,(+1.23)}$ & $1.33_{-0.71\,(-1.07)}^{+0.76\,(+1.19)}$ & $1.25_{-0.71\,(-1.00)}^{+0.76\,(+1.16)}$ & $1.67_{-0.95\,(-1.39)}^{+0.52\,(+0.80)}$ \\
        $\grMP$ & [0.2, 3] & $0.52_{-0.17\,(-0.26)}^{+0.36\,(+0.95)}$ & $0.77_{-0.27\,(-0.47)}^{+0.42\,(+0.88)}$ & $1.45_{-0.41\,(-0.79)}^{+0.41\,(+0.77)}$ & $0.88_{-0.28\,(-0.44)}^{+0.41\,(+0.88)}$ \\
        $\gzMP$ & [0.2, 3] & $1.86_{-0.36\,(-0.96)}^{+0.24\,(+0.50)}$ & $1.70_{-0.45\,(-0.92)}^{+0.32\,(+0.61)}$ & $1.10_{-0.31\,(-0.57)}^{+0.33\,(+0.67)}$ & $1.49_{-0.36\,(-0.68)}^{+0.31\,(+0.51)}$ \\
        $\MMMP$ [dex] & [-3, -2.2]* & $-2.37_{-0.08\,(-0.17)}^{+0.07\,(+0.13)}$ & $-2.28_{-0.16\,(-0.41)}^{+0.12\,(+0.23)}$ & $-2.52_{-0.11\,(-0.26)}^{+0.11\,(+0.19)}$ & $-2.58_{-0.10\,(-0.19)}^{+0.10\,(+0.16)}$ \\
        $\sigmaMP$ [dex] & [0, 1.5] & $0.18_{-0.05\,(-0.10)}^{+0.05\,(+0.09)}$ & $0.37_{-0.07\,(-0.20)}^{+0.07\,(+0.16)}$ & $0.29_{-0.11\,(-0.23)}^{+0.11\,(+0.25)}$ & $0.28_{-0.07\,(-0.14)}^{+0.07\,(+0.17)}$ \\
        $\log\Mdm$ [M$_\odot$] & [8, 11] & $9.36_{-0.51\,(-0.93)}^{+0.51\,(+1.05)}$ & $9.41_{-0.53\,(-0.94)}^{+0.60\,(+1.18)}$ & $9.63_{-0.43\,(-0.80)}^{+0.42\,(+0.79)}$ & $9.12_{-0.34\,(-0.63)}^{+0.35\,(+0.71)}$ \\
        $\rs$ [$\kpc$] & [0.2, 5] & $3.04_{-1.41\,(-2.19)}^{+1.31\,(+1.87)}$ & $3.05_{-1.27\,(-1.99)}^{+1.24\,(+1.85)}$ & $3.61_{-1.02\,(-1.84)}^{+0.91\,(+1.31)}$ & $3.54_{-1.13\,(-2.01)}^{+0.98\,(+1.37)}$ \\
        $\gamma$ & [0, 2] & $0.98_{-0.26\,(-0.53)}^{+0.29\,(+0.51)}$ & $0.82_{-0.31\,(-0.64)}^{+0.30\,(+0.59)}$ & $0.37_{-0.24\,(-0.35)}^{+0.31\,(+0.56)}$ & $0.74_{-0.30\,(-0.60)}^{+0.22\,(+0.48)}$ \\
        $\alpha$ & [0.5, 6] & $3.40_{-1.63\,(-2.45)}^{+1.67\,(+2.46)}$ & $3.45_{-1.49\,(-2.43)}^{+1.69\,(+2.42)}$ & $4.02_{-1.44\,(-2.39)}^{+1.35\,(+1.89)}$ & $3.25_{-1.38\,(-2.03)}^{+1.54\,(+2.58)}$ \\
        $\beta$ & [3, 7] & $4.92_{-1.13\,(-1.60)}^{+1.43\,(+1.98)}$ & $5.07_{-1.27\,(-1.73)}^{+1.34\,(+1.85)}$ & $5.09_{-1.10\,(-1.59)}^{+1.29\,(+1.82)}$ & $5.12_{-1.13\,(-1.66)}^{+1.30\,(+1.79)}$ \\
        $\log\BH$ [M$_\odot$] & [2, 6] & $2.93_{-0.64\,(-0.89)}^{+1.25\,(+2.58)}$ & $2.86_{-0.56\,(-0.81)}^{+0.93\,(+2.52)}$ & $2.58_{-0.41\,(-0.56)}^{+0.48\,(+0.89)}$ & $2.63_{-0.42\,(-0.60)}^{+0.68\,(+2.30)}$ \\
        $i$ & [0, 90] & $74.3_{-11.7\,(-20.5)}^{+10.5\,(+14.9)}$ & $74.1_{-12.9\,(-22.3)}^{+11.1\,(+15.3)}$ & $82.3_{-6.2\,(-10.4)}^{+5.0\,(+7.3)}$ & $82.8_{-5.9\,(-10.7)}^{+4.7\,(+6.8)}$ \\
        $\wMR$ & [0, 1] & $0.41_{-0.13\,(-0.22)}^{+0.13\,(+0.31)}$ & $0.55_{-0.13\,(-0.30)}^{+0.13\,(+0.26)}$ & $0.68_{-0.07\,(-0.15)}^{+0.06\,(+0.11)}$ & $0.58_{-0.05\,(-0.11)}^{+0.05\,(+0.12)}$ \\
\bottomrule
\bottomrule
    \end{tabular}
    \tablefoot{The table reports the model parameters, the adopted priors, and the corresponding inferred estimates for Draco and Ursa Minor. Parameter constraints are given in the form $\mathrm{median}^{+1\sigma\,(+2\sigma)}_{-1\sigma\,(-2\sigma)}$, where the $1\sigma$ and $2\sigma$ uncertainties are computed as described in Appendix~A.\\
    * The priors for the mean metallicities of the MR and MP populations differ between the MCMC runs. The priors reported in the table refer to the G26 dataset for Draco. For the MR population, we adopt the ranges [-2.1, -1], [-2.3, -1], and [-2.4, -1] for the fits to Draco W23, Ursa Minor G26, and Ursa Minor W23, respectively. Similarly, for the MP population, we adopt the ranges [-3, -1.9], [-3, -2.3], and [-3, -2.4] for the fits to Draco W23, Ursa Minor G26, and Ursa Minor W23, respectively.}
    \label{tab:placeholder}
\end{table*}

\end{document}